\documentclass[twocolumn,trackchanges]{aastex63}
\usepackage{graphicx}
\usepackage{subfigure}
\usepackage[all]{hypcap}
\usepackage{symboldef}
\usepackage{xcolor}
\usepackage{amssymb}
\usepackage{amsmath}
\usepackage{numdef}

\num\def\m0451{m0451}
\num\def\m2129{m2129}
\num\def\m1423{m1423}
\num\def\m1341{e1341}

\received{September 2, 2020}
\revised{May 31, 2021}
\accepted{June 10, 2021}
\submitjournal{ApJ}

\shorttitle{An exquisitely deep view of quenching galaxies through the gravitational lens}
\shortauthors{Man, Zabl, Brammer et al. 2021}

\begin{document}

\title{An exquisitely deep view of quenching galaxies through the gravitational lens: \\ Stellar population, morphology, and ionized gas}

\correspondingauthor{Allison W. S. Man}
\email{aman@phas.ubc.ca, allisonmanws@gmail.com}

\author[0000-0003-2475-124X]{Allison W. S. Man}
\affiliation{Dunlap Institute for Astronomy and Astrophysics, University of Toronto, 50 St George Street, Toronto ON, M5S 3H4, Canada}
\affiliation{Department of Physics \& Astronomy, University of British Columbia, 6224 Agricultural Road, Vancouver BC, V6T 1Z1, Canada}

\author[0000-0002-9842-6354]{Johannes Zabl}
\affiliation{Univ Lyon, Univ Lyon1, Ens de Lyon, CNRS, Centre de Recherche Astrophysique de Lyon UMR5574, F-69230 Saint-Genis-Laval, France}
\affiliation{Institute for Computational Astrophysics and Department of Astronomy \& Physics, Saint Mary's University, 923 Robie Street, Halifax NS, B3H 3C3, Canada}

\author[0000-0003-2680-005X]{Gabriel B. Brammer}
\affiliation{Cosmic Dawn Center (DAWN)}
\affiliation{Niels Bohr Institute, University of Copenhagen, Jagtvej 128, DK-2100 Copenhagen, Denmark}

\author[0000-0001-5492-1049]{Johan Richard}
\affiliation{Univ Lyon, Univ Lyon1, Ens de Lyon, CNRS, Centre de Recherche Astrophysique de Lyon UMR5574, F-69230 Saint-Genis-Laval, France}

\author[0000-0003-3631-7176]{Sune Toft}
\affiliation{Cosmic Dawn Center (DAWN)}
\affiliation{Niels Bohr Institute, University of Copenhagen, Jagtvej 128, DK-2100 Copenhagen, Denmark}

\author[0000-0001-5983-6273]{Mikkel Stockmann}
\affiliation{Cosmic Dawn Center (DAWN)}
\affiliation{Niels Bohr Institute, University of Copenhagen, Jagtvej 128, DK-2100 Copenhagen, Denmark}

\author[0000-0002-9656-1800]{Anna R. Gallazzi}
\affiliation{INAF - Osservatorio Astrofisico di Arcetri, Largo Enrico Fermi 5, I-50125 Firenze, Italy}

\author[0000-0003-1734-8356]{Stefano Zibetti}
\affiliation{INAF - Osservatorio Astrofisico di Arcetri, Largo Enrico Fermi 5, I-50125 Firenze, Italy}

\author[0000-0001-8249-2739]{Harald Ebeling}
\affiliation{Institute for Astronomy, University of Hawai`i, Honolulu, HI 96822, USA}

\begin{abstract}

This work presents an in-depth analysis of four gravitationally lensed red galaxies at $z=1.6-3.2$.
The sources are magnified by factors of 2.7 -- 30 by foreground clusters,
enabling spectral and morphological measurements that are otherwise challenging.
Our sample extends below the characteristic mass of the stellar mass function
and is thus more representative of the quiescent galaxy population at $z>1$ than previous spectroscopic studies. 
We analyze deep VLT/X-SHOOTER spectra and multi-band \textit{Hubble Space Telescope} photometry that cover the rest-frame UV-to-optical regime.
The entire sample resembles stellar disks as inferred from lensing-reconstructed images.
Through stellar population synthesis analysis we infer that the targets are young (median age = 0.1 -- 1.2\,Gyr) and formed 80\% of their stellar masses within 0.07 -- 0.47\,Gyr.
\mgii\,$\lambda\lambda\,2796,2803$ absorption is detected across the sample.
Blue-shifted absorption and/or redshifted emission of \mgii\ is found in the two youngest sources,
indicative of a galactic-scale outflow of warm ($T\sim10^{4}$\,K) gas.
The \oiii\,$\lambda5007$ luminosity is higher for the two young sources (median age less than 0.4\,Gyr) than the two older ones, perhaps suggesting a decline in nuclear activity as quenching proceeds.
Despite high-velocity ($v\approx1500$\,\kms) galactic-scale outflows seen in the most recently quenched galaxies, warm gas is still present to some extent long after quenching.
Altogether our results indicate that star formation quenching at high-redshift must have been a rapid process ($<1$\,Gyr) that does not synchronize with bulge formation or complete gas removal. Substantial bulge growth is required if they are to evolve into the metal-rich cores of present-day slow-rotators.

\end{abstract}

\section{Introduction}\label{sec:intro}

Understanding how and when galaxies form stars is a cornerstone of galaxy evolution studies.
Much effort has been dedicated to studying the most massive galaxies,
as they are the brightest and thus easiest to observe.
The most massive galaxies (with stellar masses a few times the characteristic mass of the stellar mass function) by nature have complex formation and accretion histories.
The early, rapid growth of massive galaxies is driven by high gas accretion rates, 
fueling in-situ star formation at $z\gtrsim2$,
followed by a later, longer phase of mass growth characterized by frequent merging with satellite galaxies
\citep{Oser2010,Lackner2012,Hirschmann2015,RGomez2016,Wellons2016}.
This two-phase scenario provides a decent explanation for a broad set of observations of nearby, massive, early-type galaxies,
including their stellar light profiles \citep{DSouza2014},
element abundance ratios \citep{Thomas2005}, 
stellar metallicity and age gradients \citep{Greene2013,Ferreras2019b,Zibetti2020},
and diverse stellar velocity fields \citep{Emsellem2011,Krajnovic2013,Veale2017a}.

The evolution of intermediate-mass galaxies (with stellar masses near the characteristic mass of the stellar mass function), on the other hand, is less constrained by observations than massive galaxies.
The number density of quiescent galaxies declines steeply with stellar mass beyond the characteristic mass \citep[e.g.,][]{Muzzin2013b,Davidzon2017},
i.e., intermediate-mass galaxies are orders of magnitude more numerous than the most massive ones.
The relative dominance of in-situ star formation and ex-situ mass accretion in intermediate-mass galaxies is a matter of contention amongst various simulations \citep{Oser2010,Lackner2012,RGomez2016,Moster2019}.
Intermediate-mass quiescent galaxies at $z>2$,
observed shortly after these galaxies shut down their star formation,
can thus be used to discern between different star formation and feedback models.

Related to understanding how galaxies form stars,
an outstanding question in galaxy evolution studies is why galaxies experience a decline in star formation,
a process loosely referred to as ``quenching''.
Spectroscopic confirmation of massive, quiescent galaxies at $z\gtrsim2$ provides evidence for fast quenching of \textit{individual} galaxies \citep[e.g.,][]{Glazebrook2017,Forrest2020,Stockmann2020,Valentino2020}.
Their high stellar masses (log(\mstar/\msun)$\gtrsim 11$) imply that they had a much higher star formation rate in the past,
followed by a rapid decline in star formation rate that creates the absorption lines typical of post-starburst galaxies without emission of OB-type stars.
Active galactic nucleus (AGN) feedback is a common explanation for fast quenching, at least in theory:
supermassive blackhole accretion creates winds that expel gas from galaxies (momentum injection) and/or heat gas (energy injection) leaving little cold gas available for further star formation \citep[][and references therein]{Alexander2012,Fabian2012}.
Because of the complex, multi-scale nature of blackhole-gas interaction,
AGN feedback is amongst the most debated topic in galaxy studies and consequently its implementations vary drastically across simulations \citep{DMatteo2005,Springel2005,Dubois2012,Sijacki2015,Croton2016,Bower2017}.

On the other hand,
a \textit{population} of galaxies is said to have quenched as the fraction of red (passive or quiescent) galaxies becomes higher towards lower redshifts \citep[e.g.,][]{Peng2010,Muzzin2013b,Davidzon2017}.
While there is certainly a causal relation between the star formation histories and quenching of \textit{individual} and a \textit{population} of galaxies,
they are not equivalent to each other as the latter involves also population changes (e.g., formation of blue galaxies that cross a mass threshold, mergers).
The two different methodologies result from the empirical nature of galaxy evolution studies:
galaxies evolve on such long timescales that astronomers must infer their evolution either by reconstructing their star formation and assembly histories through spectral features or population changes.
The former approach requires deep spectroscopy and is thus limited to small samples,
while the latter approach is less accurate in separating star-forming galaxies from quiescent ones but bears merit of larger samples that span a range of environments.
The two are different yet complementary approaches to understanding how galaxies form stars and quench.

The question of what quenches galaxies can be posed another way:
what explains the diversity of star formation histories among galaxies?
Massive, quiescent galaxies formed the bulk of their stars earlier on than less massive ones in general \citep{Thomas2005,Thomas2010,Gallazzi2006,DSanchez2016,Chauke2018,Wu2018}.
Star-forming galaxies have a more extended duration of star formation than quiescent galaxies at later times at a given mass \citep{Ferreras2019a}.
It remains debated whether quenching is triggered by an event (e.g., AGN) or it is simply due to the exhaustion of star-forming fuel.
Cosmological simulations in general require an energy source like AGN to quench star formation and to prevent galaxies from forming too many stars too early and efficiently \citep[][and references therein]{Su2019}.
Some studies suggest that feedback can be provided by other sources like old stellar populations, i.e., Type Ia supernovae \citep{Matthews1990,Ciotti1991} or asymptotic giant branch stars \citep{Conroy2015}.
Alternatively gas can be prevented from cooling sufficiently by virial shocks \citep{Birnboim2003} or gravitational heating \citep{Khochfar2008}.
Other models suggest that galaxy star formation can be made inefficient in presence of a stellar bar or bulge \citep{Martig2009, Khoperskov2018}.
A large number of quenching mechanisms have been proposed in literature \citep[see][for an extensive list]{Man2018}.
The answer as to what quenches star formation in galaxies likely depends on the timescale (early or late cosmic epochs, fast or slow) as well as the conditions of the galaxies (mass, environment, age of stellar populations).
To distinguish between the many quenching mechanisms proposed, we need more constraints on quiescent galaxies beyond what is already known from correlations. 

The most stringent constraints of distant quiescent galaxies come from deep spectroscopy.
Absorption line strengths and the overall continuum shape provide valuable information on the age and metallicity of the integrated stellar population of galaxies, and their star formation histories.
Several hours of integration time with eight-meter class telescopes are generally needed to secure detections of multiple absorption lines for constraining stellar populations.
The advent of sensitive near-infrared spectrographs has enabled absorption line studies for several tens of $z>1.5$ quiescent galaxies \citep{vDokkum2009,Onodera2010,Onodera2012,vdSande2011,vdSande2013,Toft2012,Bezanson2013,Belli2014b,Belli2017,Kriek2009,Kriek2016,Glazebrook2017,KFong2017,Schreiber2018,ECarpenter2019,DEugenio2020,Forrest2020,Stockmann2020,Valentino2020}. 
These results have confirmed the existence of distant quiescent galaxies with low star formation rates (through weak or non-detections of \halpha\ emission and \oii), evolved stellar ages ({$\approx0.5-1$\,Gyr), and in some cases high stellar metallicity.
Despite the representative nature of intermediate-mass galaxies and their potential to constrain galaxy evolution models,
no deep spectroscopic observations exist for quiescent galaxies at $z>2$ around the characteristic masses to date.

Gravitational lensing is a promising way to deepen our understanding of the faint, compact quenching galaxies that are elusive.
Foreground galaxy clusters can magnify background galaxies and amplify their fluxes,
enabling us to obtain higher signal-to-noise (S/N) spectra within reasonable integration time for these otherwise faint galaxies.
This allows us to probe intrinsically fainter galaxies that are more representative of the quiescent galaxy population than the most luminous ones \citep[e.g.,][]{Stockmann2020}.
Lensed galaxies can also be magnified,
so we can achieve a more resolved view of the starlight of these compact galaxies that are barely resolvable with the  \textit{Hubble Space Telescope (HST)}.
The increasing rarity of quiescent galaxies towards higher redshift implies that cluster-lensed quiescent galaxies are uncommon, typically less than one per galaxy cluster.
Only nine of such galaxies have been discovered to date \citep{Muzzin2012,Geier2013,Newman2015,Newman2018a,Newman2018b,Hill2016,Toft2017,Ebeling2018,Akhshik2020}.
The most remarkable finding is the confirmation of a rotating stellar disk that has ceased star formation \citep{Toft2017,Newman2018b}.

We have systematically searched for lensed quiescent galaxies behind galaxy clusters for spectroscopic follow-up.
This Paper presents the analysis of a pilot sample and is structured as follows.
\S\ref{sec:data} presents the target selection and provides a description of each source.
\S\ref{sec:phot} presents the details of the photometric measurements.
\S\ref{sec:lensmodel} presents the details on the lens model and source reconstructions.
In \S\ref{sec:reduction} we present the procedures for reducing the VLT/\xs\ spectra.
\S\ref{sec:specfit} describes the procedures for the stellar population synthesis fitting.
In \S\ref{sec:results} we present the results of our analysis,
while in \S\ref{sec:discuss} we discuss the broader implications of our findings.
\S\ref{sec:conclusions} summarizes our conclusions.

All magnitudes are quoted in the AB system \citep{Oke1983}, unless otherwise stated.
The common logarithm with base 10 is used unless otherwise stated.
We assume a \citet{Kroupa2001} initial mass function.
A cosmology of $H_{0}$ = 70 km s$^{-1}$ Mpc$^{-1}$, $\Omega_\mathrm{M}$ = 0.3 and $\Omega_{\Lambda}$ = 0.7 is adopted throughout the Paper.

\section{Sample} \label{sec:data}

The objective of our survey is to identify and to conduct spectroscopic follow-up of lensed, distant quiescent galaxies.
We searched the \textit{HST} archive for galaxy cluster imaging, and identified red galaxies with broad-band colors consistent with quiescent stellar populations at $z>1$, that are gravitationally lensed by foreground clusters and bright enough for spectroscopic follow-up.
Our spectroscopic observation campaign spanned over 12 semesters and we refined our selection criteria over time.
In general we use infrared imaging to identify red sources bright enough for spectroscopic follow-up and to construct reliable lens models.
\m2129\ was selected using a distant red galaxy criterion $(J-K_{\mathrm{S}})>1.36$ \citep{Franx2003} based on VLT/ISAAC imaging as described in \citet{Geier2013}.
\m0451\ was identified to be red in its \hst/WFC3 color ($J_{\mathrm{F110W}}-H_{\mathrm{F160W}}=1.35$) and known to be at $z\approx3$ based on the lensing configuration.
\m1423\ was selected because of its high photometric redshift (\zphot\,$>3$) and low specific star formation rate (log(sSFR)\,$\approx-10$) based on spectral energy distribution fitting of the CLASH data,
although these numbers have been revised with the spectroscopic analysis.
\m1341\ was discovered from an \hst\ Snapshot program of massive clusters (PI: Ebeling) as a spectacular candidate:
the target is red based on its Gemini/GMOS ($g'$, $r'$, $i'$) and \hst/WFC3 (F110W, F140W) images. Its spectral energy distribution is consistent with being a quiescent galaxy at $z\approx1.35$ that is triply lensed with one image close to the critical line and has extremely high magnification \citep{Ebeling2018}.
This Paper presents the analysis of our pilot sample,
comprising four lensed quiescent galaxies with sufficiently high S/N spectra for absorption line analysis.
The color images of the sample are shown in Figure~\ref{fig:colorim}. 
The targets were observed with the X-SHOOTER spectrograph \citep{Vernet2011} at the Very Large Telescope.
The observations and data reduction procedures are described in \S\ref{sec:reduction}.
Below we briefly discuss the individual properties of each target, 
and provide an overall summary of the lensing properties in Table~\ref{table:lens}.

\begin{table*}[htbp!]
	\caption{Gravitational lensing properties}
	\label{table:lens}
    \centering
	\begin{tabular}{lccccc}
		\hline
		& \m1341 & \m2129 & \m0451 & \m1423 \\
		\hline
        \zspec & 1.5954 & 2.1487 & 2.9223 & 3.2092 \\
		Lensing cluster & eMACSJ1341.9-2442 & MACSJ2129.4-0741 & MACSJ0454.1-0300 & MACSJ1423.8-2404  \\
        & & & (a.k.a. MS0451.6-0305) & \\
        \zlens & 0.835 & 0.5889 & 0.5377 & 0.5431 \\
        Magnitude & \underline{18.5}, 19.8, 20.4 & \underline{20.1} & \underline{20.9}, 21.8, 22.0 & \underline{22.1} \\
        $\mu$ & \underline{$30\pm8$}, $12.6\pm1.5$, $8.2\pm0.9$ & \underline{$4.6\pm0.2$} & \underline{$10.9\pm2.1$}, $6.1\pm0.6$, $3.8\pm0.3$ & \underline{$2.7\pm0.2$} \\
        Mass model reference & \citet{Ebeling2018} & \citet{Toft2017} & \citet{Jauzac2020} & \citet{Limousin2010} \\
        rms & $0\farcs45$ & $0\farcs60$ & $0\farcs73$ & $0\farcs82$ \\
		\hline
	\end{tabular}
	\tablecomments{The total AB magnitudes as measured in \hst\ F160W (or F140W for \m1341) are reported.
	The lensing magnification factors ($\mu$) scale linearly with flux.
	The underlined values indicate the spectroscopic targets.
	The rms refers to the root-mean-square of the observed image positions of the multiple images for each lensing system.
	}
\end{table*}

\begin{figure*}[!htbp]	
   \begin{minipage}{0.5\linewidth}
	\centering
   \includegraphics[width=\linewidth]{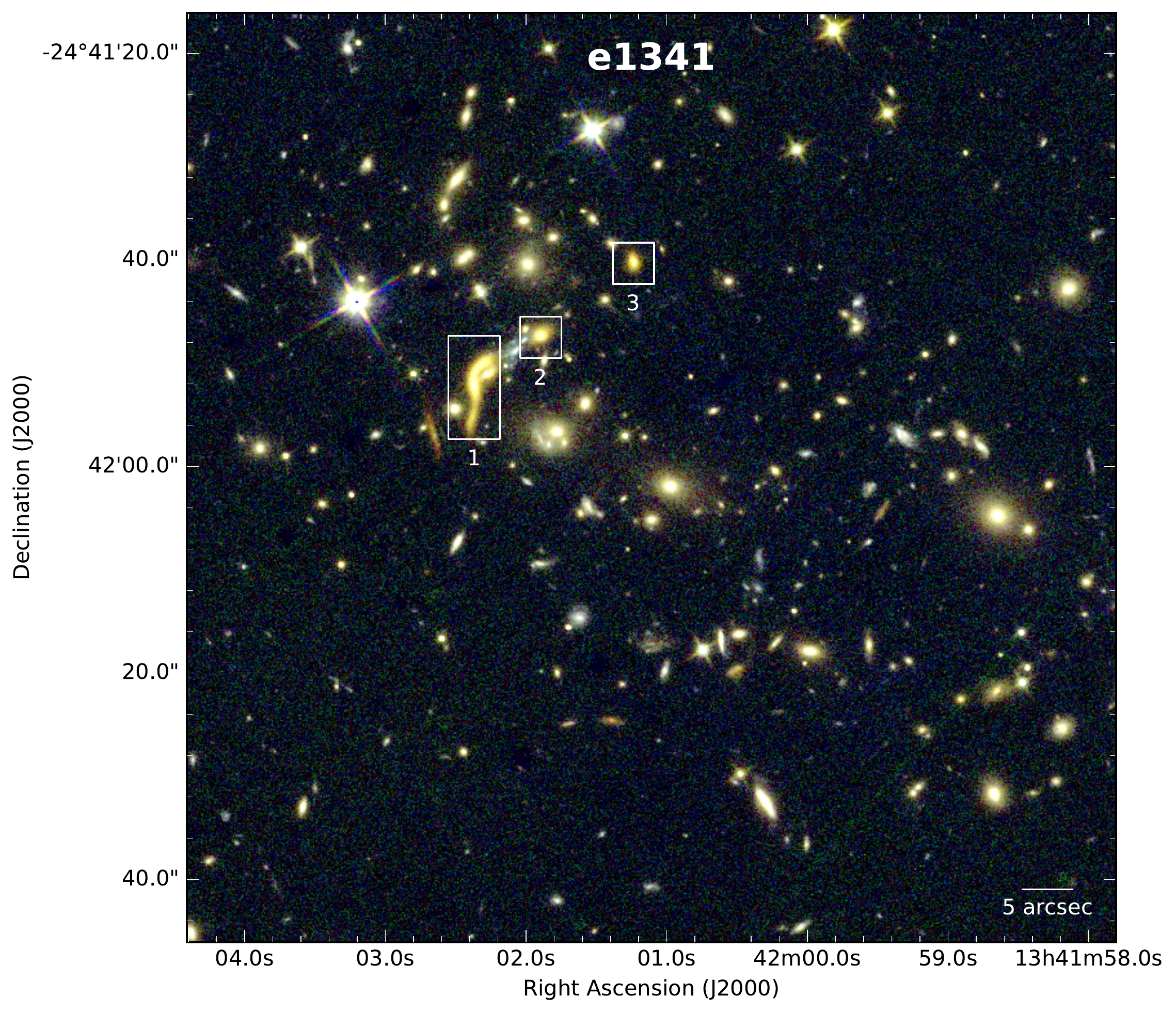}
	\end{minipage}
   \begin{minipage}{0.49\linewidth}
	\centering
   \includegraphics[width=\linewidth]{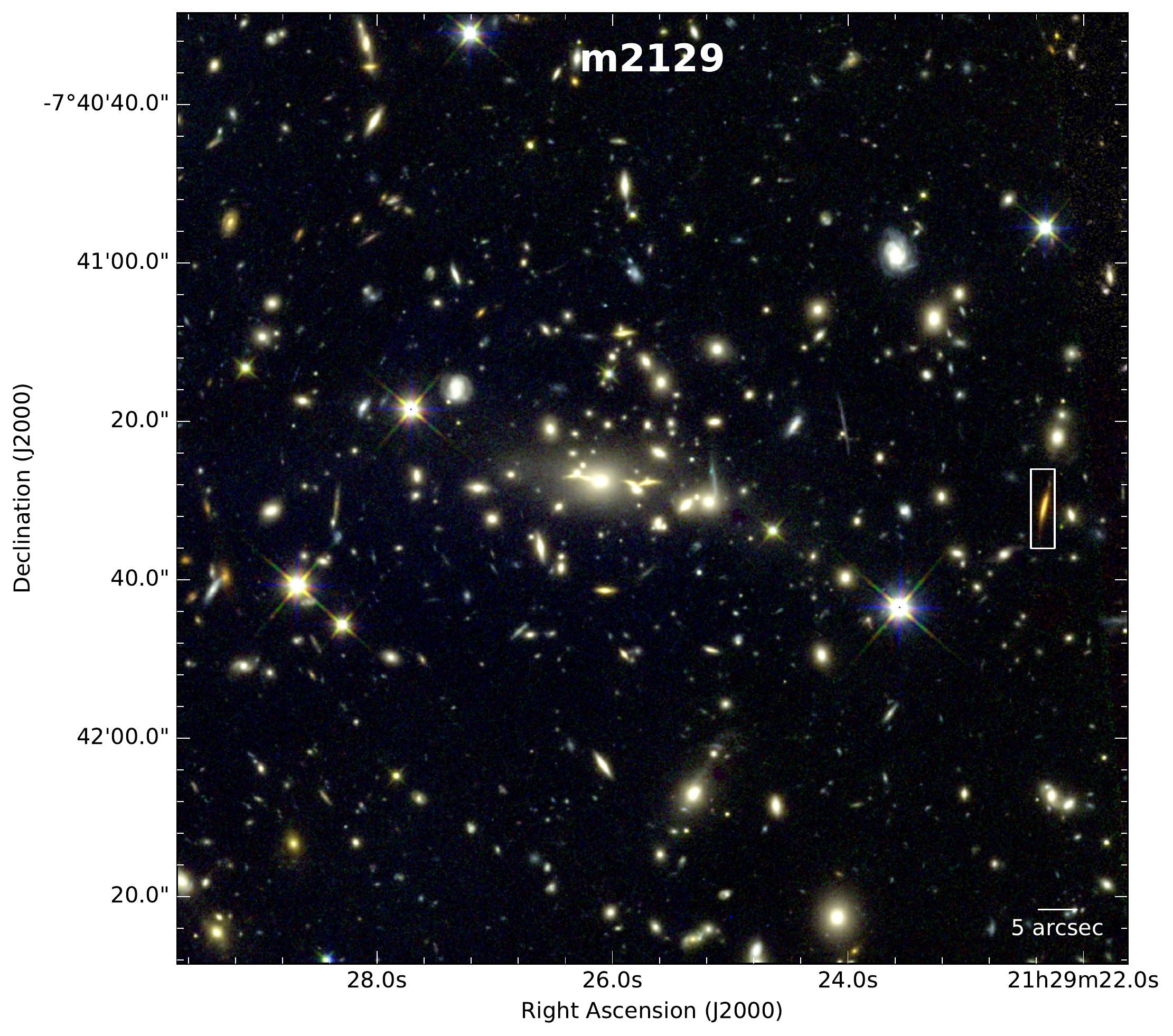}
	\end{minipage}
   \begin{minipage}{0.48\linewidth}
	\centering
   \includegraphics[width=\linewidth]{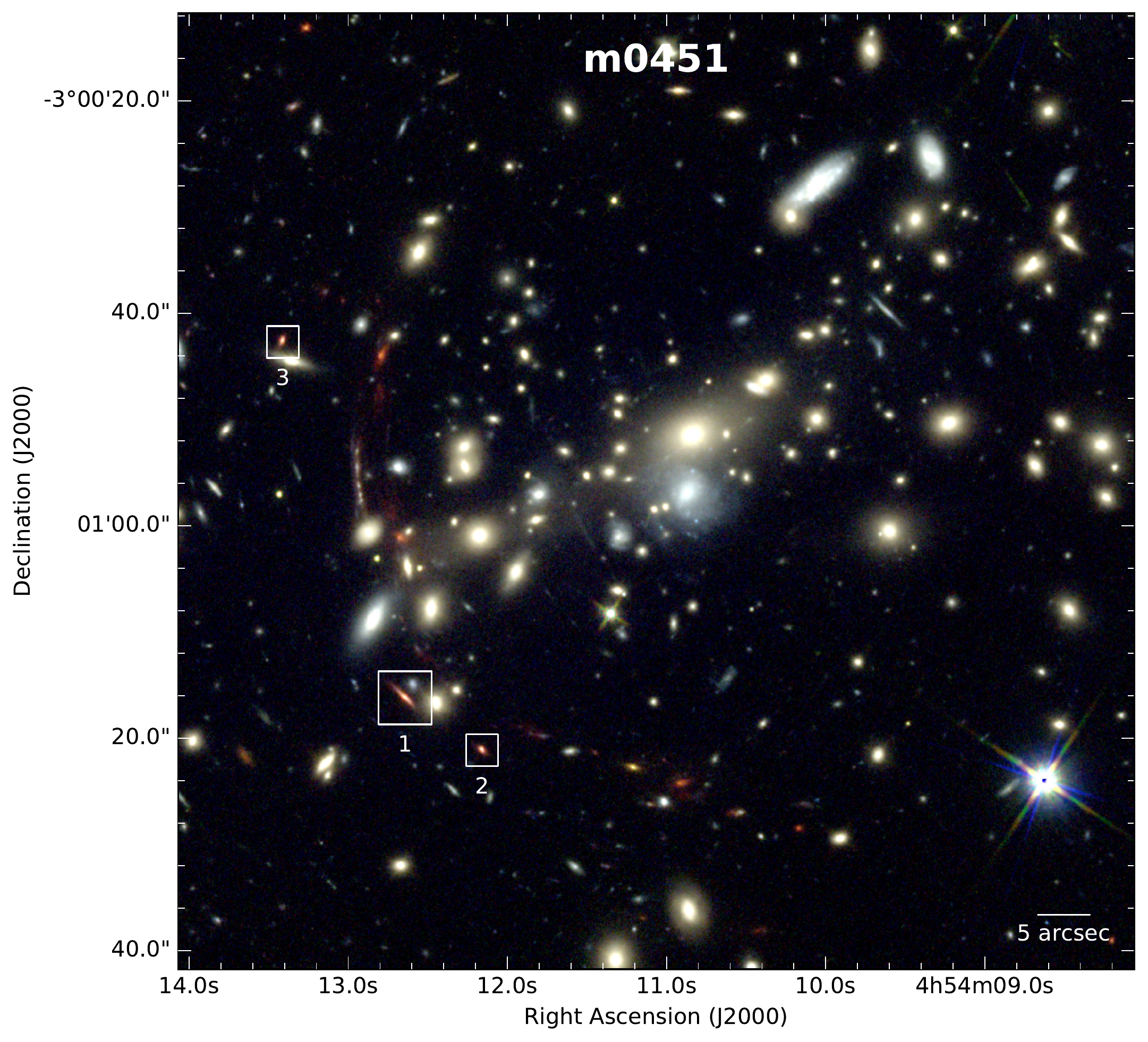}
	\end{minipage}
    \begin{minipage}{0.5\linewidth}
	\centering
   \includegraphics[width=\linewidth]{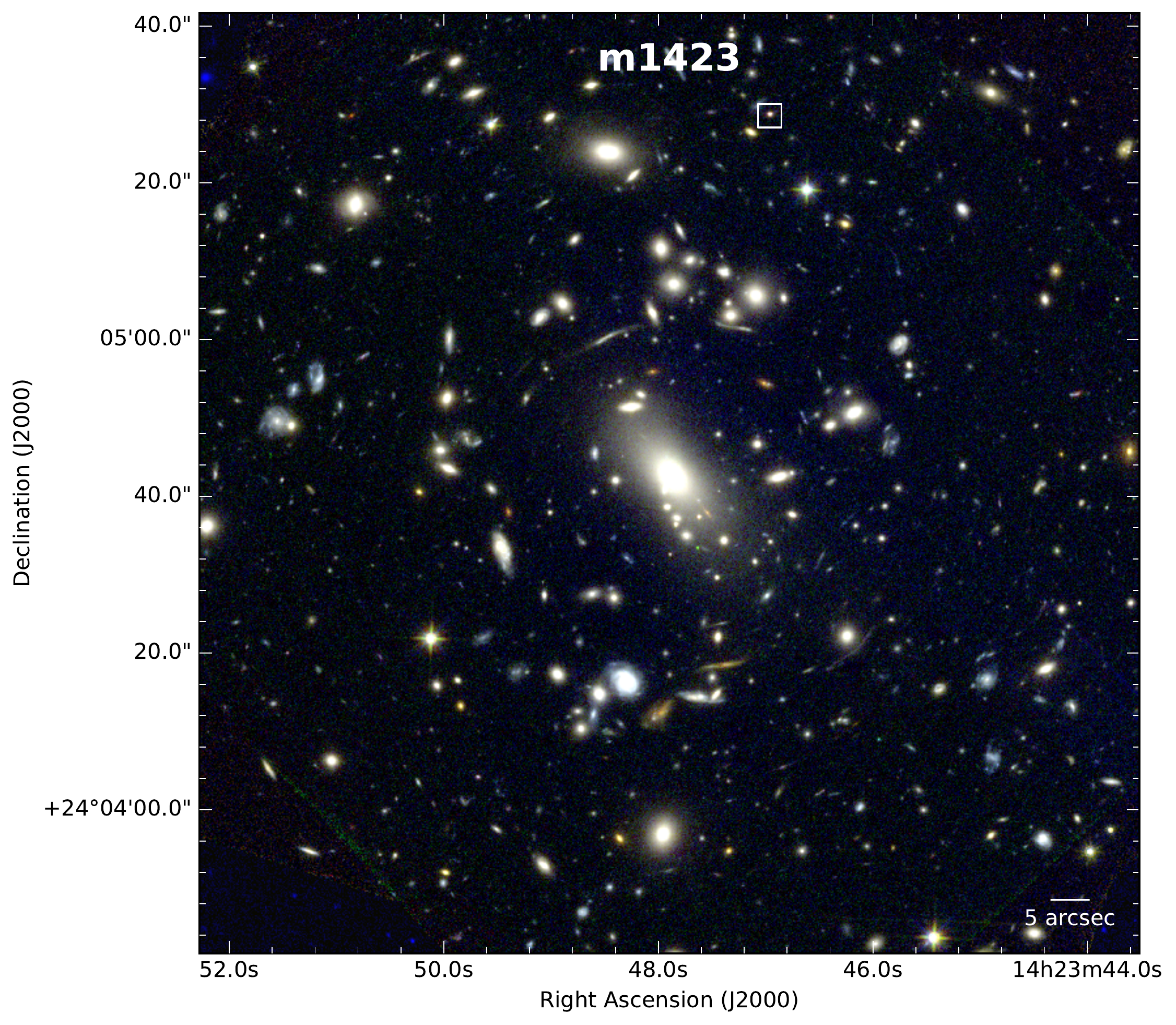}
	\end{minipage}
    \caption{\hst\ color images of the four spectroscopic targets and their respective foreground lensing clusters. The lensed quiescent galaxies are shown with white rectangles. The composite images are made using F140W for \m1341\ or F160W for other targets (red), F110W (green), and F814W (blue) on a logarithmic color scale. The images measure 90$\arcsec$ (\m1341\ and \m0451) or 120$\arcsec$ (\m2129\ and \m1423) on each side.}
    \label{fig:colorim}
\end{figure*}

\subsection{Targets}\label{sec:targets}

\subsubsection{e1341}
\m1341\ is a triply imaged quiescent galaxy with the highest magnification factor known to date \citep{Ebeling2018}.
The \xs\ spectrum presented in this paper is based on observations conducted on 2017 Jul 24, 25, 26, 27, Aug 6, 2018 Mar 21, 28 under ESO programme ID 099.B-0912 (PI: M. Stockmann).

\subsubsection{m2129}
\m2129\ is a singly lensed massive galaxy at $z=2.1$ with quiescent SFR.
Follow-up deep absorption line spectroscopy with VLT/X-SHOOTER was first presented in \citet{Geier2013}.
A more recent analysis of the stellar kinematics shows that m2129 is the first quiescent galaxy at $z>2$ confirmed to have rotation-dominated stellar kinematics \citep{Toft2017,Newman2018b}.
The analysis presented here is based on the \xs\ observations conducted on 2011 May 12, 14 and Sep 5, 14 under ESO programme ID 087.B-0812 (PI: S. Toft) as also used in \citet{Geier2013} and \citet{Toft2017}.

\subsubsection{m0451}
\m0451\ is situated in a compact group at $z\approx2.9$ with at least eight members separated within $<100$\,kpc in projection \citep{MacKenzie2014,Shen2021}.
The quenched galaxy studied in this work,
with star-forming companion galaxies,
are triply lensed and together they form a giant submillimeter arc with a total unlensed SFR of $\approx(450\pm50)$\,\msun\peryr.
The system was originally identified by \citet{Takata2003}, 
and the lensing model was presented in \citet{Zitrin2011} as multiple system 4. 
We adopt a magnification factor of $\mu=10.9\pm2.1$ for our spectroscopic target as estimated by the latest lens model \citep[][ID G.2]{Jauzac2020}.
The \xs\ spectrum analyzed in this work is based on the observations conducted on 2011 Sep 11, 13, 2015 Dec 5, 9, and 2016 Jan 9 under ESO programmes ID 087.B-0812 and 096.B-0994 (PI: S. Toft).

\subsubsection{m1423}
\m1423\ is a singly lensed system.
The \xs\ spectrum presented here is based on observations conducted on 2014 Apr 18, May 9, 10, June 5, 8, 2017 Mar 26, Apr 12, Jul 1, 2, Aug 5, 6, 2018 Feb 21, 22, 23, Mar 12, under ESO programmes ID 093.B-0815 (PI: A. Man) and 097.B-1064 (PI: M. Stockmann).

\section{Method} \label{sec:method}
\begin{table*}[t]
	\centering
	\caption{VLT/\xs\ observation log}
	\label{table:obs_log}
	\begin{tabular}{lccccc}
		\hline
	     & \m1341\ (im1) & \m2129\ & \m0451\ (im1) & \m1423\ &  \\
		\hline
		RA (J2000) & $\mathrm{13^{h}42^{m}02^{s}.374}$ & $\mathrm{21^{h}29^{m}22^{s}.34}$ & $\mathrm{04^{h}54^{m}12^{s}.65}$  & $\mathrm{14^{h}23^{m}46^{s}.97}$ \\ 
		DEC (J2000) & $\mathrm{-24^{d}41^{m}52^{s}.11}$ & $\mathrm{-07^{d}41^{m}31^{s}.06}$ & $\mathrm{-03^{d}01^{m}16^{s}.50}$ & $\mathrm{+24^{d}05^{m}28^{s}.52}$ \\ 
		Integration time (h) for UVB/VIS/NIR &  4.14/4.43/4.80 & 2.29/1.74/2.67 & 4.12/4.36/7.87 & 11.64/12.04/13.33 \\ 
		S/N (\AA$^{-1}$) & 11.6 & 6.3 & 15.5 & 9.4 \\
		Slit angle (North to East) & 10\degree & $-10$\degree & 35\degree & $-80$\degree, $-10$\degree \\
		\hline
	\end{tabular}
	\tablecomments{
	Positions are based on F160W, or F140W if the former is unavailable.
	The listed integration times only include useful data and not discarded data.
	The S/N per \AA\ (rest-frame) is estimated from the rest-frame wavelength region between 4110 and 4210\,\AA\ just redward of the \hdelta\ feature, with clipping against strong skylines and assuming the same correction for correlated noise following the procedures as done in \citet{Toft2017}. 
	}
\end{table*}

\subsection{Photometry}\label{sec:phot}
\subsubsection{HST Photometry}

All targets are selected from the clusters identified from the (extended) MAssive Cluster Survey \citep[MACS and eMACS;][]{Ebeling2001,Ebeling2007,Ebeling2010} targeting X-ray luminous, high-redshift clusters.
Two of the clusters, MACS2129 and MACS1423, are part of the Cluster Lensing And Supernova survey with Hubble \citep[CLASH;][]{Postman2012}.
MS0451-03 has 6-band \textit{HST} imaging available through the \textit{Herschel} Lensing Survey \citep{Egami2010,Jauzac2020}.

The \hst\ imaging observations are processed with the \texttt{grizli} analysis software \citep{Brammer2019}\footnote{\url{https://github.com/gbrammer/grizli}}.  Briefly, we first retrieve all available \hst\ Advanced Camera for Surveys (ACS) and Wide-Field Camera 3 (WFC3) images described above that cover a given target.  These observations are divided into associations defined as sets of exposures of a target in a given instrument and filter combination obtained at a single epoch (i.e., a ``visit'' in the \textit{Hubble} observation planning parlance).  We align the absolute astrometry of each visit to stars in the GAIA DR2 catalog \citep{GAIADR2} accounting for proper motions between the GAIA and \hst\ observation epochs.  Finally, we create mosaics in each instrument/filter combination using the \texttt{AstroDrizzle} software package, where we use the dithered exposures to identify additional cosmic rays and hot pixels with \texttt{AstroDrizzle} parameters similar to those used by \cite{Momcheva2016}.  All filters of a given target are drizzled to a common $0\farcs06$ pixel grid.

The total magnitude of the reddest \hst\ filter is measured from the automatic aperture with the \texttt{SExtractor} code \citep{Bertin1996}.
Color magnitudes are measured from small apertures of diameter $1\farcs2$ to optimize S/N, and then scaled to match the total magnitude of the reddest filter.
We include the correlated noise in the noise budget introduced by \texttt{MultiDrizzle}, following the procedure described in \citet{Casertano2000}.
The photometry is listed in Table~\ref{table:phot}.

\subsubsection{IRAC and WISE Photometry}

All but one target have \textit{Spitzer}/IRAC 3.6\,\um\ and 4.5\,\um\ observations from the SURFSUP survey \citep{Bradac2014,Huang2016}.
Given the significantly coarser spatial resolution of IRAC (point spread function full-width half-maximum PSF FWHM $\approx1\farcs7$ for channels 1 and 2) compared to \hst,
and that source blending is severe due to the high source density in cluster environments,
aperture photometry is inadequate to measure IRAC fluxes.
Here we use the \texttt{TPHOT} code version 2.0.8\footnote{\url{https://tphot.wordpress.com}} \citep{Merlin2015,Merlin2016} to obtain deblended IRAC flux measurements.
We use the reddest \hst\ images, F160W, as prior for the morphology and brightness to deblend the IRAC emission.
We first derive a convolution kernel to match the F160W images to the IRAC images based on their PSFs.
PSFs for in-flight IRAC observations are used.
The F160W images are then convolved with the respective kernels,
which are then fitted to the observed IRAC emission maps.
The best-fitting multiplicative scaling factor for each object is the deblended IRAC fluxes.
The IRAC fluxes and errors are listed in Table~\ref{table:phot}.
For \m1341\ we retrieve Wide-field Infrared Survey Explorer \citep[WISE; ][]{Wright2010} images from the AllWISE data release. 
WISE has an angular resolution of $6\farcs1$, $6\farcs4$, $6\farcs5$, and $12\farcs0$ at 3.4, 4.6, 12, and 22\,\um, with point-source sensitivity of 0.08, 0.11, 1, and 6\,mJy in unconfused regions ($5\sigma$).
We perform the deblending procedure with \texttt{TPHOT} as was done for IRAC,
using the reddest image, F140W, as prior.
The deblended WISE flux measurements are listed in Table~\ref{table:phot}.
We convert the deblended WISE intensities to magnitudes and fluxes in Jy using the zero points derived for a flat spectrum (i.e., $F_{\nu}$ is constant). 
In the AllWISE source catalog,
W3 is shallow and shows only a weak detection (S/N\,=\,3.3) detection at the position of im1 prior to deblending.
After running \textsc{TPHOT}, the W3 fluxes of all three multiple images have S/N\,$<1.5$ and are therefore individually undetected.
W4 has no significant detection at all image positions, and the \texttt{TPHOT} fit did not converge.
Thus we consider W3 and W4 non-detections given their shallow sensitivity.

\subsection{Magnifications and source reconstructions}\label{sec:lensmodel}

We use detailed mass models of each cluster to derive the intrinsic properties (magnification and source morphology) of each target. 
These parametric models are constrained by the location of strongly-lensed background systems using the software \texttt{Lenstool} \citep{Jullo2007}. For each cluster the optimization of the mass model is performed through a Monte-Carlo Markov Chain (MCMC) minimizing the root-mean-square (rms) between the observed and the predicted locations of multiple systems.
The magnification and error bars for the targets provided in Table~\ref{table:lens} are derived using both the best model (providing the best reproduction of the multiple images) as well as the full set of MCMC realizations sampling the posterior distribution of model parameters.
Details on these mass models are presented in the reference papers.
Specifically, 
eMACSJ1341.9-2442 has one multiple image system with five images \citep{Ebeling2018}. The lensed target \m1341\ is multiply imaged. 
MACSJ2129.4-0741 has two systems with nine images \citep{Toft2017}. The lensed target \m2129\ is just outside the region of constraints, located $11^{\prime \prime}$ away from the multiply imaged region.
MS0451.6-0305 has 17 systems with 49 images \citep{Jauzac2020}. The lensed target \m0451\ is multiply imaged.
MACSJ1423.8-2404 has 3 systems with 12 images \citep{Limousin2010}. The lensed target \m1423\ is $20^{\prime \prime}$ away from the multiply imaged region.
The rms of the observed locations of multiple systems is listed in Table~\ref{table:lens}.

The same mass models are used to reconstruct the intrinsic source morphology of each target. We directly ray-trace the \hst/WFC3 F160W image of each image onto a regular grid on the source plane, and perform the same reconstruction for a \hst/WFC3 PSF model to derive our spatial PSF on the source plane. These reconstructions were obtained for the full set of MCMC realizations to estimate the statistical errors in the source reconstructions.

The exception is for target \m1341: we used the closest \hst/WFC3 band in wavelength (F140W) as F160W imaging data are not available.
The lensing arc, im1, of \m1341\ is additionally galaxy-galaxy lensed so that the reconstructed image does not cover the entire source once reconstructed.
Therefore only im2 and im3 are reconstructed for \m1341.

\subsection{Spectroscopic observations and data reduction} \label{sec:reduction}

\begin{figure*}[htbp!]
	\includegraphics[width=\textwidth]{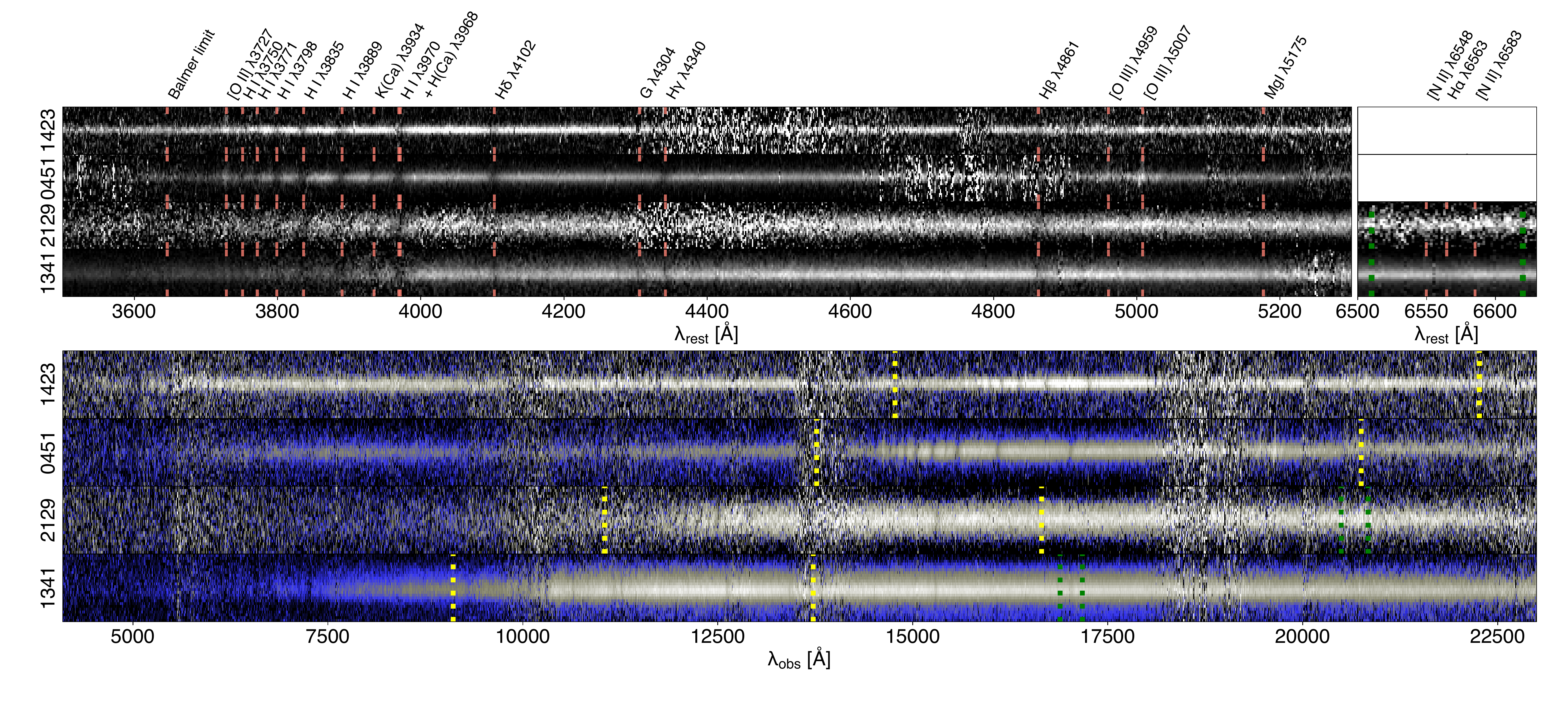}
    \caption{Rectified 2D \xs\ spectra in rest-frame (top panel) and observed frames (bottom panel), ordered by descending redshifts from top to bottom. 
    The top panel shows the spectra in linear grayscale for the rest-frame wavelength range from 0.35 to 0.53\,\micron\ (and for \m1341\ and \m2129\ also from 0.650 to $0.663\,\micron$). 
    The bottom panel shows the spectra for the observed wavelength range from 0.41 to 2.3\,\micron. 
    The color-scaling has three separate linear parts (black-blue / blue-grey / grey-white) such that both high-and low surface brightness parts are visualized. 
    Important lines are annotated in the top panel and shown as vertical dashed orange lines. 
    The ranges shown in the top panel are indicated in the bottom panel by the vertical yellow and green dotted lines. 
    All spectra are shown for a spatial extent of $\pm2\farcs5$. 
    The outermost parts are affected by the nodding reduction and therefore do not fully represent the actual light profile there.}
    \label{fig:spec2d}
\end{figure*}

\begin{figure*}[htbp!]
	\includegraphics[width=\textwidth]{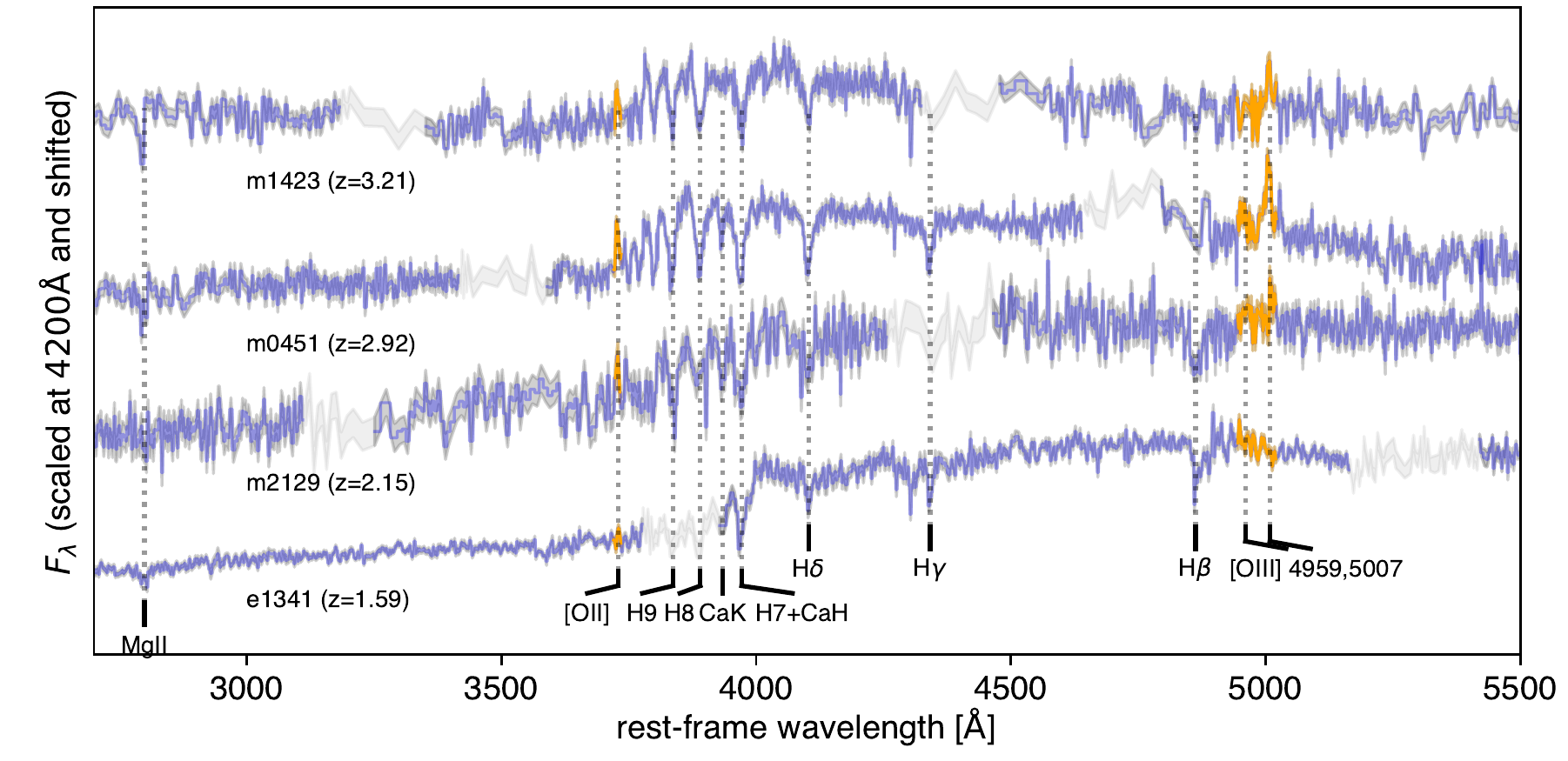}
    \caption{Extracted 1D \xs\ spectra of our sample,
    sorted by descending redshifts and increasing stellar population ages from top to bottom.
    Relevant features are denoted by vertical dotted lines.
    The regions near the \oii\ and \oiii\ emission lines are colored orange.
    The gray shades denote the flux errors.
    The spectra are shifted to their rest-frame, and scaled to the flux at 4200\,\AA\ and arbitrarily shifted upwards for visualization.
    The shown spectra are binned using a variance-weighted mean. The width of each bin is chosen so that the S/N per bin is similar over the full spectral range. The median bin size is rest-frame 2.9, 1.5, 1.9, 2.3$\,\textnormal{\AA}$ for \m1423, \m0451, \m2129, \m1341, respectively.
    \label{fig:spec1d}
    }
\end{figure*}

We used the panchromatic VLT/\xs\ Echelle spectrograph \citep{Vernet2011} to obtain stellar continuum spectra for our targets.
All spectra were taken with slit widths of $1\farcs0$ in the UVB arm, and $0\farcs9$ for the VIS and NIR arms.
This corresponds to nominal spectral resolution\footnote{The quoted $R$ values are strictly lower limits as these values are derived for sources whose emission completely fill the slit. Our sources are generally compact so $R$ can be slightly higher than stated. The improvement in $R$ is very small compared to the intrinsic velocity dispersion of our targets as stated in Table~\ref{table:stellarprop} and does not affect our conclusions in any significant way.} of $R \approx$ 5400, 8900, and 5600 for the UVB, VIS, and NIR arms, respectively.
The observation log is provided in Table~\ref{table:obs_log}.

The reduction procedures for \m2129\ are described in \citet{Toft2017}.
The other three targets were reduced using the \xs\ pipeline \citep[version 2.6.8; ][]{Modigliani2010}, supplemented by our customized scripts.
The reduction steps were as follows.
We used standard pipeline recipes up to the creation of the rectified Echelle orders (ORDER2D), with a few exceptions:
First, we removed in the UVB frames fixed pattern noise with a Fourier analysis, similar to the method described by \citet{Schonebeck2014}. Second, we removed a temporarily and spatially changing read-out pattern from the NIR frames, which manifests itself as a large-scale parabola-like pattern in the rectified frames. This pattern can be efficiently removed by subtracting the median of unexposed pixels in each detector row\footnote{Only detector rows covering the bluest NIR orders have enough unexposed pixels for this self-calibration to work. Fortunately, these are the orders most strongly affected by the issue.}. Third, the used pipeline version creates small-scale ring-like features around bad pixels during rectification. To avoid problems with these rings, we increased the masks around detected cosmic ray hits, a modification that we directly implemented in the pipeline \citep{Zabl2015}.

Instead of processing all exposures simultaneously, we created ORDER2D files from only a single pair of nodding positions at a time.
We stacked all nodding pairs for each order with our scripts and, subsequently, merged the order stacks to a single 2D spectrum.
In both steps, we weighted during the combining at each wavelength with the spatial median of the inverse pipeline variance.
This weighting is different from the pipeline in two ways: First, the pipeline does not apply weights in the combining of the exposures.
Second, during the order merging, the pipeline uses a pixel-based inverse variance weight, which we found to be somewhat problematic, as the pipeline variance estimate is a biased estimator.

Before stacking the nodding pairs, we corrected each of the 2D frames to a heliocentric standard and a vacuum wavelength scale. 
Flux calibration was done based on response functions determined from standard stars taken during the same nights as the science observations, or if no standard was taken during the same night, from the nearest available night.
Telluric corrections were derived for each science exposure using observations of telluric standard stars observed close in time and airmass.
In practice, we first determined an atmospheric model using \texttt{molecfit} \citep[version 1.5.1; ][]{Kausch2015,Smette2015} from the telluric standard stars and subsequently evaluated this model at the exact airmass of the science observations. 
Finally, a substantial part of our data was observed without a working atmospheric dispersion corrector.
Therefore, we implemented a software-based correction, which applies for the relevant exposures the appropriate wavelength-dependent spatial offset to the 2D frames and corrects approximately for the increased slit-loss. 

The 1D spectra were extracted directly from the stacked ORDER2D files and were subsequently merged with inverse variance weighting.
We did the extraction with a simple aperture extending from $-0\farcs5$ to $+0\farcs5$ around the pointing center.

\subsection{Stellar population fitting} \label{sec:specfit}

\begin{table*}[ht]
    \tiny
	\centering
	\caption{Stellar population properties}
    \label{table:stellarprop}
    \begin{tabular}{lcccccccccc}
	\hline
          & \zspec & log($\mu$\mstar/\msun) & log(sSFR/\peryr) & \AV & log(\Zstar/\Zsun)& 
          \sigmastar & $t_{0}$ & $\tau$ & \thalf & \twidth \\
	 & & & & & & (\kms) & (Gyr) & (Gyr) & (Gyr) & (Gyr) \\
	\hline
\m1341 &    1.5954 $\pm$ 0.0001 &     11.62 $\pm$ 0.07 &    -11.20 $\pm$ 0.46 &       0.4 $\pm$ 0.1 &     -0.12 $\pm$ 0.07 &       160 $\pm$ 9 & $1.43_{-0.08}^{+0.06}$ & $0.14_{-0.03}^{+0.01}$ & 1.19 $\pm$ 0.05 &      0.47 $\pm$ 0.07 \\
\m2129 &    2.1487 $\pm$ 0.0002 &     11.86 $\pm$ 0.05 &    -12.77 $\pm$ 1.83 &       1.1 $\pm$ 0.1 &     -0.10 $\pm$ 0.11 &       318 $\pm$ 28 & $0.68\pm0.06$ & $0.04_{-0.01}^{+0.02}$ & 0.61 $\pm$ 0.05 &      0.14 $\pm$ 0.05 \\
\m0451 &    2.9223 $\pm$ 0.0001 &     11.65 $\pm$ 0.06 &     -9.77 $\pm$ 0.08 &       0.9 $\pm$ 0.1 &     -0.30 $\pm$ 0.07 &       170 $\pm$ 12 & $0.42\pm0.03$ & $0.051\pm0.005$ & 0.33 $\pm$ 0.02 &      0.17 $\pm$ 0.01 \\
\m1423 &    3.2092 $\pm$ 0.0002 &     11.01 $\pm$ 0.07 &     -8.43 $\pm$ 0.10 &       0.8 $\pm$ 0.1 &     -0.01 $\pm$ 0.10 &       277 $\pm$ 19 & $0.15\pm0.01$ & $0.021_{-0.001}^{+0.002}$ &  0.12 $\pm$ 0.01 &      0.07 $\pm$ 0.01 \\
\hline 
Prior   & $\mathcal{N}(z_0, 0.001(1+z))$ &  $\mathcal{N}(11.3, 0.5)$  & --- & Uniform(0,2) & $Z/Z_\odot : \mathcal{N}(1,0.5)$ & $\mathcal{N}(\sigma_0, 50)$ & $\mathcal{N}(0.8, 0.3)$ & $\mathcal{N}(0.1, 0.2)$  & --- & --- \\
Limits &  --- & (10, 12.8) & --- & --- & $Z/Z_\odot$: (0.2, 2) & (100, 450) & (0.05, $t_\mathrm{univ}$) & (0.02, 0.3) & --- & --- \\
    \hline
      \end{tabular}
	\tablecomments{
	We report the median values and quote the average difference of 16 and 84 percentiles of the posterior distributions from the median values as uncertainties.
        See \S\ref{sec:specfit} for details on the derivation of these parameters.
    	Note that the uncertainties are propagated from photometric and spectral flux errors and do not include systematic uncertainties inherent to stellar population synthesis.
        A comparison of the derived parameters of \m2129\ with values published in \citet{Toft2017} and \citet{Newman2018a} is presented in Table~\ref{table:m2129}. The priors used in the population synthesis fits are indicated in the bottom two rows, where $\mathcal{N}(\mu, \sigma)$ indicates a Gaussian distribution with mean $\mu$ and standard deviation $\sigma$ .  Numbers $z_0$ and $\sigma_0$ are initial guesses based on basic template fits to the individual \xs\ spectra and the broad prior width is adopted there to allow substantial flexibility around those values.  There are implicit priors on sSFR, \thalf\ and \twidth\ arising from the priors SFH parameters $t_0$ and $\tau$ as specified.
		}
\end{table*}

\begin{table}[!htb]
	\centering
	\label{table:delensed_param}
    \caption{De-lensed sellar population properties}
	\begin{tabular}{lcccc}
		\hline
	    & log(\mstar) & log(SFR$_{100}$) & \mstar/\mchar$_{\mathrm{a}}$ & \mstar/\mchar$_{\mathrm{q}}$ \\
		\hline
    \m1341 & $10.15\pm0.14$ & $-1.0_{- 0.7}^{+ 0.3}$ & $0.2\pm0.1$ & $0.3\pm0.1$ \\ 
    \m2129 & $11.20\pm0.05$ & $-1.6_{- 2.5}^{+ 1.4}$ & $2.7\pm0.5$ & $2.3_{-0.7}^{+1.0}$ \\ 
    \m0451 & $10.61\pm0.10$ & $ 0.8\pm0.1$ & $0.4\pm0.1$ & $0.7_{-0.5}^{+0.4}$ \\ 
    \m1423 & $10.58\pm0.08$ & $ 2.2\pm0.1$ & $0.1\pm0.1$ & $0.6_{-1.0}^{+0.6}$ \\ 
		\hline
	\end{tabular}
	\tablecomments{
    Stellar population parameters for our sample after correcting for lensing magnification. The errors are propagated from the spectral fitting and the lens model.
	Masses and star formation rates are quoted in logarithmic units of \msun\ and \msun\peryr, respectively.
	We use the stellar mass function presented in \citet{Muzzin2013b} to estimate the characteristic masses of all galaxies (\mchar$_{\mathrm{a}}$) and those of quiescent galaxies only (\mchar$_{\mathrm{q}}$) at their respective redshifts.
	}
\end{table}

We model the \xs\ spectra using synthetic models from the \bagpipes\ population synthesis fitting code \citep{Carnall2018}, which uses the updated versions of the \cite{BC03} population synthesis models. 
Prior to the fitting,
we correct the spectra and photometry for galactic reddening following the approach of \citet{Schlafly2011},
using the \citet{Fitzpatrick1999} reddening law and $R_{V} = 3.1$ appropriate for high Galactic latitudes.

We adopt the delayed exponentially declining parameterization of the star formation history (SFH), 
\begin{equation}
    \mathrm{SFR} (t) \propto (t-t_{0}) e^{-(t-t_{0})/\tau}
    \label{eqt:sfh}
\end{equation}.  
To facilitate comparisons with literature measurements,
we also provide the median stellar age and star formation duration calculated from the SFH following \citet{Pacifici2016}, which are shown to be more robust than the age since the onset of star formation in the delayed-$\tau$ model \citep{Belli2019}.  The median stellar age, \thalf, is defined as the lookback time from the spectroscopic redshift \zspec\ at which 50\% of the \mstar\ has been assembled.
\twidth\ is defined as the duration between 10\% and 90\% of the \mstar\ being assembled.
The provided uncertainties of \thalf\ and \twidth\ are derived from posterior distributions of the SFH parameters. 

In addition to the SFH parameters, we fit for the systemic redshift, stellar velocity dispersion (\sigmastar), stellar mass normalization (\citealt{Kroupa2001} initial mass function), stellar metallicity (\Zstar/\Zsun), and dust attenuation (\AV) of a dust screen following the \cite{Calzetti2000} reddening law.  The adopted priors on these fitting parameters are indicated in Table~\ref{table:stellarprop}. \bagpipes\ includes nebular emission lines calculated with the \texttt{Cloudy} photoionization code \citep{cloudy2017}, where we assume the line-emitting gas has the same metallicity and dust attenuation as the stars.

The \xs\ spectra of the four lensed targets at different redshifts are masked such that all targets have similar rest-frame spectral coverage from $\approx2500\AA$ to just beyond \oiii\,$\lambda$5007, though the targets have different gaps within this range due to the redshifted NIR atmospheric windows.  To fit the combined observed data of the multiple \xs\ channels and broad-band photometry, we allow for an additional low-order ``calibration'' function multiplied to the spectra to reduce the effect of normalization differences between the spectra and photometry \citep{Carnall2019b}.  Therefore, the photometry constrains the overall shape of the spectral energy distribution (SED) and the spectra constrain features on wavelength scales smaller than the calibration function (i.e., stellar absorption lines).  Finally, we allow a final parameter that scales the \xs\ spectrum uncertainties and applies the appropriate normalization penalty to the likelihood \citep{Carnall2019b}.

The posterior probability distribution function of the fit parameters is sampled within \bagpipes\ using the \texttt{MultiNest} algorithm \citep{multinest}.  The maximum a posteriori parameters of the chain are presented in Table~\ref{table:stellarprop}, and the 
parameter covariances are shown in Figure~\ref{fig:corner_spop}.
The best-fit models are shown in Figures~\ref{fig:sed_m1341}--\ref{fig:sed_m1423}.
The de-lensed stellar masses and SFR are presented in Table~\ref{table:delensed_param}.

\section{Results} \label{sec:results}

\subsection{Rapid star formation \& Old ages} \label{sec:stellarpop}

   \begin{figure}[htbp!]
   \centering
   \includegraphics[width=\linewidth]{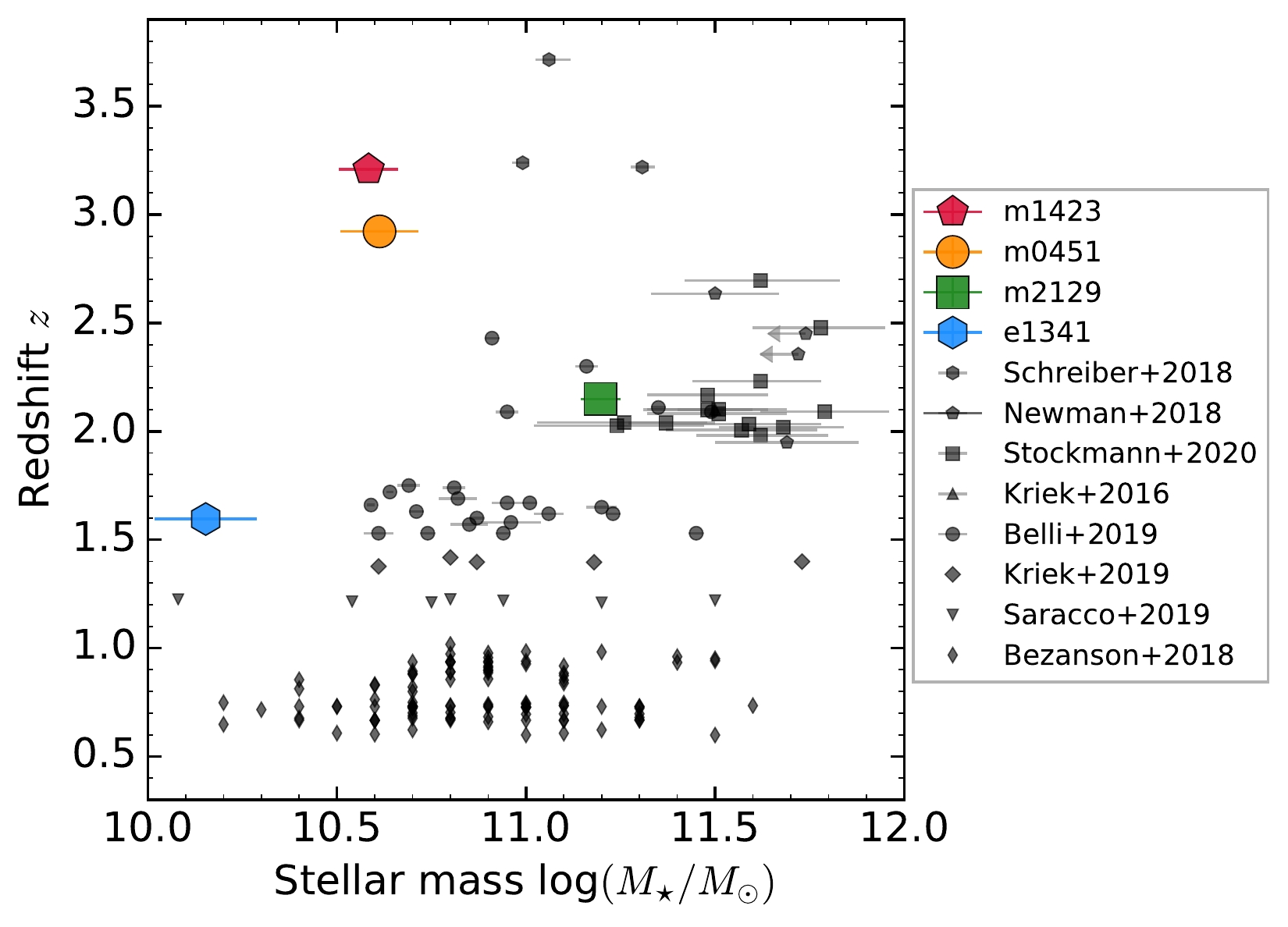}
      \caption{
      Stellar masses and redshifts of our sample (large colored symbols),
       compared with samples of quiescent galaxies with high-quality spectra for absorption line analysis \citep{Kriek2016,Kriek2019,Belli2019,Bezanson2018,Newman2018a,Schreiber2018,Saracco2019,Stockmann2020}.
       The stellar masses of lensed samples of this work and \citet{Newman2018a} are intrinsic, i.e., de-lensed.
        }
    \label{fig:mass_z}
   \end{figure}

Our sample probes galaxies near the characteristic mass of the stellar mass function of quenched galaxies.
Three out of four galaxies in the sample are intrinsically less massive than the characteristic mass of all galaxies and only quiescent ones according to the stellar mass function presented in \citet{Muzzin2013b}, as listed in Table~\ref{table:delensed_param}. 
Intrinsic stellar masses are derived by dividing the lensed stellar mass by their magnification factors, 
propagating the respective uncertainties.
Thanks to lensing we reach, for the first time, less extreme quenched galaxies at $z\gtrsim1.5$ that are more representative of the overall galaxy population, as illustrated in Figure~\ref{fig:mass_z}.

\begin{figure}[tb!]
   \centering
   \includegraphics[width=\linewidth]{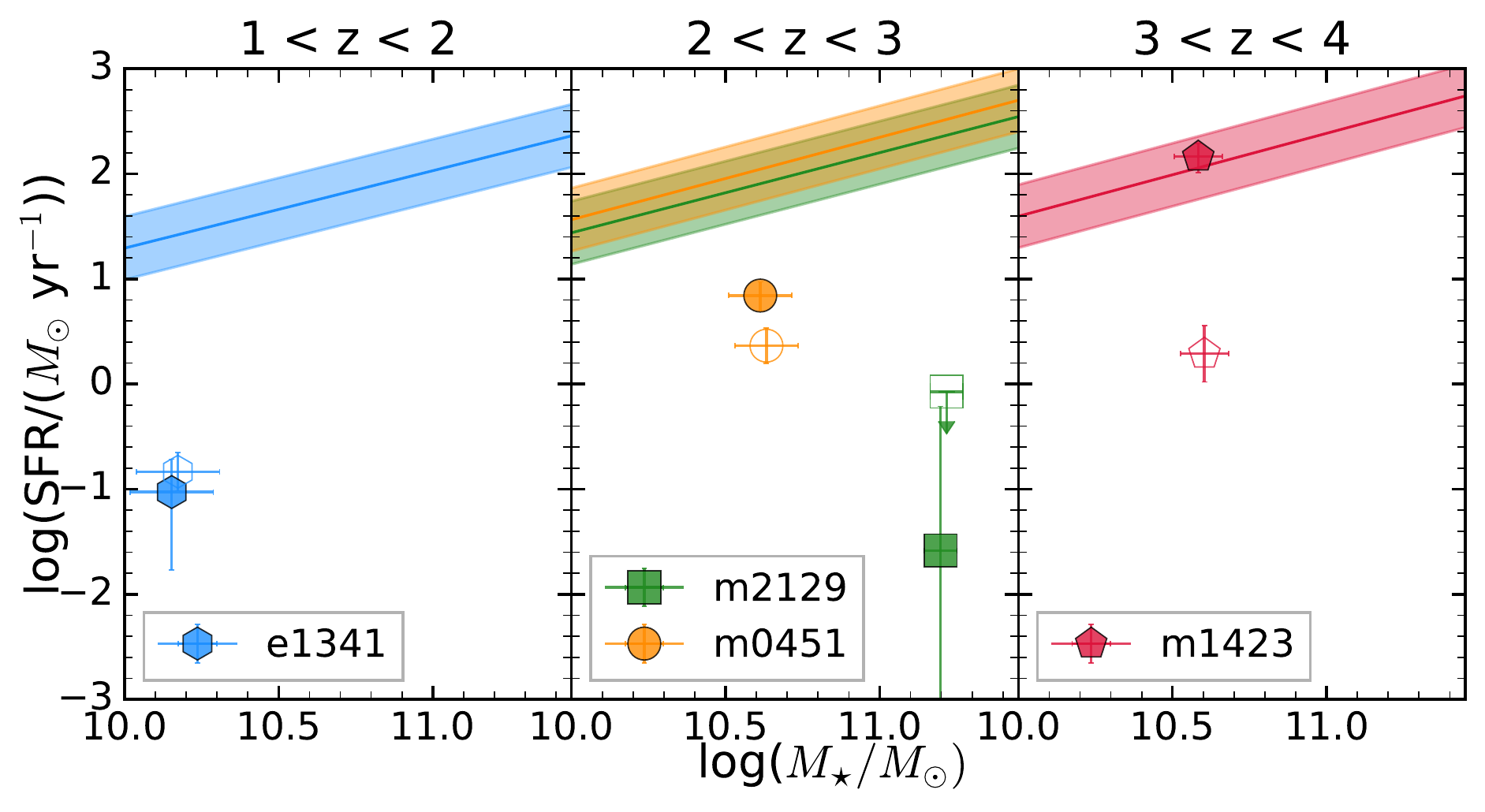}
      \caption{
      Star formation rate (SFR) and stellar mass (\mstar) of our sample, split in three redshift bins. Corrections for lensing magnifications have been applied. Colored symbols denote our targets as labeled in the legend. Filled symbols show the SFR from stellar population fitting (\S\ref{sec:specfit}). Open symbols show the SFR inferred from the \oii\ emission line (\S\ref{sec:oii}), slightly shifted along the x-axis for clarity.
       The colored lines show the best-fitting SFR-\mstar\ relation presented in \citet{Speagle2014}, calculated at the redshift of each target following the respective color scheme.
       The color shades show the observed scatter of 0.3\,dex around these relations.
        }
    \label{fig:ms}
   \end{figure}
   
\begin{figure*}[htb!]
   	\centering
   	\begin{minipage}[b]{0.49\linewidth}
	\centering
	\includegraphics[width=\linewidth]{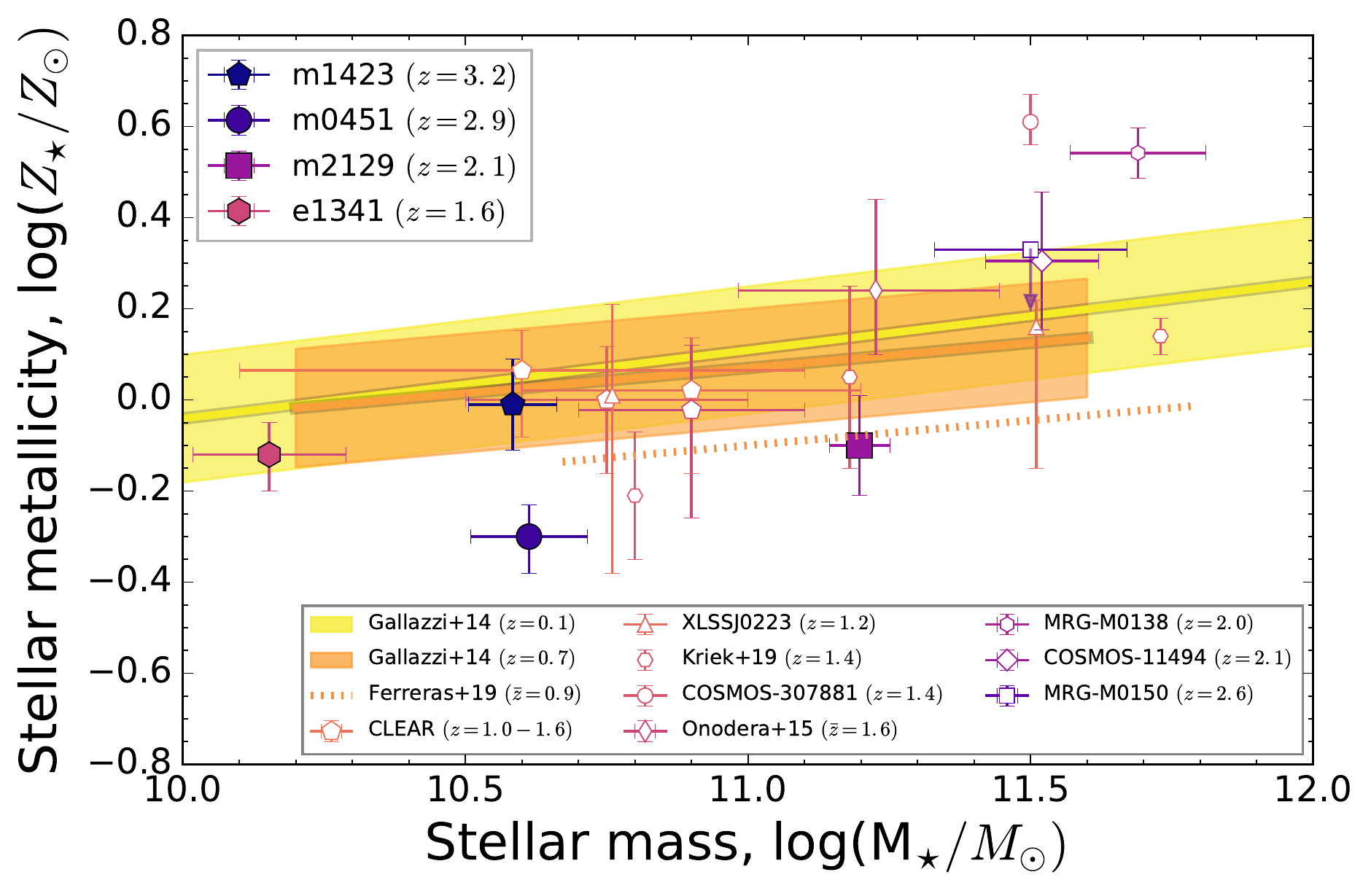}
	\end{minipage}
	\begin{minipage}[b]{0.45\linewidth}
	\centering
	\includegraphics[width=\linewidth]{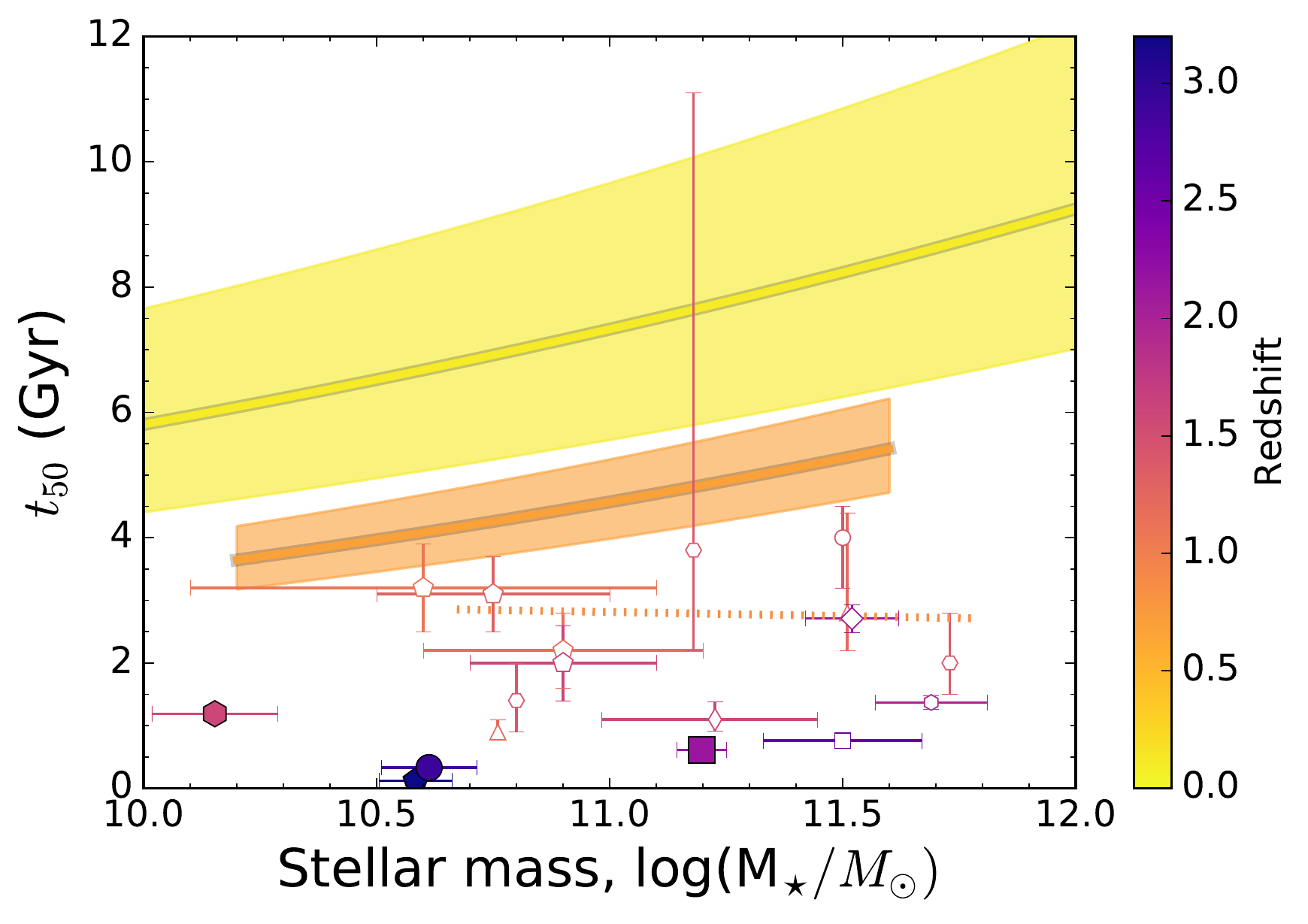}
	\end{minipage}
	\caption{
	    Stellar mass-metallicity relation (left) and stellar mass-age relation (right) of quiescent galaxies.
	    Datapoints are colored by redshifts as indicated in the color bar.
	    Filled symbols denote our sample, and \thalf\ is plotted as the y-axis on the right panel.
	    Open symbols show literature measurements of galaxies with individual measurements:  COSMOS-307881 \citep{Lonoce2015}, MRG-M0138 \citep{Jafariyazani2020}, COSMOS-11494 \citep{Kriek2016}, and MRG-M0150 \citep{Newman2018a}.
	    We plot samples of massive quiescent galaxies \citep{Kriek2019} and spheroidal galaxies in the cluster XLSSJ0223 \citep{Saracco2019}.
	    Also included are measurements made on binned grism spectra from the CLEAR survey \citep{ECarpenter2019} and stacked spectra \citep{Onodera2015}.
	    The solid yellow and orange lines show $z\approx0.1$ and $z\approx0.7$ scaling relations to quiescent galaxies from \citet{Gallazzi2014} and the total scatter is shown as colored shades.
	    The dotted line shows the best-fit relation to the quiescent galaxy sample of \citet[][$\bar z$=0.9]{Ferreras2019a}.
        When error bars overlap we apply a small shift of less than 0.03 along the x-axis to facilitate visualization. 
	   	\label{fig:MZR}
	    }
   \end{figure*}

The spectral fitting results presented in \S\ref{sec:specfit} indicate that our sample of quiescent galaxies have undergone rapid star formation histories.
The reported uncertainties are propagated from photometric and spectral flux errors, and do not include systematic uncertainties such as variation in IMF, SFH, etc.
Their star formation timescales, \twidth, are at most a few hundred Myr and are short compared to their median stellar population ages and the ages of the Universe at the observed redshifts.
As shown in Figure~\ref{fig:ms},
their specific star formation rates (sSFR) are lower than typical star-forming galaxies by at least an order of magnitude except for \m1423.
Its SFR from spectral fitting is higher than that derived from \oii, suggesting that the discrepancy is due to recent rapid quenching.
We argue in \S\ref{sec:oii} that \m1423\ is likely caught in transition from being star-forming to quiescent.

Our spectral fitting assumes, like many other works, a parameterized SFH.
Specifically, our work assumes the commonly used delayed-$\tau$ parameterization.
Although these assumptions are quite standard,
and are likely reasonable assumptions for high-redshift galaxies,
it may still be insufficient to capture the individual SFH of some galaxy types as argued by some works \citep[e.g., ][]{Maraston2010,Behroozi2013,Gladders2013,Abramson2016,Pacifici2016,Ciesla2017}. 
We note however that most of the concerns raised about the use of (delayed)-$\tau$ models do not apply to high-redshift quiescent galaxies like our sample.
The short time between their formation and observation implies that there is little effect for model degeneracies, at least compared to low-redshift galaxies that are much more evolved.
Possible alternatives are log-normal SFHs \citep{Gladders2013} or non-parameteric models \citep{Iyer2019,Leja2019,Akhshik2020}.
The diversity of SFH models used in different studies makes a direct comparison of resulting parameters difficult,
and certainly has a non-negligible effect on scaling relations derived \citep{Carnall2019}.  
With \bagpipes\ we find similar results for the stellar masses, metallicities and timescales when adopting ``double powerlaw'' and ``log-normal'' SFHs as defined by \cite{Carnall2018}.  The only parameter with a significant dependence on the SFH model is the recent star formation rate, which is significantly lower for these other parameterizations that fall off more sharply than the delayed-$\tau$ model at late times for a given \thalf\ and \twidth.  Despite this systematic uncertainty on the absolute value of the recent star formation rate, all of the parameterized fits agree that the specific star formation rate for this sample is well below that of typical star-forming galaxies at similar redshifts.
Investigating the impact of assuming a non-parametric SFH on derived properties will be explored in a future work. 

\subsection{Stellar mass-metallicity relation}\label{sec:MZ}

The metal content of galaxies is a powerful probe of their enrichment histories.
Stellar metallicity correlates with stellar mass in quiescent, early-type galaxies at $z=0.1-0.7$ \citep{Gallazzi2005,Gallazzi2014}.
As it requires deep spectra to accurately measure stellar metallicity, only a handful of quiescent galaxies at $z\gtrsim1$ have robust stellar metallicity measurements thus far.
Lensing magnification offers a unique opportunity to obtain such a measurement in galaxies with stellar mass below the characteristic mass out to $z>3$.

In Figure~\ref{fig:MZR} we show the stellar mass-metallicity relation (MZR) and the stellar mass-age relation.
Only literature measurements with errors smaller than $\Delta$log(\Zstar/\Zsun)~$<0.3$ are shown in Figure~\ref{fig:MZR}.
The only works presenting stellar metallicity measurements of quiescent galaxies at $z>1.9$ are \citet{Kriek2016}, \citet{Newman2018a}, and \citet{Jafariyazani2020}.
Only one galaxy, MRG-M0150, in the sample of \citet{Newman2018a} is plotted because we exclude those without lensing models which are needed to constrain the intrinsic \mstar.
As for MRG-M0138 we obtain a metallicity\footnote{MRG-M0138 has [Mg/Fe]~$=0.51\pm0.05$, which is unusually high given that the abundance of other $\alpha$-elements follow closely the values of nearby early-type galaxies \citep{Jafariyazani2020}. For our purpose of estimating [Z/H], we thus adopt the median [$\alpha$/Fe] value of the cores of nearby massive early-type galaxies \citep{Greene2019} rather than the measured value of [Mg/Fe], i.e., [Z/H] = [Fe/H] + 0.94 [$\alpha$/Fe] = $(0.26\pm0.04) + 0.94\times(0.30\pm0.04) = (0.54\pm0.05)$.} based on the abundance measurement presented in \citet{Jafariyazani2020} which provided a more detailed analysis than in \citet{Newman2018a}.
As \m2129\ is part of the sample in this work as well as in \citet{Newman2018a},
to avoid duplication we only show our measurements here\footnote{Table~\ref{table:m2129} provides a comparison of the properties of \m2129\ presented in this work, \citet{Toft2017}, and \citet{Newman2018a}.}.
We also include measurements of individual quiescent or early-type galaxies at $z=1-2$ from \citet{Lonoce2015} and \citet{Kriek2019} as well as composite spectra \citep{Onodera2015,ECarpenter2019,Saracco2019}\footnote{We multiplied the stellar mass presented in \citet{Onodera2015} by a factor of 0.67 to bring their Salpeter IMF to the Kroupa IMF assumed throughout this work following the value used in \citet{Madau2014}. No correction was applied to stellar masses obtained using Chabrier IMF, as the correction (0.04\,dex) is negligible and does not affect our conclusions here.}.
We only consider literature measurements of \Zstar\ when they are based on detection of metal absorption features and determined to within 0.3\,dex accuracy.
We take the MZR for quiescent galaxies from \citet{Gallazzi2014} for $z\approx0.1$ (SDSS) and $z\approx0.7$, as well as the best-fit relation presented in \citet{Ferreras2019a} for 19 quiescent galaxies at $z=0.6-1.3$ (median $z=0.85$).

Our sample provides new insights on the stellar MZR relation, 
because they probe the most distant and least massive, quenching galaxies among all $z>1$ literature studies.
\m1341\ and \m1423\ have stellar metallicities on par with literature measurements extrapolated to higher redshifts and lower stellar masses,
while \m0451\ and \m2129\ lie lower.
The stellar metalliticies of our sample span from $\approx$0.5 to 1 solar value,
somewhat lower than those of more massive, quiescent galaxies at comparable redshifts \citep{Lonoce2015,Kriek2016,Newman2018a,Jafariyazani2020},
but are still higher those of star-forming galaxies at $z=2.5-5$ \citep{Halliday2008,Cullen2019}.
While these differences may be due to diverse chemical enrichment histories (and thus dependence on redshift and stellar masses),
variations in metallicity calibrations and stellar populations probed are also plausible explanations.
Indeed, existing stellar metallicity measurements of \m2129\ range from subsolar \citep[log(\Zstar/\Zsun)=$-0.6\pm0.5$;][]{Toft2017} to mildly supersolar \citep[log(\Zstar/\Zsun)=$0.16\pm0.13$;][]{Newman2018a}, as detailed in Appendix~\ref{sec:m2129}.
This work provides an intermediate measurement of log(\Zstar/\Zsun)=$-0.10\pm0.11$.
Systematic effects in deriving stellar metallicities are thus a significant source of uncertainty that needs to be better quantified in future works.
We will discuss the implications of these results in \S\ref{sec:discuss}.

\subsection{Magnesium features at rest-frame UV}\label{sec:mgii}

Singly ionized magnesium \mgii\ $\lambda\lambda\,2796,2803$ (hereafter \mgii) absorption in the rest-frame UV spectrum can arise from warm atomic gas ($T\sim10^{4}$\,K) or stellar photospheres \citep[late B-, A-, F- and G-type stars;][]{Fanelli1992,Snow1994}.
The adjacent \mgi\ $\lambda2853$ traces neutral magnesium and has lower equivalent widths than \mgii\ in both the interstellar medium and stars \citep{Maraston2009,Coil2011,Zhu2015}.
When present in stars \mgi\ absorption is stronger in later type stars (F- and G-types) and absent in O- and B-type stars \citep{Fanelli1992,RMerino2005}.

\begin{figure}[htbp!]
   \centering
   \includegraphics[width=\linewidth]{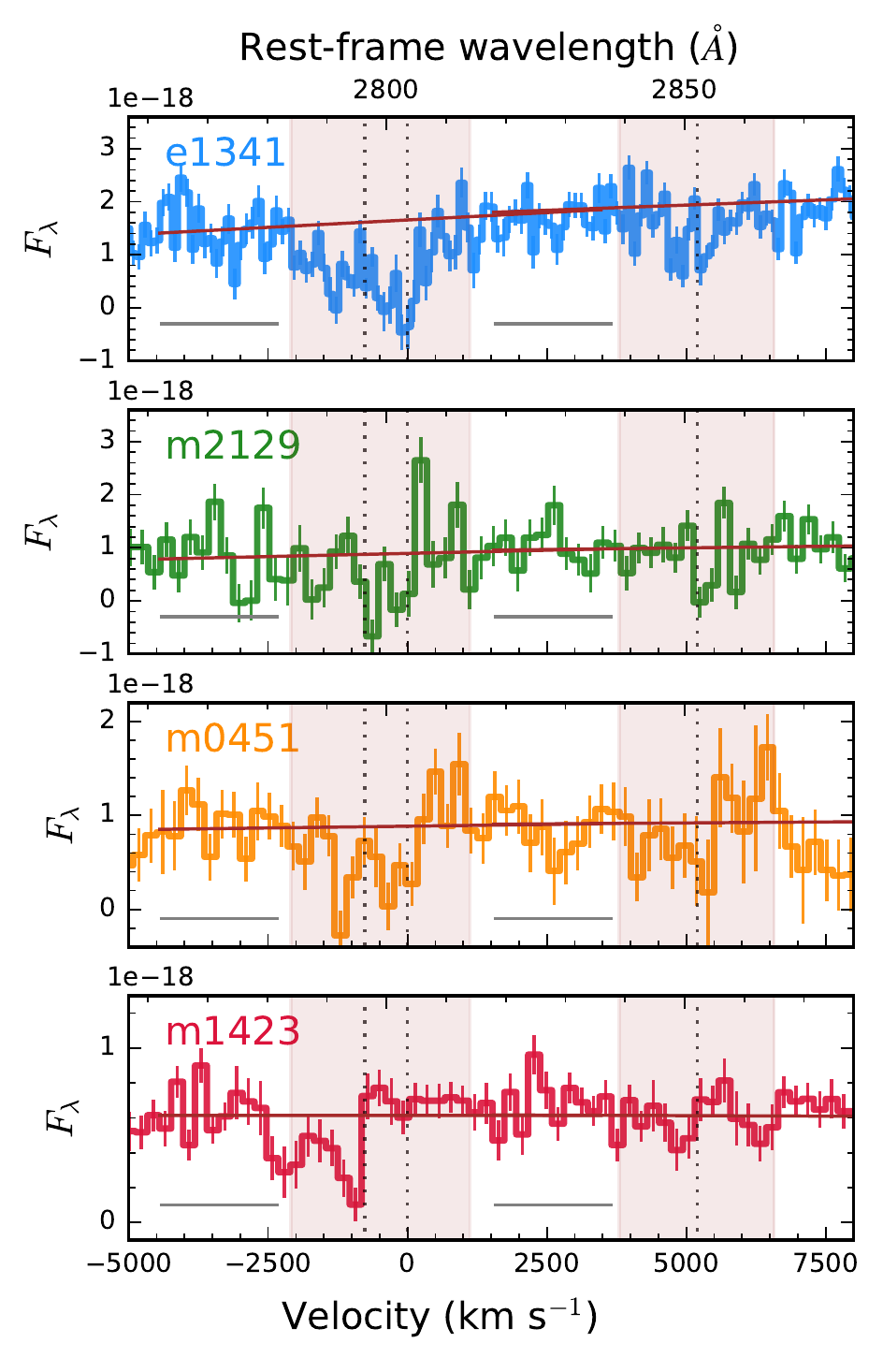}
      \caption{
       The rest-frame UV spectra zoomed in on \mgii\,$\lambda\lambda$\,2796,\,2803 and \mgi\,$\lambda$\,2853 as indicated by the vertical dotted lines.
       The bottom x-axis refers to the velocity with respect to the \mgii\,$\lambda$\,2803 at the systemic redshifts.
       The y-axis is in unit of erg\,s$^{-1}$\,cm$^{-2}$\,\AA$^{-1}$.
       The vertical red shades show the central bandpasses for \mgii\ and \mgi.
       The continua used for the EW and line flux measurements are shown as brown straight lines.
       The horizontal gray lines at the bottom of each panel show the continuum bandpasses used for fitting the continuum.
       The red continuum bandpass for \mgi\,$\lambda$\,2853 is beyond the axis limit and is not shown in this figure.
       Refer to \S\ref{sec:mgii} for more details.
        }
    \label{fig:MgII}
\end{figure}

To facilitate spectral line measurements, 
we model the local stellar continua as a straight line using the model spectra within the line-free bandpasses specified in \citet[][Table~1]{Maraston2009}.
Equivalent widths (EW) and line fluxes of \mgii\ and \mgi\ are then measured by summation of the \xs\ spectra normalized by the local continua over the central bandpasses as shown in Figure~\ref{fig:MgII}.
We caution that emission line infilling might lead to underestimated absorption line EW and fluxes.
Errors are propagated from the noise on the \xs\ spectra.
The spectral line measurements are provided in Table~\ref{table:EW}.

\mgii\ absorption is ubiquitous in our sample with different profile shapes,
sometimes accompanied with redshifted emission, as shown in Figure~\ref{fig:MgII}.
\mgi, being a weaker feature than \mgii, is detected in some of the galaxies as well.
\m1341\ has deep \mgii\ and \mgi\ absorption at the systemic velocity,
as well as a broad blue-shifted \mgii\ absorption wing that extends to $\approx -2000$\,\kms\ that is not apparent in \mgi.
Although \m2129\ has the lowest S/N spectrum within the sample,
a comparison of the \mgii\ and \mgi\ profiles shows tentative absorption at systemic velocity accompanied by redshifted emission.
\m0451\ shows a pair of slightly blueshifted \mgii\ doublet absorption near systemic velocity accompanied by redshifted emission.
Its \mgii\,$\lambda$\,2796 and \mgii\,$\lambda$\,2803 absorptions are shifted by $\approx-200$\,\kms\ from the systemic redshift.
In addition to gas associated with \m0451\ there might also be contribution from its close companion, tidal interaction and/or intragroup environment:
the quiescent galaxy has a star-forming companion galaxy located $\approx23$\,kpc away on the source plane \citep[gal\,2;][]{MacKenzie2014} with the same redshift as measured from its CO J=3$\rightarrow$2 emission \citep[$z_\mathrm{CO,gal\,2}=2.921\pm0.001$;][]{Shen2021}.
This pair of galaxies is embedded in a $z=2.9$ group as discussed in \S\ref{sec:targets}.
\m1423\ has a highly blueshifted \mgii\ absorption by up to $\approx -1500$\,\kms\ from the systemic velocity.
In all cases, the \mgii\ absorption at its trough is consistent with having zero flux, i.e., black.
This suggests that the absorbing gas has high column density and covering fraction.

How do we interpret the origins of the Mg features in our sample?
One way to discern between these origins is to examine the line profile.
Photospheric absorption takes places at the systemic velocity.
Deep UV spectroscopy in fact provided the first confirmations of evolved stellar populations dominated by stars older than $\gtrsim0.5$\,Gyr at $z\approx1.5-2$ with optical spectrographs \citep{Dunlop1996,Spinrad1997,Cimatti2004,Cimatti2008,McCarthy2004,Daddi2005}.
On the other hand, interstellar or circumgalactic medium absorption is not restricted to the systemic velocity and can be blueshifted (outflow) or redshifted (infall).
While part of the absorption can be attributed to stellar photospheres given the stellar metallicities of our sample (\S\ref{sec:MZ}),
gas absorption is required to account for the absorption depth as well as the blueshifted absorption and redshifted emission components.
The \mgii\ profiles are unlikely to be dominated by circumstellar winds:
velocity of \mgii\ in stellar winds are modest, of order $\approx100$\,\kms\ \citep{Praderie1980,Snow1994} except for supergiants and giants.
The \mgii\ absorption in stars is not accompanied by redshifted \mgii\ emission \citep{Snow1994}.

Galactic outflow is a likely explanation for the \mgii\ profiles in \m0451\ and \m1423.
Gas outflowing into the line-of-sight obscures light from the stellar continuum creating the blueshifted absorption.
Gas outflowing away from us scatters photons back into our line-of-sight to create the redshifted emission \citep{Weiner2009,Rubin2011}.
Galactic outflow are commonly seen in massive post-starburst galaxies at $z<1.5$ \citep{Tremonti2007,Coil2011,Sell2014,Maltby2019}.
In a galactic outflow, \mgi\ shows weaker but similar absorption profile as \mgii\ and has EW of a quarter or less than that of \mgii\ \citep{Coil2011}.
The similarities of the \mgi\ and \mgii\ profiles in \m1423, \m0451, and perhaps \m2129\ suggest that outflows are present,
as \mgi\ can only have photospheric absorption but not emission in stars \citep{Fanelli1992}.
It is unlikely that the gas outflows reported in the youngest targets here are due to circumnuclear outflow.
The rest-frame UV continua are well-modeled by the stellar populations without the need to invoke an AGN continuum (Figures~\ref{fig:sed_m1341}--\ref{fig:sed_m1423}).
The \mgii\ profiles of our targets are distinct from those of quasar winds in broad absorption line quasars \citep[e.g.,][]{Trump2006}.
In \S\ref{sec:ionized} we rule out the presence of type-1 AGN in our sample.
Thus the Mg features in our sample provide unambiguous evidence for galactic-scale gas outflow in addition to evolved stellar populations.

The \mgii\ profile encapsulates information about the structure of a galactic outflow,
i.e., covering fraction, opacity, orientation, etc.
Detailed modeling of the \mgii\ profile is beyond the scope of this Paper and will be deferred to a future work.
Observations of rest-frame optical emission lines could further constrain the warm gas in quiescent galaxies, 
as we shall explore in \S\ref{sec:ionized}.
The implications of galactic-scale outflows are discussed in a broader context in \S\ref{sec:quench}.

\subsection{Ionized gas emission}\label{sec:ionized}

\begin{table*}[htbp!]
	\footnotesize
	\centering
	\caption{Spectral line measurements}
	\label{table:EW}
	\begin{tabular}{l|cccccc|ccc}
		\hline
		 & \mgii & \mgi & \oii & \oiii$\lambda4959$ & \oiii$\lambda5007$ & \halpha & O32 & \oiii$\lambda5007$/$\lambda4959$ & log(sSFR(\oii)\,yr$^{-1}$)\\
		\hline
		& \multicolumn{6}{c}{Equivalent widths, $W$ (\AA)} \\
\m1341 & $ 16.4\pm  1.0$ & $  4.8\pm  0.6$ & $ -3.3\pm  0.4$ & $  0.2\pm  0.3$ & $  0.2\pm  0.3$ & $  0.4\pm  0.1$ & \nodata & \nodata & \nodata \\ 
\m2129 & $  6.3\pm  4.0$ & $  6.6\pm  2.6$ & $ -1.5\pm  2.2$ & $ -0.2\pm  0.6$ & $ -3.3\pm  0.5$ & $  4.2\pm  0.8$ & \nodata & \nodata & \nodata \\ 
\m0451 & $  5.9\pm  2.9$ & $  2.3\pm  3.9$ & $ -8.0\pm  1.4$ & $ -1.5\pm  0.5$ & $ -8.0\pm  0.7$ & \nodata & \nodata & \nodata & \nodata \\ 
\m1423 & $  2.6\pm  1.4$ & $  0.7\pm  1.8$ & $ -2.7\pm  1.5$ & $ -0.7\pm  0.9$ & $ -4.5\pm  0.8$ & \nodata & \nodata & \nodata & \nodata \\ 
		& \multicolumn{6}{c}{Line fluxes ($\times10^{-18}$\,erg\,s$^{-1}$\,cm$^{-2}$)} \\
\m1341 & $-69.1\pm  4.3$ & $-24.4\pm  3.0$ & $ 28.0\pm  3.4$ & $ -4.2\pm  5.2$ & $ -4.2\pm  5.7$ & $  6.3\pm  2.0$ & $<0.2$ & \nodata & $-11.0\pm0.4$\\ 
\m2129 & $-17.5\pm 11.2$ & $-20.7\pm  8.1$ & $  8.7\pm 11.9$ & $  3.3\pm  8.4$ & $ 41.6\pm  5.8$ & $ 38.5\pm  7.2$ & $>3.5$& $>5.0$ & $<-11.3$ \\ 
\m0451 & $-20.2\pm  9.9$ & $ -8.1\pm 14.2$ & $ 37.1\pm  6.3$ & $ 13.2\pm  4.4$ & $ 67.6\pm  5.7$ & \nodata & $  1.8\pm  0.3$ & $  5.1\pm  1.8$ & $-10.2\pm0.4$\\ 
\m1423 & $ -6.8\pm  3.7$ & $ -1.7\pm  4.7$ & $  6.1\pm  3.3$ & $  1.8\pm  2.6$ & $ 11.8\pm  2.2$ & \nodata & $  1.9\pm  1.1$ & $>4.5$ & $-10.3\pm0.5$ \\
		\hline
	\end{tabular}
	\tablecomments{
	Rest-frame equivalent widths and observed frame line fluxes are provided. Emission lines have negative equivalent widths and positive line fluxes, and vice versa for absorption lines.
	No correction for lensing magnification or extinction has been applied.
	Note that some of the \mgii\ and \mgi\ measurements are likely affected by emission line infilling (Section~\ref{sec:mgii}).
	\oii\ refers to the blended \oii\,$\lambda\lambda$\,3726,3729 doublet.
	No correction has been applied to deblend \halpha\ from the \nii\ doublet that brackets it.
	O32 refers to the \oiii\,$\lambda$\,5007/\oii\,$\lambda\lambda$\,3726,3729 flux ratio.
	sSFR(\oii) is the specific SFR inferred from the \oii\ emission. Upper limits are 1$
	\sigma$.
	}
\end{table*}

Warm ($T\sim10^{4}$\,K) gas in galaxies can be photoionized by AGN or hot stars (both young and old), or excited by shocks.
The strengths and ratios of emission lines can inform us of their production mechanism \citep[][and references therein]{Kewley2019}.
Locally, massive quiescent galaxies sometimes have 
low-ionization emission regions (LIER)\footnote{Spatially resolved observations show that emission lines in massive, quiescent galaxies are not necessarily confined to the nuclear region \citep{Singh2013,Belfiore2016}. Therefore we adopt the acronym LIER rather than LINER without reference to the nuclear region as used in earlier studies.}.
Characterizing the gas conditions of quenching galaxies will provide clues to understand how they become quiescent.

The brightest emission lines within our spectral coverage are \oii\,$\lambda\lambda$\,3726,3729\ (hereafter \oii), \oiii\,$\lambda\lambda$\,4959,5007, and \halpha.
The oxygen emission lines are shown in velocity space along with the normalized \mgii\ spectra in Figure~\ref{fig:ism}, and highlighted in orange in Figure~\ref{fig:spec1d}.
Emission line EWs and line fluxes are provided in Table~\ref{table:EW}.
The measurements are computed as the excess emission to the best-fitting stellar continua (\S\ref{sec:stellarpop}) using the observed spectra without binning or smoothing,
measured within the passband range specified in \citet[][Table~1]{Westfall2019}. 

\begin{figure}[htb!]
   \centering
   \includegraphics[width=\linewidth]{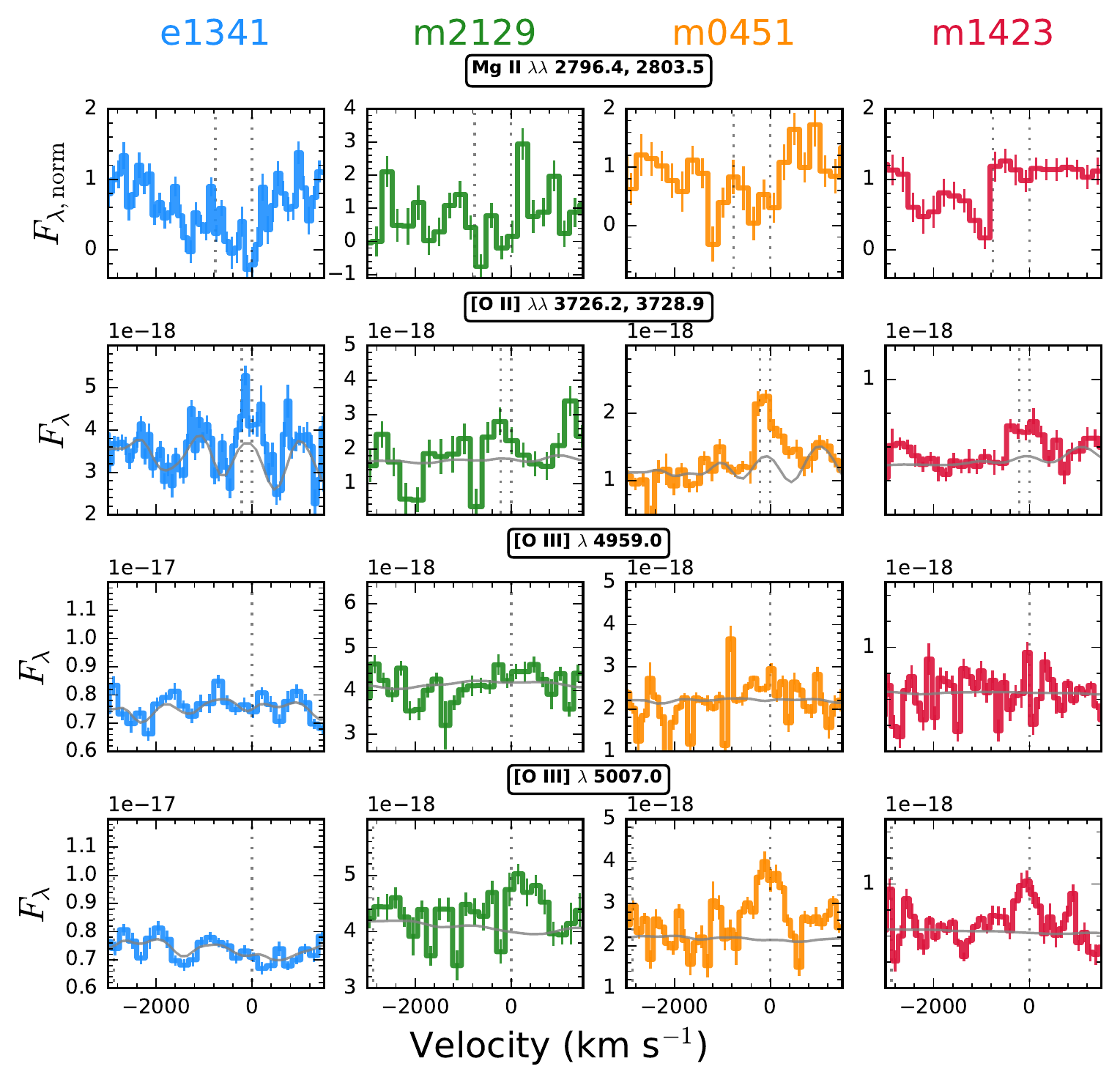}
      \caption{
    The binned spectra are zoomed in on the \oii\ and \oiii\ emission lines (middle and bottom panels) in velocity space, together with the \mgii\ absorption (top panel).
    The \texttt{Bagpipes} model continua are shown as gray lines. 
      }
    \label{fig:ism}
\end{figure}

\subsubsection{SFR inferred from \oii}\label{sec:oii}

The \oii\ luminosity is sometimes used as a SFR indicator in high-redshift galaxies via an empirical calibration to \halpha\ luminosity.
Its excitation is sensitive to the oxygen abundance and the ionization state of the gas \citep[][and references therein]{Kennicutt1998,Kennicutt2012}.
The use of \oii\ as a SFR indicator in post-starburst galaxies has been brought to question as \oii\ can arise from AGN, shocks, post-asymptotic giant branch stars, and cooling flows \citep{Yan2006}.
We use the \oii\ luminosity as a consistency check for the SFR provided by the stellar population fitting.
We apply the \citet[][Equation~3]{Kennicutt1998} conversion adjusted to the \citet{Kroupa2001} IMF to infer the SFR from the \oii\ line fluxes and provide them in Table~\ref{table:EW}.
The inferred log(sSFR) range from $-10.2$ to $<-11.3$, on par with the low values obtained from the full spectral fitting (Table~\ref{table:stellarprop}).
Reassuringly the SFRs inferred from \oii\ are fully consistent with being quiescent stellar populations.

As \hii\ regions may not be the sole supplier of ionizing photons,
the SFRs inferred from \oii\ are upper limits.
Although the inferred SFRs are not corrected for extinction,
we expect the correction to be minor in these galaxies if the nebular extinction is similar to that of the stellar continuum.

The most discrepant measurement is for \m1423\ (Figure~\ref{fig:ms}).
\m1423\ is most likely caught in a rapidly quenching phase as emission line SFR indicators are sensitive to more recent SF than the UV continuum \citep{Kennicutt2012}. 
This is in line with the transitional nature suggested by its U--V and V--J colors as discussed in \S\ref{sec:quench} and shown in Figure~\ref{fig:uvj}.
Another possibility is that the \oii\ emission is heavily dust-obscured in \m1423.
Its stellar continuum has \Av$=0.8\pm0.1$ implying an attenuation of 1.2\,mag at the wavelength of \oii, which is too small to fully account for the discrepant SFR estimate from spectral fitting.
Unless the nebular emission is substantially more dust-obscured than the stellar continuum, this is unlikely the cause for discrepancy.

\subsubsection{\oiii-to-\oii\ flux ratio}

The \oiii\,$\lambda$\,5007/\oii\,$\lambda\lambda$\,3726,3729 flux ratio, hereafter O32, is commonly used as an ionization parameter diagnostic.
Type~1 quasars have a distribution of O32 that peaks at $\approx5$,
while type~2 quasars peak at O32 $\approx2$ \citep[][Figure~7]{Zakamska2003}.
LIERs are expected to have O32 near unity \citep{Netzer1990}.
Non-active galaxies with stellar populations aged $\gtrsim$1\,Gyr have O32\,$\lesssim$\,0.6 \citep{Johansson2016}.
The variation of O32 reflects the different ionization parameters as well as the oxygen abundance and ISM pressure across galaxies \citep{Kewley2002,Kewley2019}.

Our sample spans a range of O32 as shown in Table~\ref{table:EW}.
O32 is higher than unity in the youngest three targets, \m1423, \m0451\ and \m2129, 
in line with the expectation that they host type~2 AGN as discussed in \S\ref{sec:oiii}.
Their O32 and \oiii\ are comparable to type-2 AGN at $z\approx0$ \citep{Kauffmann2003,Silverman2009}.
O32 is poorly constrained for the oldest two targets, \m1341\ and \m2129,
but their low values along with low \oiii\ luminosities suggest that they are unlikely to host type~1 AGN.

Discerning the source of ionizing photons provides additional clues to quenching mechanisms.
If a past, luminous AGN episode was responsible for star formation quenching,
one might expect to see signatures in the structure and ionization state of the ionized gas \citep{King2011,Zubovas2014}.
We discuss the implications of these findings in \S\ref{sec:discuss}.

To unambiguously discern the source of ionizing photons,
it is imperative to access more emission lines to place them on diagnostic plots.
In particular the \oi-to-\halpha\ ratio is an important addition to characterize the hardness of the radiation field,
to distinguish between \hii\ regions, Seyferts and LIERs \citep[][Figure~5]{Kewley2006,Johansson2016}.
Observations of the \sii\ and \nii\ doublets will help to quantify the importance of shocks \citep{Yan2012}.
Spatially resolved emission line ratios will help distinguish between different ionizing sources, 
as commonly done in nearby galaxies \citep{Singh2013,Belfiore2016}.

\subsubsection{\oiii\ as a tracer for nuclear activity}\label{sec:oiii}

\oiii\,$\lambda$\,5007 luminosity, \Loiii, is a proxy of the AGN accretion rate \citep[][and references therein]{Heckman2014}.
The line is detected in our sample except for the oldest target \m1341.
The \oiii\ FWHM line width is $<1000$\,\kms, ruling out the presence of Seyfert 1 AGN in the sample \citep[][and references therein]{Padovani2017}.
While \nev\,$\lambda$\,3426 is a reliable tracer for AGN activity \citep[e.g.,][]{Feltre2016},
it is typically an order of magnitude fainter than \oiii\,$\lambda$\,5007 in AGN \citep{Zakamska2003} and is below our detection limit even if present.
Only \m0451\ has both of the \oiii\,$\lambda\lambda$\,4959,\,5007 emission lines detected and the ratio is within $1.5\sigma$ of the theoretical value of 2.98 \citep{Storey2000}.

The intrinsic (de-lensed) \Loiii\ are shown in Table~\ref{table:oiii}.
These \Loiii\ are more than an order of magnitude below those of powerful radio galaxies at $z\approx2$ during the quasar feedback phase \citep{Nesvadba2017}.
In absence of an alternative tracer of the AGN bolometric luminosity (\Lbol) such as the IR SED or X-ray observations,
we estimate the \Lbol\ from \Loiii\ (without dust extinction correction) using a mean bolometric correction of 3500 assuming the same factor applies to type~1 and type~2 AGNs \citep{Heckman2004}.
Attributing the entirety of the \oiii\ emission to AGN, this implies that \Lbol$~=1.1-1.6\times10^{45}$\,\ergs\ for the three targets with detected 
\oiii,
on par or below those at the faint end of AGN samples at $z\approx2$ \citep{Circosta2018,Leung2019}.
These \Lbol\ imply a mass outflow rate of $<10$\,\msun\,\peryr\ following the best-fit relation between the \Lbol\ and mass outflow rate \citep[][Figure~9]{Leung2019}.
The mass outflow rates are likely upper limits as this derivation assumes that all the \oiii\ emission is photoionized by the AGN.
The outflowing ionized mass are thus insignificant and unlikely to escape the deep gravitational potentials of these compact, massive galaxies.

\begin{table}[htbp!]
	\caption{Inferred supermassive blackhole properties}
    \begin{center}
	\begin{tabular}{lcccc}
		\hline
		& log(\Loiii) & log(\Lbol) & log(\mbh) & \Lbol/\Ledd (\%) \\
		\hline
m1341 & $<5.9$ & $<9.5$ & $7.8\pm0.1$ & $ <0.1$ \\ 
\m2129 & $ 7.9\pm 0.1$ & $11.5\pm 0.1$ & $9.5\pm0.2$ & $   0.3\pm   0.2$ \\ 
\m0451 & $ 8.1\pm 0.1$ & $11.6\pm 0.1$ & $7.9\pm0.2$ & $  14.4\pm   6.2$ \\ 
\m1423 & $ 8.0\pm 0.1$ & $11.6\pm 0.1$ & $9.1\pm0.2$ & $   0.8\pm   0.3$ \\ 
	\hline
	\end{tabular}
	\end{center}
	\tablecomments{
	The first three columns are quoted in logarithmic units of solar values.
	The uncertainties are propagated from measurement errors only.
	Upper limits are $1\sigma$ for the \oiii\ non detection of \m1341.
	\label{table:oiii}
	}
\end{table}

The Eddington ratio, \Lbol/\Ledd, is a useful quantity to examine the accretion mode of supermassive blackholes (SMBH).
Radiative-mode SMBHs accrete at 1\%--10\%\ of \Ledd, whereas jet-mode SMBHs accrete much less efficiently at less than 1\% of \Ledd\ \citep{Best2012}.
The luminosity of the classical Eddington limit for each target is derived as \Ledd~=~$3.3\times10^{4}$~\mbh\  \citep[][Equation~4]{Heckman2014}.
Blackhole masses are inferred from the stellar velocity dispersions (Table~\ref{table:stellarprop}) using the local \mbh-\sigmastar\ relation presented in
\citet{McConnell2013}: log(\mbh) = 8.32 + 5.64 log(\sigmastar/200\,\kms).
The resulting Eddington ratios are listed in Table~\ref{table:oiii}.
Based on this calculation, the older two targets, \m2129\ and \m1341, host jet-mode SMBH. \m0451\ hosts a radiative-mode SMBH, while \m1423\ straddles the threshold.
Unavoidably there are systematic uncertainties involved in the calculation of \Ledd, such as the fraction of \Loiii\ photoionized by the AGN, the AGN bolometric correction, the contribution of rotation in the measured stellar velocity dispersions, and the scatter in the \mbh-\sigmastar\ relation.
Accurately constraining these quantities is beyond the scope of this Paper.
If our sample evolves along the \mbh-\sigmastar\ relation,
their supermassive blackholes must have already grown considerably.
The weak \oiii\ emission suggests that our sample are well past their peak AGN episode and do not harbour copious amount of outflowing, line-emitting ionized gas.
Altogether our results suggest a transition of AGN accretion from radiative-mode to jet-mode throughout the star formation quenching process.
We discuss the implications of these findings in \S\ref{sec:discuss}.

\subsection{Morphological analysis} \label{sec:morphology}

\begin{table*}[!htbp]
	\centering
	\caption{Morphological constraints of reconstructed images}
	\label{table:galfit}
	\begin{tabular}{lcccccccc}
		\hline
		 & Filter & $\lambda_{\mathrm{rest}}$ & Angular scale & \sersic\ index ($n$) & Axis ratio ($b/a$) & Position angle & \re & \rc \\
	    & & (\AA) & (kpc/$"$) & & & ($^{\circ}$) & (kpc) & (kpc) \\ 		
	    \hline
\m1341\ (im2) & F140W & 5366.1 & 8.471 & $1.68\pm0.02$ & $0.45\pm0.09$ & $35.0\pm 0.7$ & $2.82\pm0.07$ & $1.89\pm0.20$ \\ 
\m1341\ (im3) & \ditto & \ditto & \ditto & $1.47\pm0.01$ & $0.40\pm0.06$ & $14.8\pm 5.6$ & $2.64\pm0.11$ & $1.67\pm0.15$ \\ 
\m2129\ & F160W & 4881.7 & 8.294 & $1.17\pm0.01$ & $0.42\pm0.01$ & $-35.2\pm 0.5$ & $2.12\pm0.06$ & $1.37\pm0.04$ \\ 
\m0451\ (im1) & F160W & 3919.7 & 7.762 & $1.09\pm0.01$ & $0.44\pm0.03$ & $-16.2\pm 2.3$ & $0.99\pm0.04$ & $0.66\pm0.03$ \\ 
\m0451\ (im2) & \ditto & \ditto & \ditto & $1.28\pm0.90$ & $0.68\pm0.06$ & $58.8\pm22.8$ & $1.05\pm0.18$ & $0.87\pm0.15$ \\ 
\m0451\ (im3) & \ditto & \ditto & \ditto & $0.95\pm0.03$ & $0.63\pm0.09$ & $-37.3\pm 5.6$ & $0.79\pm0.04$ & $0.63\pm0.05$ \\ 
\m1423\ & F160W & 3651.9 & 7.542 & $1.22\pm0.01$ & $0.56\pm0.03$ & $-26.3\pm 1.6$ & $1.09\pm0.03$ & $0.81\pm0.03$ \\ 
		\hline
		\end{tabular}
	\tablecomments{
	Half-light effective radii (\re) and circularized radii (\rc) are provided in kpc.
    The quoted values and errors refer to the mean and standard deviation of 100 Monte-Carlo realizations perturbed by lensing model errors.
    Reconstruction of the im1 of \m1341\ is not possible because the lensing arc is only a partial view of the source.
	}
\end{table*}

\begin{figure*}[!htbp]
	\begin{minipage}[b]{0.49\linewidth}
	\centering
	\includegraphics[angle=0,width=\textwidth]{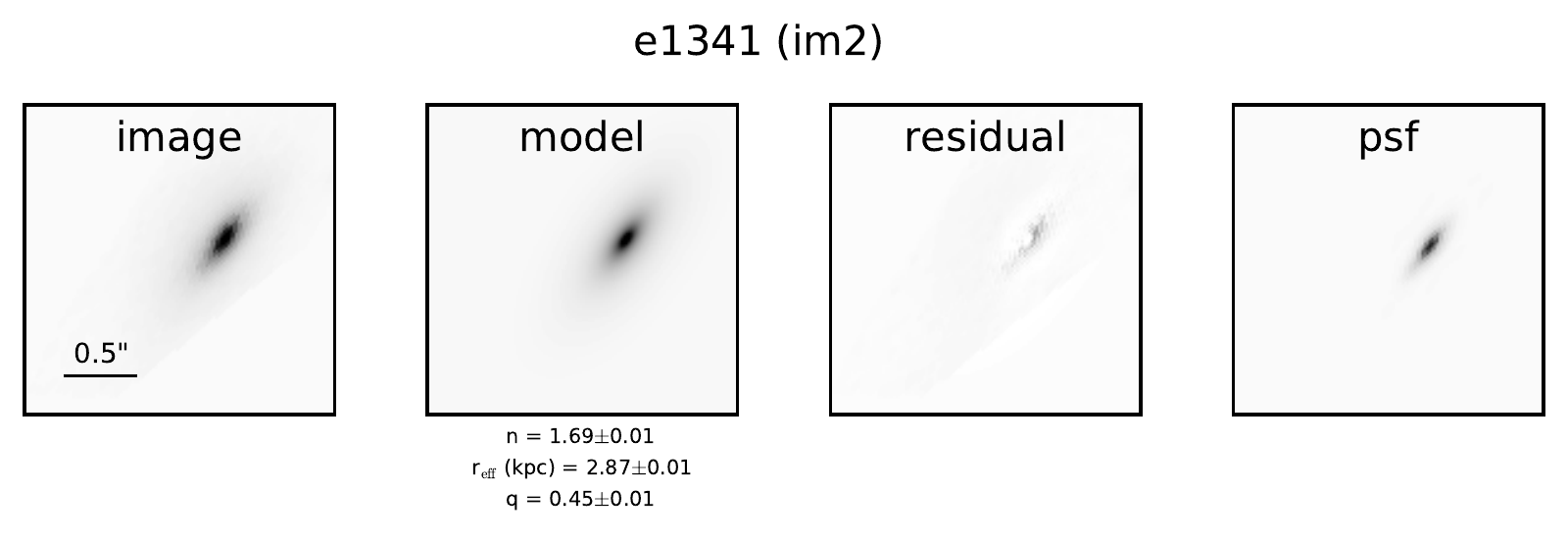}
	\end{minipage}
	\begin{minipage}[b]{0.49\linewidth}
	\centering
	\includegraphics[angle=0,width=\textwidth]{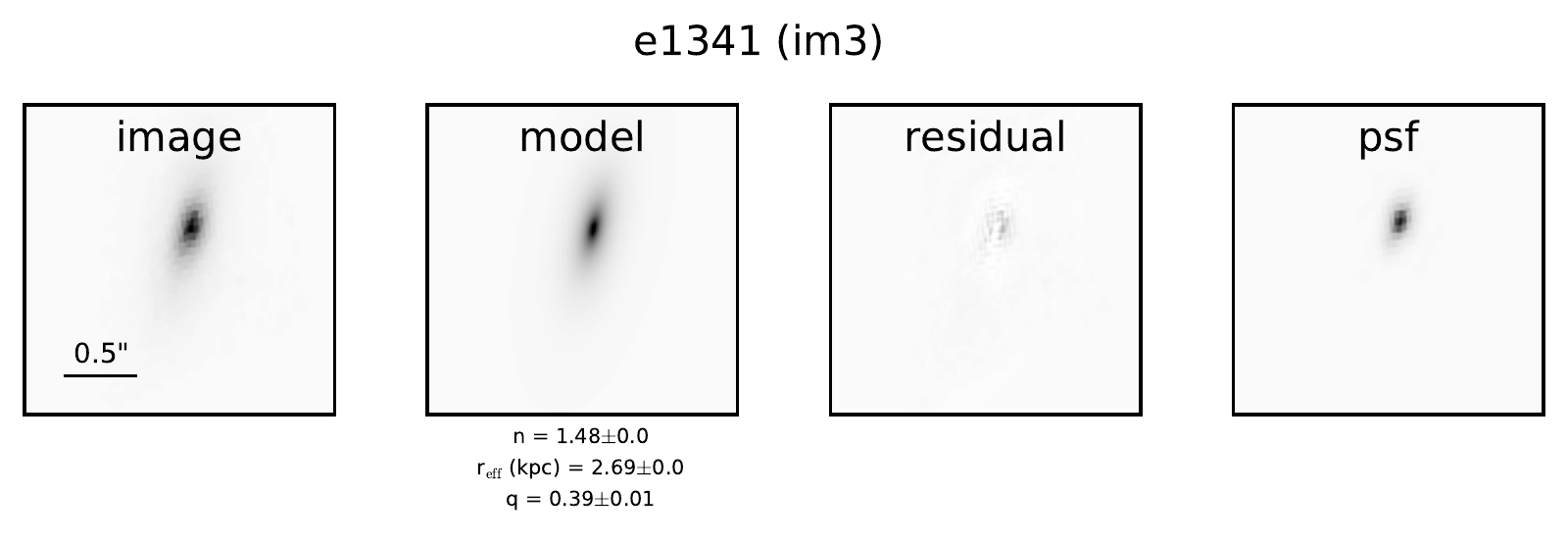}
	\end{minipage}
	\hfill
	\begin{minipage}[b]{0.49\linewidth}
	\centering
	\includegraphics[angle=0,width=\textwidth]{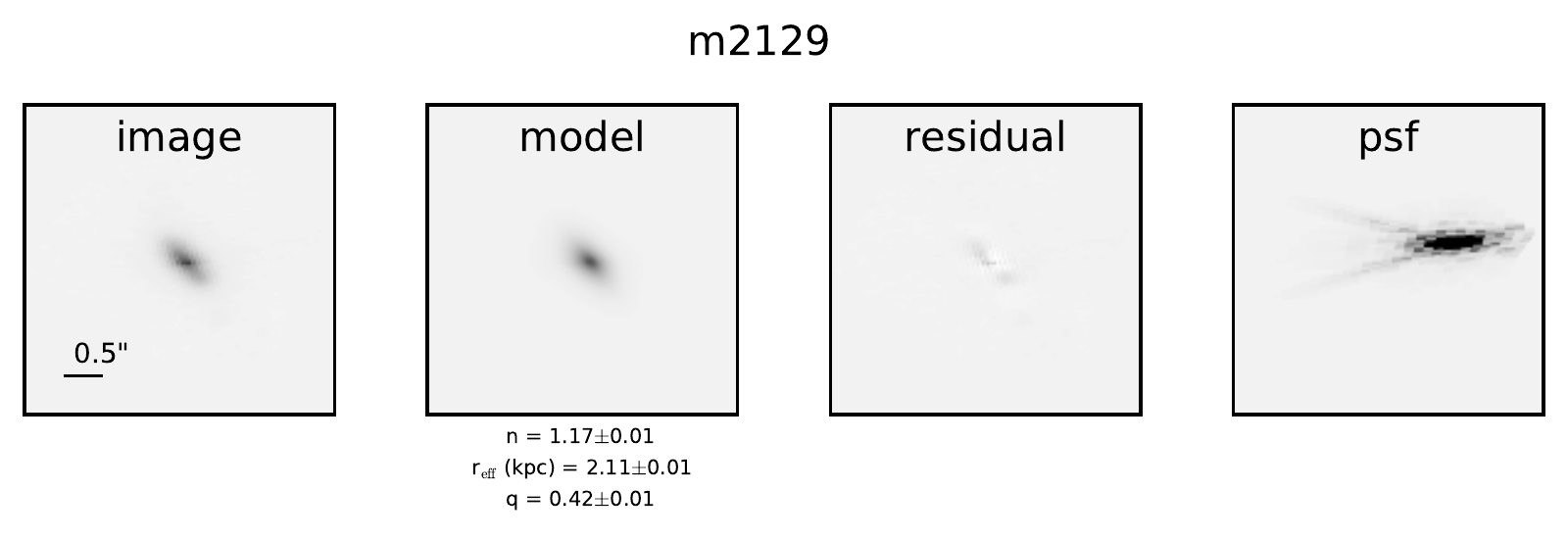}
	\end{minipage}
	\begin{minipage}[b]{0.49\linewidth}
	\centering
	\includegraphics[angle=0,width=\textwidth]{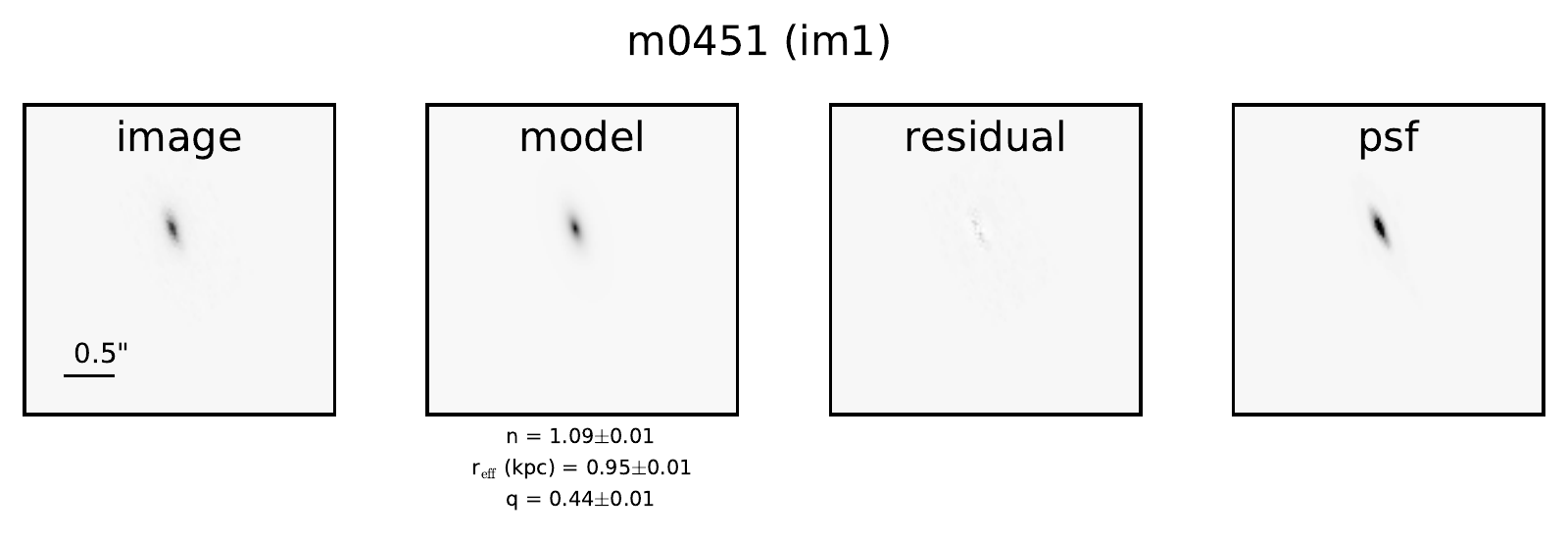}
	\end{minipage}	
	\hfill
	\begin{minipage}[b]{0.49\linewidth}
	\centering
	\includegraphics[angle=0,width=\textwidth]{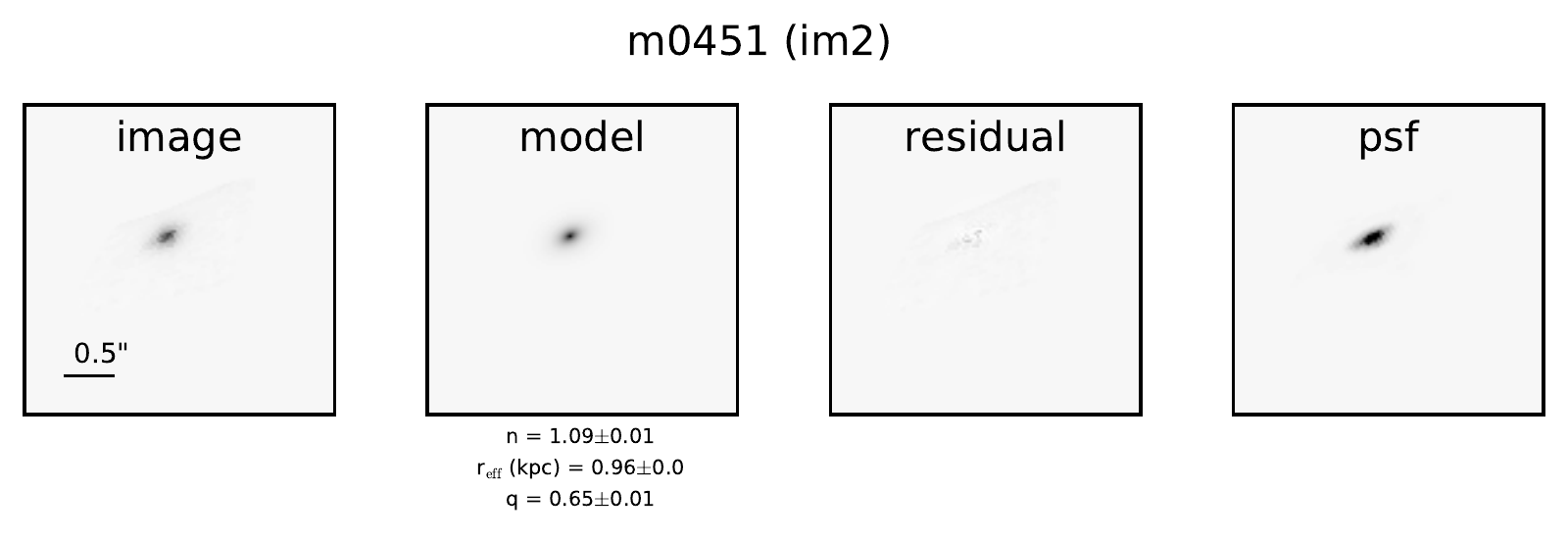}
	\end{minipage}
  	\begin{minipage}[b]{0.49\linewidth}
	\centering
	\includegraphics[angle=0,width=\textwidth]{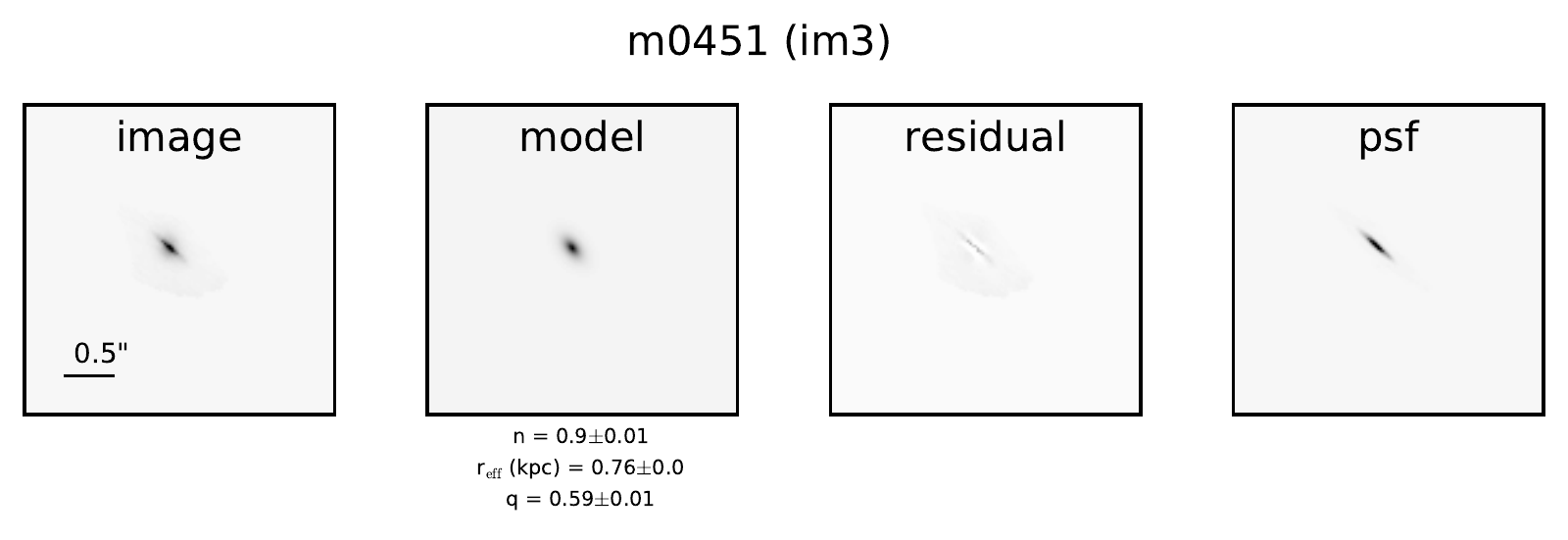}
	\end{minipage}
	\hfill
  	\begin{minipage}[b]{0.49\linewidth}
	\centering
	\includegraphics[angle=0,width=\textwidth]{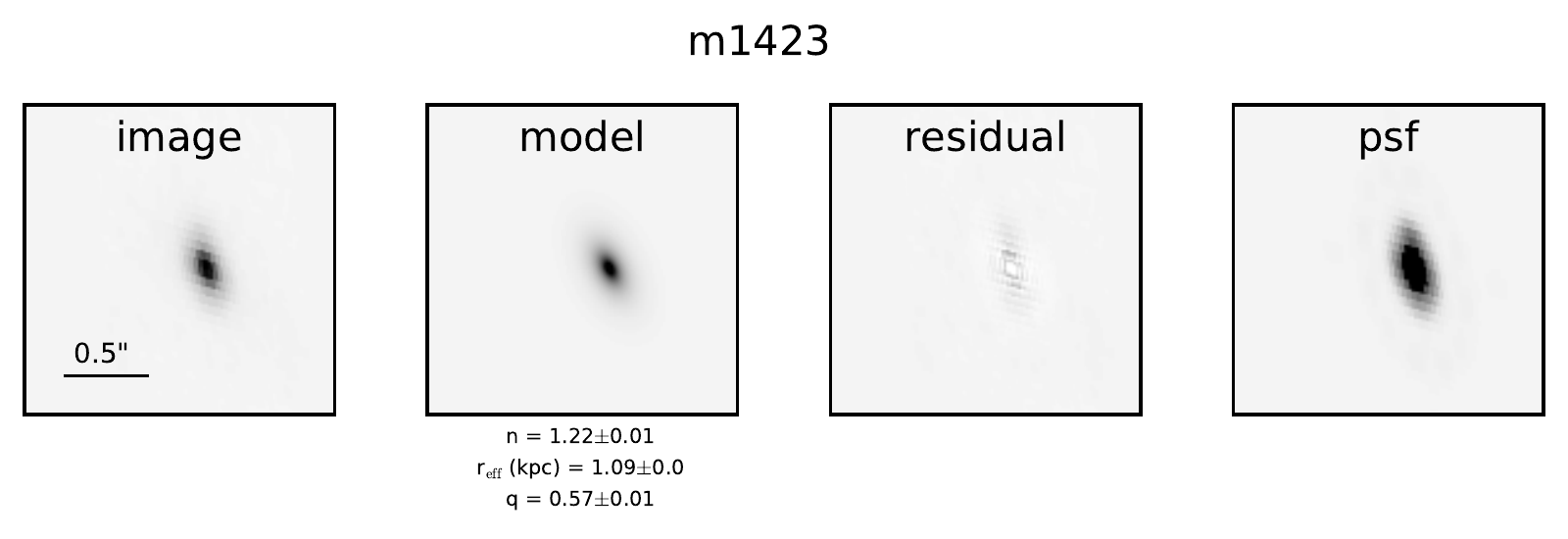}
	\end{minipage}
	\caption{
    Source-plane morphologies of our sample reconstructed from the lens models.
    The reddest \hst\ filters available are used.
    For each image, the first column shows the reconstructed image on the source plane. 
    The second column shows the best-fitting \sersic\ model as determined with \texttt{GALFIT}. 
    The third column shows the residual of the image subtracted by the model. 
    The last image shows the effective point-spread function,
    i.e., how point sources (stars) on the image plane would look like on the source plane.
    The first three columns of each multiple image are always shown with the same color scale,
    although the color scale is not identical across the images.
    }
	\label{fig:galfit}
\end{figure*}

\begin{figure}[htbp!]
	\centering
	\includegraphics[width=\linewidth]{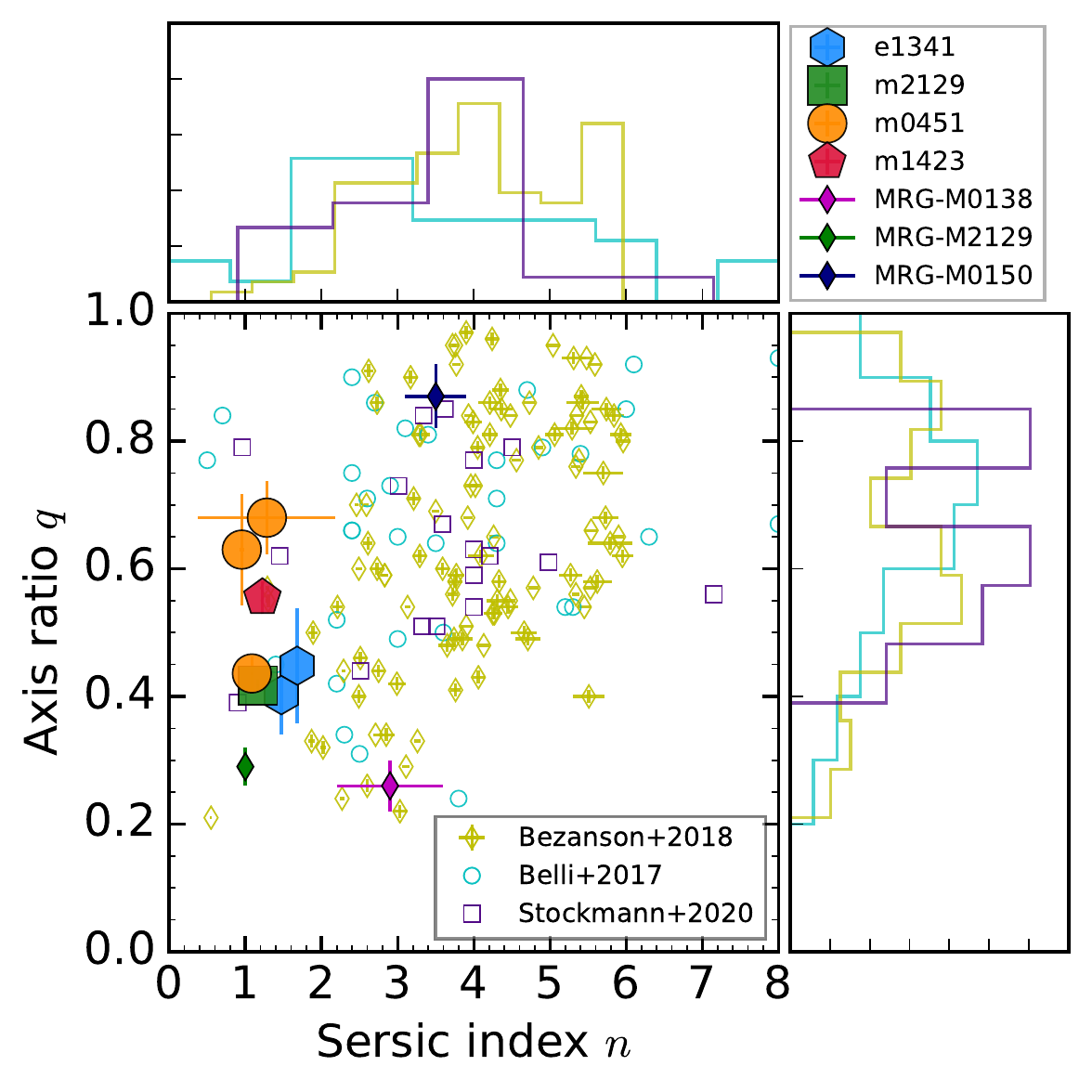}
    \caption{
    The \sersic\ indices and axis ratios of lensed quiescent galaxies from this work and \citet{Newman2018a} are individually labelled.
    They are contrasted against unlensed quiescent galaxies from the LEGA-C survey \citep[][$z=0.6-1.0$]{Bezanson2018}, from the MOSFIRE survey of the CANDELS field \citep[][$z=1.5-2.5$]{Belli2017}, and from the X-SHOOTER survey of the COSMOS field \citep[][$z=2.0-2.7$]{Stockmann2020}.
    The distribution of \sersic\ indices and axis ratios of these three samples are plotted as histograms.
    The stellar mass dependence of the \sersic\ indices and axis ratios are shown in Figure~\ref{fig:n_q_Mstar}.
    }
    \label{fig:n_q}
   \end{figure}
   
\begin{figure}[htbp!]
	\centering
	\includegraphics[width=\linewidth]{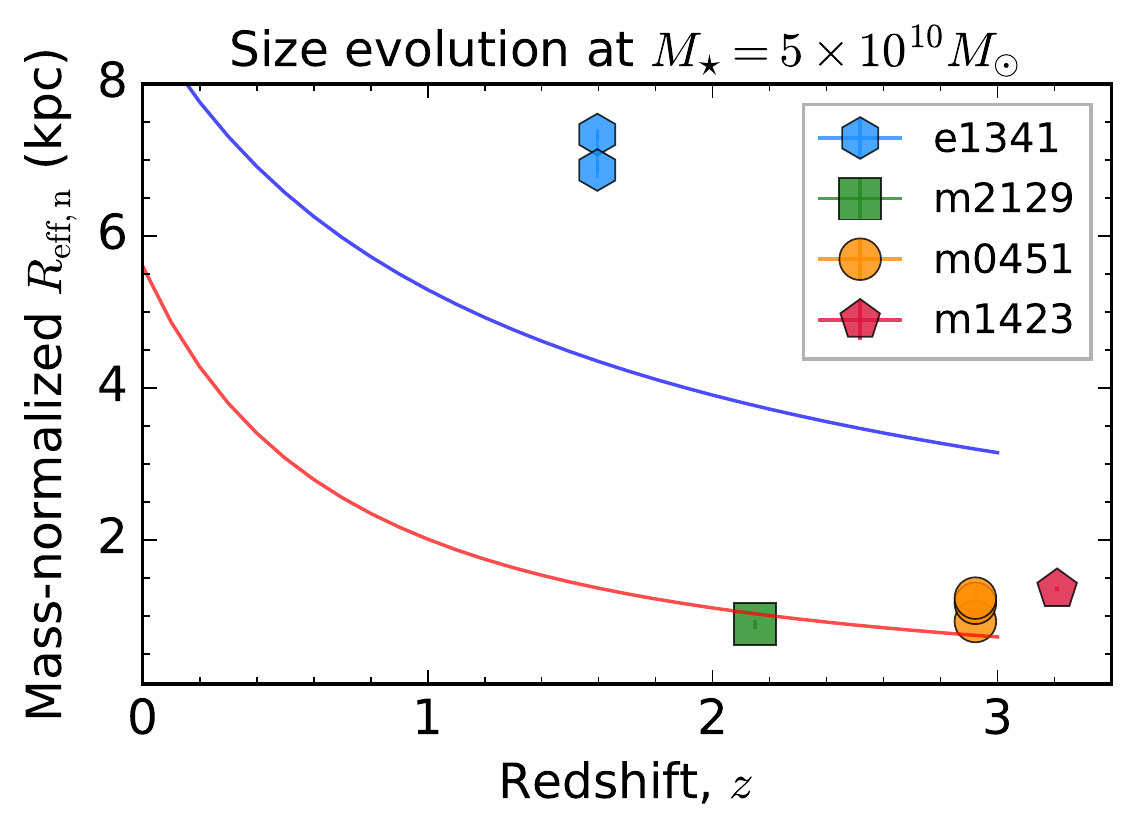}
    \caption{
    The effective radii of our sample, normalized to by their sizes as \ren\,=\,\re/(\mstar/$(5\times10^{10}M_{\odot})^{\alpha})$, are denoted as filled colored symbols.
    We adopt $\alpha$ values for early-type galaxies at the nearest redshift bins \citep[Table~1]{vdWel2014}.
    The red and blue solid curves represent the size evolution of early- and late-type galaxies with \mstar\,=\,$5\times10^{10}$\,\msun, following the parameterized redshift evolution presented in \citet[][Figure~6]{vdWel2014}.
    }
    \label{fig:size_evol}
   \end{figure}

To constrain the intrinsic morphology of our galaxies on their source planes,
we fit \sersic\ profiles \citep{Sersic1963,Sersic1968} to the lensing-reconstructed images.
Our procedures follow the approach described in \citet{Toft2017}.
In summary, to propagate the uncertainty introduced by the lensing model,
we generate 100 realizations of each reconstructed image,
and obtain the best-fitting morphological parameters to each realization using the \texttt{GALFIT} code\footnote{\url{https://users.obs.carnegiescience.edu/peng/work/galfit/galfit.html}} \citep{Peng2002,Peng2010,Peng2011}.
The best-fitting parameters are reported in Table~\ref{table:galfit}.
The quoted uncertainties are computed from the standard deviation of 100 realizations (i.e., error due to the lensing model) and \texttt{GALFIT} errors added in quadrature.
The reconstructed images, best-fitting \sersic\ models, and the residual images are shown in Figure~\ref{fig:galfit}.
Multiple image systems (\m1341\ and \m0451) provide additional constraints on the systematic uncertainties associated with the image reconstruction.
Reassuringly, the \sersic\ index $n$ and effective radii \re\ are in good agreement across the multiple images.
The axis ratio $q$, on the other hand, appears less constrained.
While the axis ratio measurements of \m1341\ are consistent across the two multiple images, those of \m0451\ are more discrepant.
The axis ratio is $q\approx0.4$ for the most magnified image (im1) compared to $q\approx0.6-0.7$ for the other two less magnified images.
Although \m0451\ has a robust lensing model based on 17 multiply imaged systems (\S\ref{sec:lensmodel}), there is possibly a systematic uncertainty in the reconstruction of the most magnified source as the position angles vary across the three images (Table~\ref{table:galfit}). 
The morphological parameters of the least magnified image (im3) should be least affected by lensing systematic uncertainties as long as it is resolved.
Overall, multiple measurements enable us to assess the lensing systematic uncertainties in the morphological parameters. The \re\ and $n$ are more robust to lens modeling uncertainties than $q$.

We examine how the stellar morphologies of lensed quenched galaxies compare to those of unlensed ones.
Figure~\ref{fig:n_q} shows a comparison of the lensed quenched galaxies in this work as well as those presented in  \citet{Newman2018a},
compared to other unlensed spectroscopic samples of quiescent galaxies \citep{Belli2017,Bezanson2018,Stockmann2020}.
The \sersic\ indices of our sample are low with $n<2$.
The apparent axis ratio have intermediate values of $q\approx0.4-0.7$,
below the median redshift relation of the apparent axis ratio presented in \citealt[][Figure 4]{Hill2019}.
These findings suggest that lensed quenched galaxies of this work clearly have lower \sersic\ indices and axis ratios compared to unlensed ones.
All but one target (\m1341) lie within the $1\sigma$ scatter of the mass-size relation of early-type galaxies at their respective redshifts \citep{vdWel2014}, as shown in Figure~\ref{fig:size_evol}.
\m1341\ has \re\,$\approx2.7$\,kpc (average of the estimate from its two reconstructed images), roughly four times larger than the mass-size relation of early-type galaxies at its redshift and \mstar,
corresponding to $\approx4.3\times$ the scatter of the relation.
This places \m1341\ above the late-type galaxy mass-size relation.

It is unclear why lensed quenched galaxies of our sample are more disk-like than unlensed ones.
The shapes of galaxies evolve with stellar mass and redshift.
In the local Universe, massive galaxies tend to have rounder stellar light profiles (higher $n$ and $q$) than lower mass galaxies \citep{Krajnovic2013}.
The trend is less pronounced at higher redshift as stellar disks are more prevalent \citep{vdWel2011,Chang2013a,Chang2013b,Hill2019}.
In Figure~\ref{fig:n_q_Mstar} we show the stellar mass dependence of $n$ and $q$.
Our lensed sample have lower $n$ than unlensed quiescent galaxies at $z=0.6-1.0$ \citep{Bezanson2018} and $z=1.5-2.5$ \citep{Belli2017}.
The lack of an unlensed comparison sample at matching $z$ and \mstar\ precludes us from concluding whether the disky profiles of our sample is due to their high redshifts and/or low stellar masses.
Furthermore, lensed arcs that have flatter light profiles are preferentially selected as spectroscopic targets in our survey for spatially resolved studies.
Another possible reason for the discrepancy is  lens model uncertainty. If the magnifications were underestimated, for example due to additional substructure missed in the model, this could lead to artificially flatter light profiles.
Lastly, \textit{HST} cannot adequately resolve the light profiles of distant compact quiescent galaxies,
so that unlensed quiescent galaxies might appear rounder than they actually are.
Higher spatial resolution imaging for a large sample of distant quiescent galaxies with \jwst\ will help address the cause of this discrepancy.

We discuss the implications of these findings in \S\ref{sec:discuss}.
A caveat is that different rest-frame wavelengths are traced for the four galaxies in our sample (Table~\ref{table:galfit}),
due to their different redshifts and filters used.
\citet[][Eqt.~2]{vdWel2014} presented a scaling relation to correct for the wavelength dependence of the effective radii. 
However this involves the use of the average size gradient, $\Delta\mathrm{log}R_{\mathrm{eff}}/\Delta\mathrm{log}\lambda$,
which is not well-characterized for early-type galaxies at our redshift and mass range.
By using their equation, we derive a correction factor of $\lesssim5\%$ in effective radii. 

\m1423\ is at the border between quiescent and star-forming (or early- and late-type) galaxies according to its position on the UVJ diagram (Figure~\ref{fig:uvj}).
We note that its small effective radius suggests that it is nearly as compact as early-type galaxies, if we extrapolate the 3DHST+CANDELS structural relation at $z=2.5-3.0$ \citep{vdWel2014} up to its redshift of 3.21.

\section{Discussion} \label{sec:discuss}

In this Section we will discuss what our results imply for the overall evolution sequence of quenched galaxies. How did they come to be, and what would they evolve into? What is responsible for quenching star formation?

\subsection{Tracking evolution by number density} \label{sec:env}

To figure out what our sample of lensed quenched galaxies would evolve into by $z\approx0$,
we estimate their halo mass (\mh) evolution in the following way.
Taking their de-lensed stellar masses,
we compute their number densities using the galaxy stellar mass function of the COSMOS field \citep{Muzzin2013b}.
We use the Number Density Evolution Redshift Code\footnote{ \url{https://code.google.com/archive/p/nd-redshift/}} \citep[\texttt{nd-redshift};][]{Behroozi2013} to calculate the halo mass evolution for a given number density of a galaxy population.
The code tracks the number density evolution due to mass accretion and mergers.
In this procedure we use a Monte Carlo simulation to propagate errors on stellar masses, the stellar mass function (due to Poisson uncertainties, photometric redshift errors and cosmic variance), as well as the number density evolution.
Figure~\ref{fig:z_Mh} illustrates the expected halo mass evolution.
To infer possible $z\approx0$ descendants we overplot the halo mass distribution of the MASSIVE sample,
as well as several well-known galaxy clusters including Coma, Persues, Fornax I, and the Local Group \citep[][and references therein]{Crook2007, Li2008, Veale2018}.

\begin{figure}[htbp!]
   \centering
   \includegraphics[width=\linewidth]{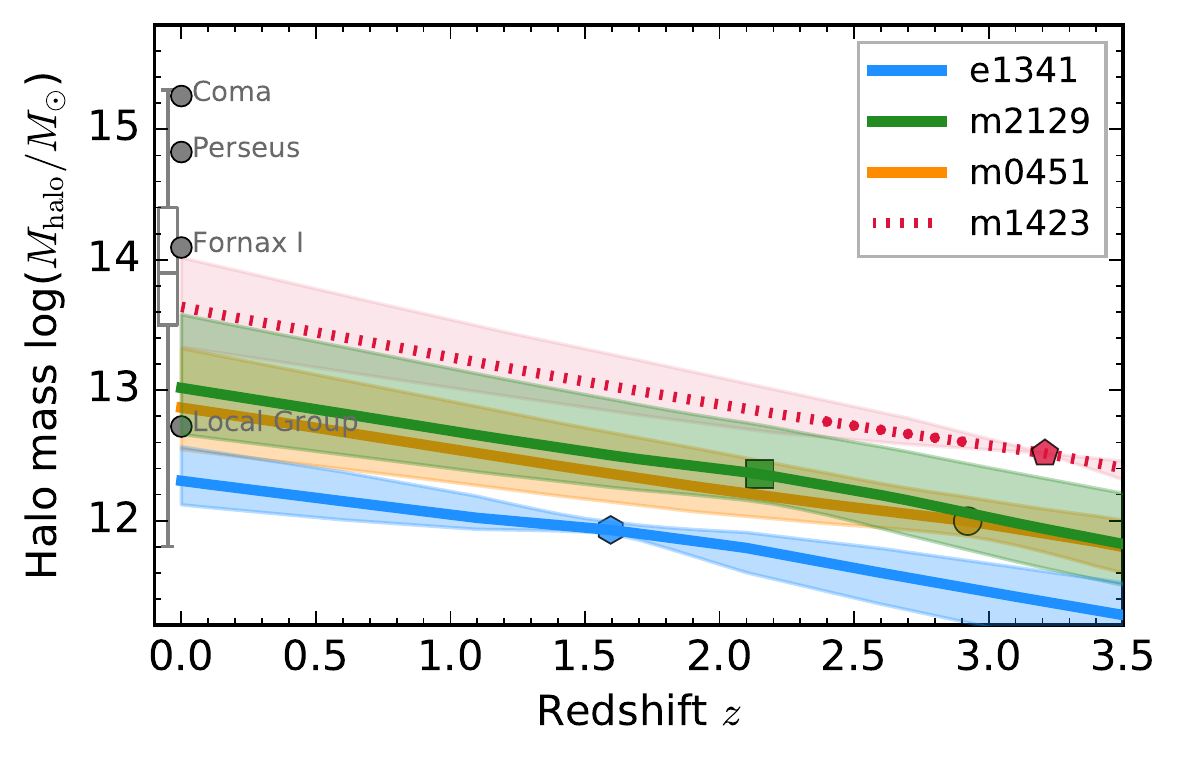}
      \caption{
       The colored lines represent the evolution of the halo mass of the lensed quenched galaxies as discussed in \S\ref{sec:env}.
       The colored shades represent the $1\sigma$ uncertainties propagated from the number densities and their evolution.
       As for \m1423, labelled as a dotted line, the shades only indicate the uncertainties for the number density evolution,
       as the lower number density error bounds reach below $10^{-5}$\,Mpc$^{-3}$ such that the \texttt{nd-redshift} does not return reliable results.
       The boxplot shows the MASSIVE sample at $z\approx0$ \citep{Veale2018}:
        the central line represents the median,
        the box represents the interquartile range,
        and the whiskers represent the full range.
        The boxplot has been offset in redshift slightly to improve visualization.
        Several well-known nearby galaxy clusters are individually labeled for comparison \citep{Crook2007,Li2008}.
        }
    \label{fig:z_Mh}
   \end{figure}

Our calculation suggests that the embedding halos of our sample of lensed quenched galaxies would evolve into intermediate-sized galaxy groups/clusters with log(\mh)\,$\approx12-14$ more like the Local Group than the Coma cluster. 
Their halo masses at $z=0$ is below the median of that of the MASSIVE sample,
and more than an order of magnitude less than that of the most massive galaxy clusters like Coma and Perseus.
The halo mass evolution of our sample lies below those of various cluster surveys like CLASH \citep{Postman2012} or GOGREEN \citep{Balogh2017}.
The findings are in line with our expectations, given that our sample is more representative and numerous than the most massive galaxies at their epoch.
An implicit assumption of this calculation is that the lensed quenched galaxies reside in field environments, 
such that the COSMOS stellar mass function provides a reliable estimate of their number densities.
It is known that \m0451\ resides in an overdense environment that resembles a compact group (\S\ref{sec:targets}; \citealt{MacKenzie2014,Shen2021}).
As for the other targets there are no indications thus far that they reside in overdensities,
although it cannot be ruled out because of the inherent challenge in quantifying the environment of a lensed volume.
So for \m0451\ the halo mass by $z=0$ could be higher than estimated.
Another limitation is that the two highest redshift targets,
\m0451\ and \m1423,
have stellar masses below the 95\%-completeness limit of the COSMOS survey,
so their number densities are less certain.
The various conversions used to estimate halo masses could add further uncertainty to this comparison.
At any rate, this exercise serves to provide an order-of-magnitude estimate of the $z=0$ descendant halo mass.
It is justified to conclude that halos containing the lensed galaxies studied in this work would not evolve into Coma-like clusters by $z=0$,
unless they are highly clustered.

\subsection{Progenitors} \label{sec:progenitors}

Our sample of galaxies has already assembled a significant mass of stars when the Universe was young.
Our analysis in \S\ref{sec:stellarpop} indicates that their star formation histories are rapid, having formed 80\% of stellar mass within 70 -- 470\,Myr.
These timescales are comparable or shorter than the median ages of the stellar populations,
and much shorter than the age of the Universe at their respective redshifts ($\approx2-4$\,Gyr).

How do our results compare with studies of other distant quenched galaxies?
A meaningful comparison can only be made if the star formation timescale is measured with the same methods (full spectral fitting vs line indices, parameterization of star formation history) and defined in the same way. 
Given the variety of methods adopted in the literature, here we only attempt to conduct an order-of-magnitude comparison to get an impression of how the derived values compare.
As near-infrared absorption line spectroscopy is time-consuming,
constraints on the star formation history of quenched galaxies are limited to the most luminous ones that are more massive than our sample.
Bearing these differences in mind,
the rapid star formation duration of our sample is in qualitative agreement with those of the massive, quiescent galaxies presented in \citet[][$\tau$ = 10 -- 80\,Myr, $z=1.5-2.1$]{vdSande2013}, \citet[][$\tau<180$\,Myr, $z=1.9-2.6$]{Newman2018a}, and \citet[][$\tau$ = 10 -- 16\,Myr, $z=3.8-4.0$]{Valentino2020},
and are similar or shorter than the quiescent galaxies at $z=1.4-2.6$ ($\tau=0.2-3.2$\,Gyr, \citealt{Zick2018}; see also \citealt{Kriek2019}).

Altogether these findings lend evidence to a rapid build-up of stars through accelerated growth.
The mere fact that massive, quiescent galaxies exist in a young Universe requires that they formed stars at a higher rate before \citep[see also][]{Pacifici2016}.
As inferred from the median SFH, 
the peak SFR of our sample ranges from 1800 to 6000\,\msun\peryr.
The most massive (log(\mstar/\msun)$>11$), quiescent galaxies are shown to be consistent with having $z=3-6$ submillimeter galaxies (SMG; with mean duty cycle of $\approx40$\,Myr) as their progenitors by means of number density and size comparison \citep{Toft2014}.
Our sample probing a lower stellar mass regime appears consistent with having experienced an SMG phase, although at lower stellar masses than those presented in \citet{Toft2017}.
It is worth noting that \m0451, together with all its companion galaxies within the lensed group, forms a submillimeter arc with total SFR $=(450\pm50)$\,\msun\peryr \citep{MacKenzie2014}.
The group members span two orders of magnitude in SFR and those with CO detections are expected to deplete their molecular gas within $<0.14-1.0$\,Gyr \citep{Shen2021}.
The lensed group in which \m0451\ resides may be a example of accelerated growth in dense environments.
Compact star-forming galaxies at $z\gtrsim2-3$ have been proposed as another progenitor for $z<2$ quiescent galaxies \citep{vDokkum2015,Barro2017a,GGuijarro2019},
given their similarly compact stellar sizes, number densities, and modest SFR ($\sim$100\,\msun\peryr).
While it is plausible that our quenched galaxies went through such a phase of compact star formation,
their star-forming progenitors is likely at $z\approx2-4$ as inferred from their best-fit SFHs.
Current work on compact star-forming galaxies are limited to $z<3$ to date as \hst\ cannot probe the rest-frame optical sizes of galaxies beyond $z=3$.
\jwst\ will soon enable a census of $z>3$ galaxies in order to identify the progenitors of these early-quenched galaxies.

Aside from number density and size comparisons,
one could also gain insights into possible progenitors of our quenched galaxies using timescales.
The star formation duration, \twidth, of our quenched galaxies is shorter than the molecular gas depletion time of ``main-sequence'' star-forming galaxies at similar $z$ and stellar mass \citep[$0.5-0.7$\,Gyr;][]{Tacconi2018} by a factor of a few to more than an order of magnitude.
These ``main-sequence'' star-forming galaxies are thus unlikely to be the \textit{immediate} progenitor to our quenched galaxy sample.
The only distant galaxies with depletion time as short as $<200$\,Myr are submillimeter galaxies \citep[e.g.,][]{Bothwell2013}, compact star-forming galaxies \citep[e.g.,][]{Spilker2016,Popping2017}, 
and starbursting radio galaxies \citep{Man2019}.
These galaxies have star formation rates above or within the ``main-sequence'' of star-forming galaxies,
but what sets them apart from the ``main-sequence'' is their high star formation rates for their molecular gas mass,
i.e., they have high star formation efficiency and equivalently short gas depletion timescale \citep{Elbaz2018}.
The star formation histories of our quenching galaxies are certainly compatible with having experienced such a rapid, efficient phase of star formation prior to quenching.

\subsection{Descendants} \label{sec:descendants}

Having discussed the possible progenitors of our sample in \S\ref{sec:progenitors}, 
we now turn to their subsequent evolution in order to fully explore their evolutionary scenario over the next $9-12$\,Gyr until the present day.
The intermediate stellar masses of our sample imply that they would evolve into relatively massive galaxies in the local Universe. 
Identifying their low-redshift descendants is like finding needles in haystacks,
as massive galaxies become more numerous and more of them quench their star formation over time \citep{Muzzin2013b,Davidzon2017}.
Thankfully the robust constraints on the stellar populations and morphologies provide clues to infer their evolution.

\begin{figure}[htbp!]
   \centering
   \includegraphics[width=\linewidth]{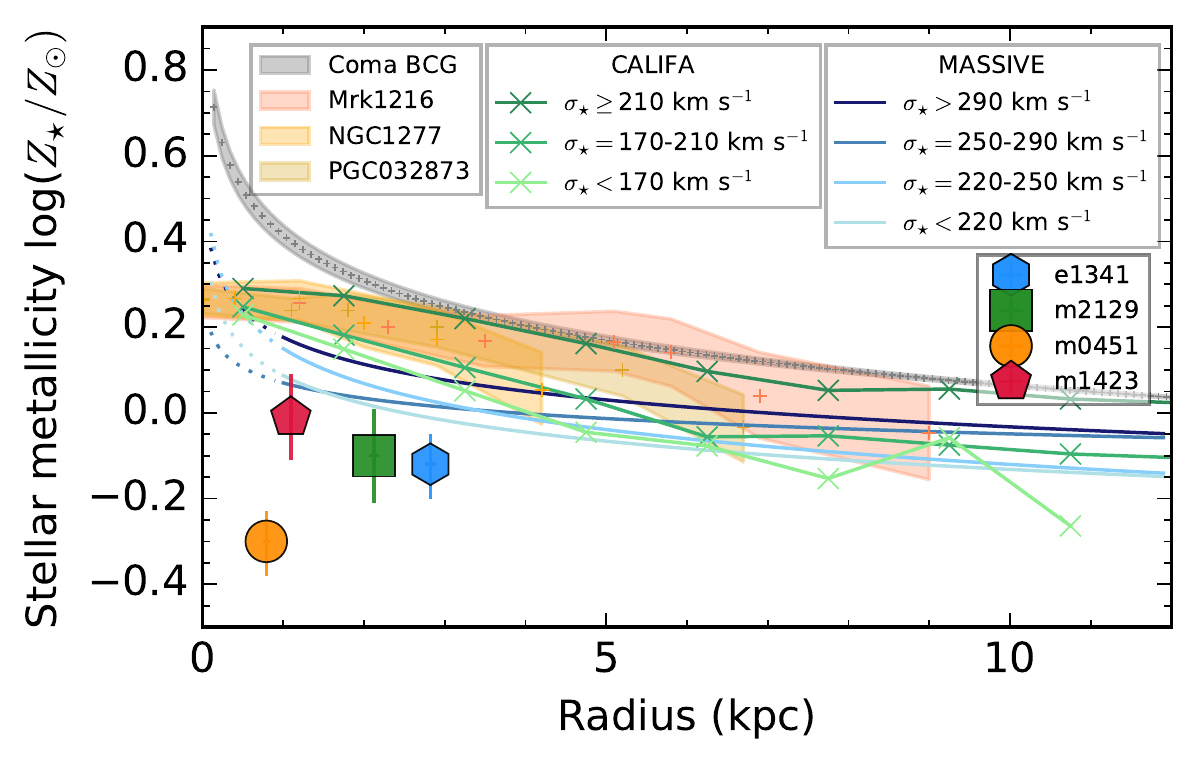}
      \caption{
       Stellar metallicities of our sample,
       compared with the metallicity gradient of nearby galaxies.
       The green curves with crosses are median metallicity gradients of early-type galaxies in the CALIFA survey \citep{Zibetti2020}.
       The blue curves are fits of massive early-type galaxies in the MASSIVE survey \citep{Greene2015}.
       The dotted portion of the lines are extrapolation of the best fits to small radii.
       Both samples are shown in bins of stellar velocity dispersions.
       The gray curve represent the best-fit to the metallicity gradient of NGC\,4889, one of the two bright central galaxies (BCG) of the Coma cluster \citep{Coccato2010}.
       The orange shades represent the metallicity gradients of local massive, compact, quiescent galaxies (or ``relic galaxies'', \citealt{FMateu2017}).
    }
    \label{fig:Z_gradient}
   \end{figure}

Archaeological studies of nearby massive, early-type galaxies suggest that the bulk of their stars formed early by $z\approx1-2$ \citep{Thomas2005,McDermid2015}\footnote{Although the stars in today's massive galaxies formed early, galaxies could have assembled later through mergers long after stars formed.}.
Is our sample of early-quenched galaxies the precursors of the metal-rich cores of nearby early-type galaxies?
We compare the spatially-integrated stellar metallicities of our sample to the resolved measurements of nearby massive early-type galaxies.
Numerous integral field spectroscopic surveys provide stellar metallicity gradient measurements.
Most such surveys of nearby massive galaxies resolve well within once or twice the effective radii \citep[e.g., SAURON, ATLAS3D, MANGA, SAMI;][]{Kuntschner2010,Krajnovic2020,GDelgado2015,MNavarro2018,Bernardi2019,Ferreras2019b,Oyarzun2019},
yet few probe beyond the cores because of the limited field of view.
Thus we compare our integrated stellar metallicities with results from the MASSIVE and CALIFA surveys that provide gradient measurements to the largest spatial extent, out to $\approx2.5$\,\re.
We use the gradient fits of the MASSIVE sample \citep[][Table~2]{Greene2015}, in units of kpc, to compute the stellar metallicity as [Z/H] = [Fe/H] + 0.94 [Mg/Fe] \citep[][Equation~4]{Thomas2003}.
As for the CALIFA survey we show the median stellar metallicity gradient of early-type galaxies presented in \citet{Zibetti2020}, in bins of semi-major axis in kpc.
We also include the stellar metallicity gradient of NGC\,4889 for comparison.
NGC\,4889 is one of the two brightest cluster galaxies in the Coma Cluster \citep{Coccato2010}.
Lastly we compare our results against the stellar metallicity gradients of the ``relic galaxies'',
i.e., local massive, quiescent galaxies that are compact in size \citep{Trujillo2014,FMateu2017}.

The comparison is shown in Figure~\ref{fig:Z_gradient},
where we label our sample at the source-plane effective radii of the spectroscopic images.
The stellar metallicities of our sample are lower than those found in the cores of local early-type galaxies. Further chemical enrichment, perhaps by gas-rich mergers and/or star formation rejuvenation, needs to take place if they are to evolve into metal-rich cores of local massive, early-type galaxies.
There are two potential caveats with the comparison shown in Figure~\ref{fig:Z_gradient}.
Firstly, it is not straightforward to compare a spatially-integrated measurement with a gradient.
A better comparison can be made by deriving a spatially-integrated measurement from the gradient fit and the luminosity or mass profile.
Secondly, the derived metallicities depend on the calibration (absorption line indices or full spectral fitting) as well as the assumptions involved (e.g., star formation history, metal yield of various stellar types).
Resolved elemental abundance analysis is required to address these issues \citep[see][]{Jafariyazani2020}.
A detailed comparison addressing these caveats is beyond the scope of this Paper.

The stellar mass range of our sample suggests that they are more likely to be fast rotators rather than slow rotators,
if we take the stellar kinematics and mass distribution of $z\approx0$ as a reference \citep{Emsellem2011,Veale2017b}.
Fast rotators are expected to be more than an order of magnitude more numerous than slow rotators at $z\approx2$ \citep{Khochfar2011}, as dry merging is a dominant mechanism to reduce the spin of galaxies \citep{Naab2006b,Lagos2018a,Lagos2018b} and should be more prevalent at later cosmic times than wet mergers \citep{Hopkins2010}.
Like fast rotators, the apparent axis ratios of our sample span a wide range as shown in Figure~\ref{fig:n_q}.
Their low sersic index ($n<2$) lends further support to them being fast rotators,
as all slow rotators have $n>2$ \citep{Krajnovic2013}.
Resolved absorption line spectroscopy is needed to confirm the kinematic nature of our sample.
Indeed \m2129\ is shown to have rotation-dominated stellar kinematics \citep{Toft2017,Newman2018b}. 
\jwst\ will enable us to obtain resolved kinematics for these compact galaxies in the near future.

Future evolution of early quenched galaxies depends on their ability to rejuvenate star formation, 
if molecular gas is made available for star formation again,
e.g., through gas accretion, mergers.
Studies of the star formation histories of $z<1$ quenched galaxies do reveal that a small fraction had evidence for rejuvenated star formation in the recent past \citep{Chauke2018}.
A recent analysis of a lensed quiescent galaxy at $z=1.9$, MRG-S0851, reveals evidence for star formation rejuvenation in the inner kpc within the past $\sim$100\,Myr. If representative of galaxies having similar spectral energy distribution, the abundance implies that $\approx$1\% of massive quiescent galaxy at $z=1-2$ are potentially experiencing star formation rejuvenation \citep{Akhshik2021}. In two future works we will report on the molecular gas content of quenching galaxies including targets of this work (Whitaker et al. submitted, Man et al. in preparation).

\subsection{Implications for quenching mechanisms} \label{sec:quench}

Gravitational lensing and deep spectroscopy have provided us an exquisitely deep view into the properties of ordinary galaxies as they quench their star formation.
What insights can we gain on the mechanism responsible for their decline in star formation activity?
Gas needs to be brought into galaxies and sufficiently cool and settle in order to form stars.
Star formation quenches, temporarily or permanently, if one or more of these necessary conditions is lacking.
In this subsection we explore how our observations enable us to constrain the cause of quenching.

An important discriminant of star formation quenching mechanisms is the timescale over which they operate.
The stellar population analysis in \S\ref{sec:stellarpop} suggests that our sample has experienced a rapid star formation history,
with $\tau < 0.2$\,Gyr.
An alternative illustration is by examining their position on the UVJ diagram,
a diagnostic first developed to separate star-forming galaxies from quiescent ones \citep{Wuyts2007,Williams2009,Muzzin2013b,Whitaker2013}.
Galaxies evolve on the UVJ diagram as they quench their star formation and become old \citep[e.g.,][]{Barro2014,Merlin2018}.
In Figure~\ref{fig:uvj} we compare the rest-frame (U--V) and (V--J) colors of our sample with the fast and slow quenching models.
The (U--V) and (V--J) colors listed in Table~\ref{table:uvj} are computed from the SED models\footnote{The conclusions of this comparison remain unchanged if we measure the colors from the observed photometry instead. Template SEDs are commonly used to interpolate between observed filters to obtain rest-frame colors in any case, see for example \citet{Taylor2009}.}.
The model tracks shown in Figure~\ref{fig:uvj} are from \citet{Belli2019},
assuming two $\tau$ models for fast quenching ($\tau=0.1$\,Gyr) and slow quenching ($\tau=1$\,Gyr), respectively.
The colors are dust-corrected by assuming an evolving \AV\ that declines with the SFR over time, starting from 2 to 0.4, and assuming \RV\ =~4.05 and the \citet{Calzetti2000} dust attenuation law.
It is apparent that the slow quenching model is incompatible with our sample:
a galaxy that follows the slow quenching track with $\tau\gtrsim1$\,Gyr would stay star-forming in the first 3\,Gyr since the onset of its star formation,
i.e., within 3 times of the e-folding timescale.
This corroborates our findings in \S\ref{sec:stellarpop} that indicates rapid star formation timescales of $\tau\lesssim0.2$\,Gyr and \twidth~=~0.07~--~0.47\,Gyr.
This conclusion stands even for higher initial values of \AV\ or adopting \RV\ ranging from 3 to 6.
The lower (V--J) values of \m1423\ and \m0451\ can be attributed to their shorter star formation histories ($\tau<0.1$\,Gyr), stronger burst strength or possibly different dust correction \citep[see also][]{Barro2014,Merlin2018,Wu2020}.
Overall, the UVJ diagram provides an independent illustration of a fast star formation quenching scenario,
in qualitative agreement with spectroscopic investigations of $z\approx2$ quiescent galaxies \citep{Kriek2016,Zick2018,Newman2018a,Belli2019}.

\begin{figure}[htbp!]
   \centering
   \includegraphics[width=\linewidth]{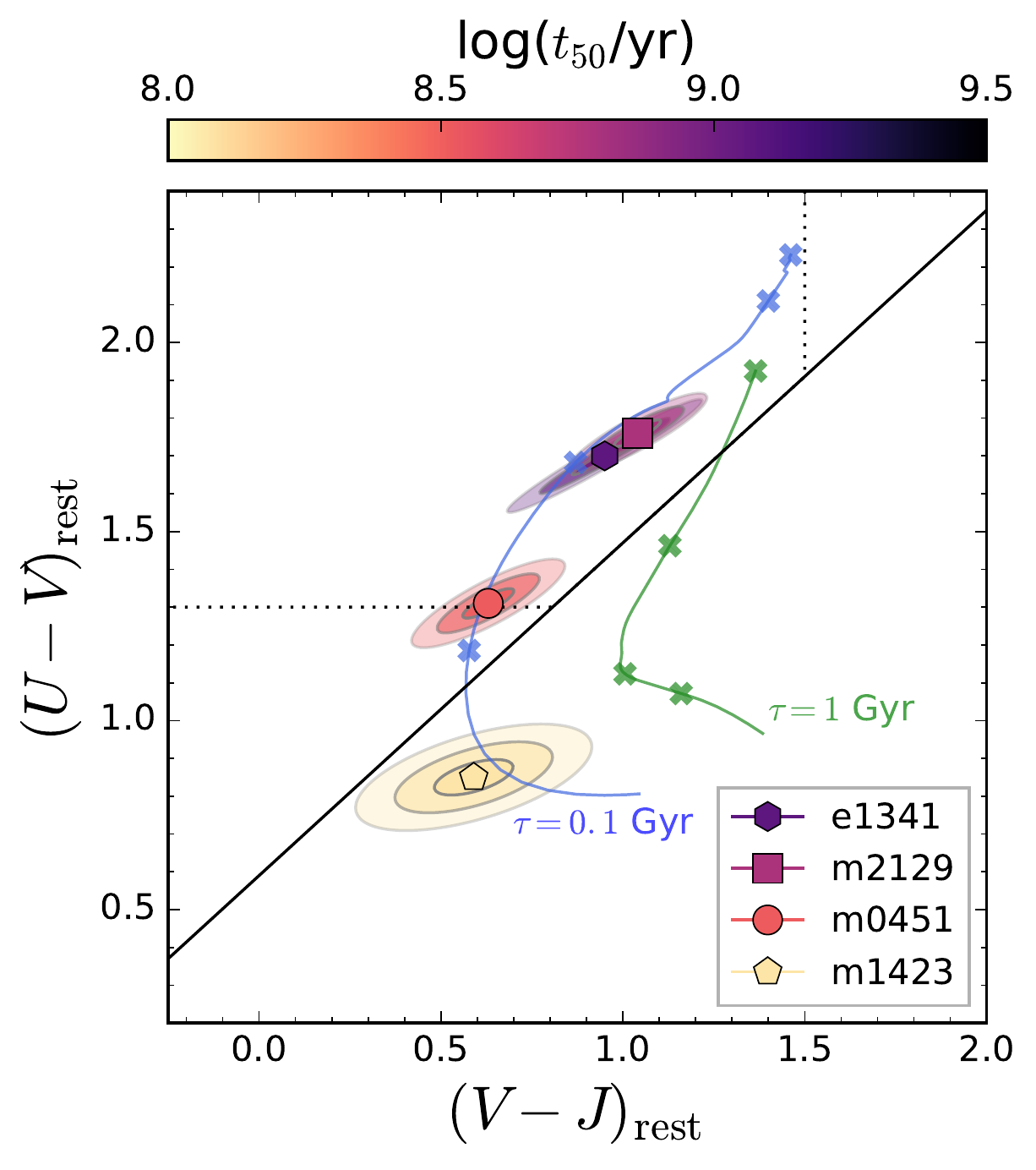}
      \caption{
       Rest-frame U--V and V--J magnitudes of our sample are denoted as filled circles, color coded by their \thalf. 
       The contours show the 1$\sigma$, 2$\sigma$ and 3$\sigma$ distributions of the colors propagated from the stellar population fitting.
       The dotted line shows the color criteria for separating quiescent (upper left corner) from star-forming galaxies as in \citet{Muzzin2013b},
       while the solid shows the modified criteria presented in \citet{Belli2019}.
       The colored curves show the evolution of fast and slow quenching model tracks presented in \citet{Belli2019}, with $\tau=0.1$\,Myr (blue) and $\tau=1$\,Gyr (green), respectively. 
       The crosses mark the time intervals at 0.5, 1, 3, and 5\,Gyr.
    }
    \label{fig:uvj}
   \end{figure}

Stellar morphology provides another clue to understanding quenching.
The lensing-reconstructed \hst\ images show that their stellar structures are disk-like (\S\ref{sec:morphology}).
This suggests that quenching does not necessarily require or coincide with bulge formation.
Similar findings have been reported in morphological analyses of photometrically-selected massive, early-type galaxies (\citealt[][$z=1-3$]{Bruce2012}; \citealt[][$z=0.6-1.8$]{Chang2013a}), as well as in the analysis of stellar kinematics of massive, quenched galaxies at $z=2-2.6$ \citep{Toft2017,Newman2018b}.
This work confirms the lack of coincidence between quenching and morphological transformation extends to intermediate-mass galaxies (log(\mstar/\msun)$=10.2-10.6$) at $z=1.6-3.2$.
Their disky morphologies are in line with the higher prevalence of flat, oblate morphology amongst photometrically selected quiescent galaxies selected at $z\approx2.5$ than their counterparts in the present day \citep{Stockton2008,vdWel2011,Chang2013a,Chang2013b,Hill2019}.
In fact, massive UVJ-selected quiescent galaxies at $z=2-3.5$ are as flat as star-forming galaxies \citep{Hill2019}.
In terms of stellar morphology they resemble local fast-rotators rather than slow-rotators.
Quenched galaxies at $z\approx2$ may thus have some level of rotation in their stellar velocity fields, as already confirmed in spatially resolved studies \citep{Toft2017,Newman2018b}. 
The compact sizes and thus high stellar mass densities, as well as their disky kinematics, all point towards a dissipative formation process such as a gas-rich merger \citep{Cox2006,Naab2006b,Naab2014,Wuyts2010,Lagos2018b}.
Although the merger fraction of massive galaxies at $z\gtrsim2$ is less than 10\% \citep{Man2016b},
there is emerging evidence that mergers may be more prevalent among quenched galaxies \citep{Glazebrook2017,Stockmann2020}.
The absence of a prominent stellar bulge or bar disfavors morphological quenching \citep{Martig2009} or bar quenching \citep{Khoperskov2018} as plausible quenching mechanisms.
These processes, however, may help to maintain the quiescence of star formation in the long run.
If our sample is to evolve into present-day slow-rotators, they would need to substantially grow their bulges later on in a process unrelated to quenching, e.g., dry major mergers.

Another clue comes from the gas conditions of quenching galaxies.
The faint emission lines in our sample suggests the presence of warm ($T\sim10^{4}$\,K), low-ionization gas as discussed in \S\ref{sec:ionized}.
Furthermore the youngest quenching galaxies show evidence for outflowing warm gas in their \mgii\ profiles (\S\ref{sec:mgii}, \S\ref{sec:ionized}).
Outflowing warm gas is detected in post-starburst galaxies at $z<1.4$ \citep{Tremonti2007,Coil2011,Sell2014,Baron2017,Baron2018}.
\citet{Maltby2019} reported tentative hint of faster outflows as seen in the \mgii\ blue-shifted absorption in younger post-starburst galaxies.
Our findings are in good agreement with the expectation that \mgii\ absorption is more likely to be detected in more recently quenched galaxies ($\lesssim 500$\,Myr as inferred by their light-weighted ages) than older ones,
as shown in a study of galactic wind in K+A galaxies \citep{Coil2011}.
Spatially-extended, redshifted \Lya\ emission is present in m0451 \citep[][$z_{\mathrm{Ly}\alpha}=2.93$]{Jauzac2020} across the three multiple images as detected with the VLT/MUSE spectrograph.
The extended \Lya\ emission in quenched galaxies may originate from photon scattering, galactic winds or unresolved satellite galaxies \citep{Taniguchi2015}. 

While these findings do not readily imply that outflows cause star formation quenching,
understanding the relation between the two phenomena could better constrain how quenching took place.
Although AGNs are commonly thought to drive fast outflows ejecting gas beyond the gravitational potential of massive galaxies \citep{Tremonti2007,Baron2017,Baron2018,FSchreiber2019},
compact starburst may also be capable of doing so \citep{Sell2014,Rupke2019}.
Does the fast outflow become less prevalent as the stellar populations age because their energy sources (young stars and/or AGN) have faded,
or is it because previous outflows have cleared them of warm gas?
The latter is unlikely the case:
the \mgii\ absorption of the entire sample is deeper than expected from photospheric absorption alone, suggesting that warm gas is present after quenching and not completely evacuated from the galaxies (see \citealt{Jafariyazani2020} for a related discussion on NaD absorption).
On the other hand,
SMBH accretion rates can vary over much shorter timescales than galaxy star formation \citep{Hickox2014}.
It is plausible that an AGN outflow persists for $\sim100$\,Myr after the AGN has turned off \citep{King2011,Zubovas2014}.
There is also ample evidence for the peak of AGN accretion to occur $\gtrsim100-250$\,Myr after the peak of star formation \citep{Schawinski2009b,Wild2010,Volonteri2015a,Volonteri2015b,Baron2018,Falkendal2019}.
The answers to these questions are fundamental to understanding how star formation quenching proceeds.
Observations with the \jwst\ and other multi-wavelength facilities will tackle these questions.

What do our results imply for how galaxies quench their star formation?
A different perspective can be gained by asking whether a mechanism is needed to stop galaxies from forming stars,
or if galaxies simply have an accelerated star formation episode and stay quiescent thereafter \citep{Abramson2016}.
It is plausible that the mechanism that quench star formation of galaxies is not the one responsible for maintaining their quiescence,
given the very different timescales considered.
Absorption line spectroscopy of $z\gtrsim2$ massive, quenching galaxies including our sample provides solid indications that they have experienced a rapid star formation history in the first few billion years after the Big Bang.
If quenching is simply a rapid decline of star formation without the need for an external actor,
the question then becomes what causes a rapid gas conversion to stars (``starburst'').
Compressive motions and effective angular momentum loss of gas can facilitate starbursts,
and so can gas-rich major mergers, both within a Gyr or so \citep{Barnes1991,Hopkins2008b}.
Gas outflows driven by AGN and/or starburst can act in concert with rapid gas consumption during starburst to rapidly clear galaxies of cold gas \citep[][and references therein]{Man2019}.
Simulations by \citet{Su2019} demonstrate that stellar feedback, morphological feedback, magnetic field, thermal conduction and stellar cosmic rays do not effectively quench the star formation of massive galaxies.
This supports the idea that other processes like AGN feedback are required,
although there is a broad variety in AGN feedback implementations in simulations
 \citep[e.g.,][]{Sijacki2007,Booth2009,Dubois2012,Su2020}.
This work presents tentative evidence that as star formation quenches, the mode of SMBH growth transitions from radiative-mode to jet-mode
(\S\ref{sec:oiii}).
This is in good agreement with low-$z$ studies \citep{Best2012},
and corroborates findings of faint AGN in massive quiescent galaxies at $z\gtrsim1$ \citep{Olsen2013,Man2016a,Barisic2017,Aird2019}.
More observations are needed to inform whether and how AGN affect star formation of host galaxies,
particularly faint AGN that are more ubiquitous than the quasar phase.
On the other hand, figuring out their future evolution until the present day would require the knowledge of whether molecular gas is present or can be made available again to form stars.
In a future work we will quantify the amount of molecular gas of our sample using ALMA observations.
To identify whether and how gas cools, 
we need measurements of the gas temperature, density, and metallicity of various gas phases to constrain the cooling functions.
These measurements are crucial to properly characterize the gas conditions in order to understand how and why massive galaxies experience a decline in their star formation.
Ultimately, we need deep, multi-wavelength spectroscopic observations to constrain the relevant timescales and trace any past and ongoing star formation activity, gas accretion and outflows, and AGN activity.
Only then can we begin to disentangle the intricate relations between stars, gas, blackholes, and how they shape the evolution of galaxies.

\section{Conclusions} \label{sec:conclusions}

This paper presents the analysis of VLT/\xs\ spectroscopic and \hst\ imaging observations of four quenched galaxies at $z=1.6-3.2$ that are gravitationally lensed by foreground clusters.
Their magnification factors range from $\approx3$ to 30, affording us an exquisitely deep view of their stellar populations and morphologies.
The photometry and spectra have been fitted with the \bagpipes\ stellar population synthesis fitting code.
Our main findings are as follows:

\begin{itemize}
    \item The four quenched galaxies have intermediate stellar masses (log(\mstar/\msun)=10.2--11.2), or $0.1-3\times$ the characteristic stellar mass at their respective redshifts. The median ages of the stellar populations range between \thalf\, $\approx$ 0.12 and 1.2\,Gyr, and they formed 80\% of their stellar masses within 0.07 -- 0.5\,Gyr. Their specific SFR span a range of log(sSFR/\peryr) from $-8.4$ to $-11.2$. 
    Three galaxies lie below the sequence of star-forming galaxies, where the youngest target is a rare example of a galaxy caught shortly after quenching as its \oii\ emission implies a SFR $\approx2$\,dex below that estimated from stellar populations fitting. 
    \item The sample has stellar metallicities of $\approx0.5-1$ solar value as inferred from \bagpipes\ fitting. However, systematic uncertainties inherent to stellar population synthesis modeling and metallicity calibration preclude a fair comparison with literature metallicity measurements across redshifts in a consistent manner at this stage. Further work is necessary to quantify such systematic effects and to self-consistently model the chemical evolution of galaxies.
    \item All targets in our sample show \mgii\ $\lambda\lambda\,2796,2804$ absorption. While stellar photospheres may partially account for the absorption, additional contribution from warm gas is required. Galactic outflows are present in the most recently quenched galaxies as evidenced by their blue-shifted absorption and/or redshifted emission.
    \item Faint emission lines are detected in some of our targets. The \oii\,$\lambda\lambda$\,3726,3729\,\AA\ fluxes of our sample are consistent with having quiescent SF. The \oiii\,$\lambda$\,5007\,\AA\ fluxes are at or below the faint end of $z\approx2$ AGN surveys.
    We rule out the presence of type-1 AGN in our sample.
    The youngest two targets may host type-2 AGN as inferred from their \oiii/\oii\ ratio and might be responsible for driving the \mgii\ outflow.
    If we attribute the \oiii\ emission to AGN only, this implies that the SMBHs of the second youngest target (\thalf$\approx0.3$\,Gyr) is accreting in radiative-mode at a few percent of the Eddington limit, where the older two targets (\thalf$\approx0.6-1.2$\,Gyr) accrete in jet-mode at less than 1\% of the Eddington limit.
    \item We use the lens models to reconstruct the rest-frame optical light profiles of our sample on their source planes.
    Their light profiles are best-fitted with low \sersic\ indices of $n<2$ and intermediate axis ratios of $q\approx0.4-0.7$. 
    The targets lie within the mass-size relation for early-type galaxies except for \m1341\ which is as extended as late-type galaxies at its mass and redshift.
    Our sample is uniformly disky ($n<2$), suggesting the need for additional morphological transformation if they are to evolve into metal-rich bulges.
    \item Altogether our results imply that star formation quenching at high-redshift is a rapid process ($<1$\,Gyr). Galactic-scale outflow of warm gas are detected in the most recently quenched targets, perhaps driven by faint, radiatively-accreting type-2 AGN or a previous starburst. Contrary to some claims, quenching does not require nor synchronize with the formation of prominent stellar bulges.
\end{itemize}

This sample forms the backbone of the REsolving QUIEscent Magnified Galaxies Survey (REQUIEM-2D).
\hst\ grism observations will enable measurements of spatially resolved stellar populations, in order to quantify age and SFR gradients \citep{Akhshik2020,Akhshik2021}.
ALMA surveys of the CO and dust continuum will provide molecular gas mass measurements (Whitaker et al. submitted, Man et al. in preparation).
With the launch of the \jwst\ we anticipate huge leaps in the characterization of distant quiescent galaxies.
NIR imaging surveys with the \jwst/NIRCam will enable the identification of $z>3$ quiescent galaxies.
Their number density evolution is vital for understanding when the first galaxies become quenched.
\jwst/NIRCam can better resolve their light profiles as they are currently barely resolved with \hst.
An unbiased census for the light distribution of $z\gtrsim2$ quiescent galaxies will enable us to confirm whether morphological transformation and bulge formation is asynchronous with star formation quenching as our work suggests.
\jwst/NIRSpec will provide resolved emission line maps of magnified quiescent galaxies,
allowing us to identify the origin of the ionizing photons.
All of these will provide valuable insights into the origins of star formation quenching.

\acknowledgments{
We thank the referee, Andrew Newman, for a timely and constructive report that improved the quality of this manuscript.
The authors are grateful to the ESO astronomers who assisted in planning and conducting the observations.
We appreciate the CLASH and SURFSUP teams for making their images publicly available. 
AM is grateful to Andrew Zirm for assistance in the early stage of this work.
AM thanks Sirio Belli and Anna Ferr\'{e}-Mateu for providing data from their published figures for comparison.
AM acknowledges helpful discussions with Alison Coil, Roberto Maiolino, Tyrone Woods during the preparation of this work.

AM was supported by a Dunlap Fellowship at the Dunlap Institute for Astronomy \& Astrophysics, funded through an endowment established by the David Dunlap family and the University of Toronto.
The University of Toronto operates on the traditional land of the Huron-Wendat, the Seneca, and most recently, the Mississaugas of the Credit River; AM is grateful to have the opportunity to work on this land.
JZ acknowledges support from ANR grant ANR-17-CE31-0017 (3DGasFlows).
JR acknowledges support from the ERC Starting Grant 336736-CALENDS.
GB, ST, and MS acknowledge support from the ERC Consolidator Grant funding scheme (project ConTExt, grant No. 648179). The Cosmic Dawn Center is funded by the Danish National Research Foundation under grant No. 140.

Based on observations collected at the European Southern Observatory under ESO programmes 087.B-0812, 093.B-0815, 096.B-0994, 097.B-1064, and 099.B-0912.
Based on observations made with the NASA/ESA Hubble Space Telescope, obtained from the data archive at the Space Telescope Science Institute. STScI is operated by the Association of Universities for Research in Astronomy, Inc. under NASA contract NAS 5-26555.
This work is based in part on observations made with the Spitzer Space Telescope, which is operated by the Jet Propulsion Laboratory, California Institute of Technology under a contract with NASA.
This publication makes use of data products from the Wide-field Infrared Survey Explorer, which is a joint project of the University of California, Los Angeles, and the Jet Propulsion Laboratory/California Institute of Technology, and NEOWISE, which is a project of the Jet Propulsion Laboratory/California Institute of Technology. WISE and NEOWISE are funded by the National Aeronautics and Space Administration.
}

\facilities{VLT:Kueyen(X-SHOOTER), HST, Spitzer, WISE}

\software{APLpy \citep{aplpy2012}, AstroPy \citep{astropy:2013,astropy:2018}, Bagpipes \citep{Carnall2018}, Cloudy \citep{cloudy2017}, GALFIT \citep{Peng2002,Peng2010,Peng2011}, grizli \citep{Brammer2019}, Lenstool \citep{Jullo2007}, Matplotlib \citep{matplotlib2007}, Molecfit \citep{Kausch2015,Smette2015}, MultiNest \citep{multinest}, ND-Redshift \citep{Behroozi2013}, Numpy \citep{numpy2006,numpy2011}, 
SExtractor \citep{Bertin1996}, TPHOT \citep{Merlin2015,Merlin2016}}

\newpage

\appendix

\section{Photometry and colors}

\begin{table}[!htbp]
	\centering
	\caption{Photometry of the targets}
	\label{table:phot}
	\begin{tabular}{lccccc}
	    \hline
	    & & \m1341\ (im1) & \m2129 & \m0451\ (im1) & \m1423 \\
	    \hline
	   \textit{HST}/ACS-WFC & F435W & \nodata & 25.809 $\pm$ 0.366 & \nodata & -99 \\ 
            & F475W & \nodata & 25.494 $\pm$ 0.218 & \nodata & 25.169 $\pm$ 0.086 \\ 
            & F555W & \nodata & 25.313 $\pm$ 0.158 & 24.810 $\pm$ 0.069 & 24.426 $\pm$ 0.036 \\ 
            & F606W & 22.645 $\pm$ 0.037 & 24.836 $\pm$ 0.272 & \nodata & 23.790 $\pm$ 0.022 \\ 
            & F625W & \nodata & 24.567 $\pm$ 0.130 & \nodata & \nodata \\ 
            & F775W & \nodata & 23.776 $\pm$ 1.095 & 23.289 $\pm$ 0.188 & 23.331 $\pm$ 0.038 \\ 
            & F814W & 20.791 $\pm$ 0.013 & 23.573 $\pm$ 0.028 & 23.125 $\pm$ 0.013 & 23.198 $\pm$ 0.013 \\ 
            & F850LP & \nodata & 22.910 $\pm$ 0.046 & 23.060 $\pm$ 0.080 & 23.007 $\pm$ 0.023 \\ 
        \textit{HST}/WFC3-IR & F105W & 19.300 $\pm$ 0.009 & 22.122 $\pm$ 0.017 & \nodata & 22.875 $\pm$ 0.014 \\ 
            & F110W & 19.000 $\pm$ 0.009 & 21.339 $\pm$ 0.009 & 22.357 $\pm$ 0.014 & 22.751 $\pm$ 0.015 \\ 
            & F125W & \nodata & 20.944 $\pm$ 0.009 & \nodata & 22.623 $\pm$ 0.013 \\ 
            & F140W & 18.493 $\pm$ 0.003 & 20.442 $\pm$ 0.005 & \nodata & 22.465 $\pm$ 0.012 \\ 
            & F160W & \nodata & 20.163 $\pm$ 0.003 & 20.893 $\pm$ 0.007 & 22.141 $\pm$ 0.006 \\ 
        \textit{IRAC} & CH1 & \nodata & 19.062 $\pm$ 0.133 & 19.937 $\pm$ 0.133 & 20.994 $\pm$ 0.133 \\ 
            & CH2 & \nodata & 18.886 $\pm$ 0.133 & 19.716 $\pm$ 0.133 & 20.730 $\pm$ 0.133 \\ 
        \textit{WISE} & W1 & 17.652 $\pm$ 0.061 & \nodata & \nodata & \nodata \\
            & W2 & 18.414 $\pm$ 0.319 & \nodata & \nodata & \nodata \\
		\hline
	\end{tabular}
	\tablecomments{
    Magnitudes are provided in the AB system.
    Correction for Galactic extinction has not been applied.
    Magnitude -99 indicates negative flux.
	}
\end{table}
\begin{table}[!htbp]
	\centering
	\caption{Rest-frame colors and covariance}
	\label{table:uvj}
	\begin{tabular}{lcccc}
		\hline
		& \m1341 & \m2129\ & \m0451 & \m1423  \\
		\hline
	$(U-V)$ & $1.70\pm0.05$ & $1.76\pm0.04$ & $1.31\pm0.04$ & $0.85\pm0.05$ \\
        $(V-J)$ & $0.95\pm0.09$ & $1.04\pm0.06$ & $0.63\pm0.07$ & $0.59\pm0.11$ \\
        cov($U-V$, $V-J$) & 0.0043 & 0.0020 & 0.0023 & 0.0029 \\
		\hline
	\end{tabular}
\end{table}

\clearpage
\section{Stellar population fits}
   \begin{figure*}[h]
   \centering
   \includegraphics[width=0.8\linewidth]{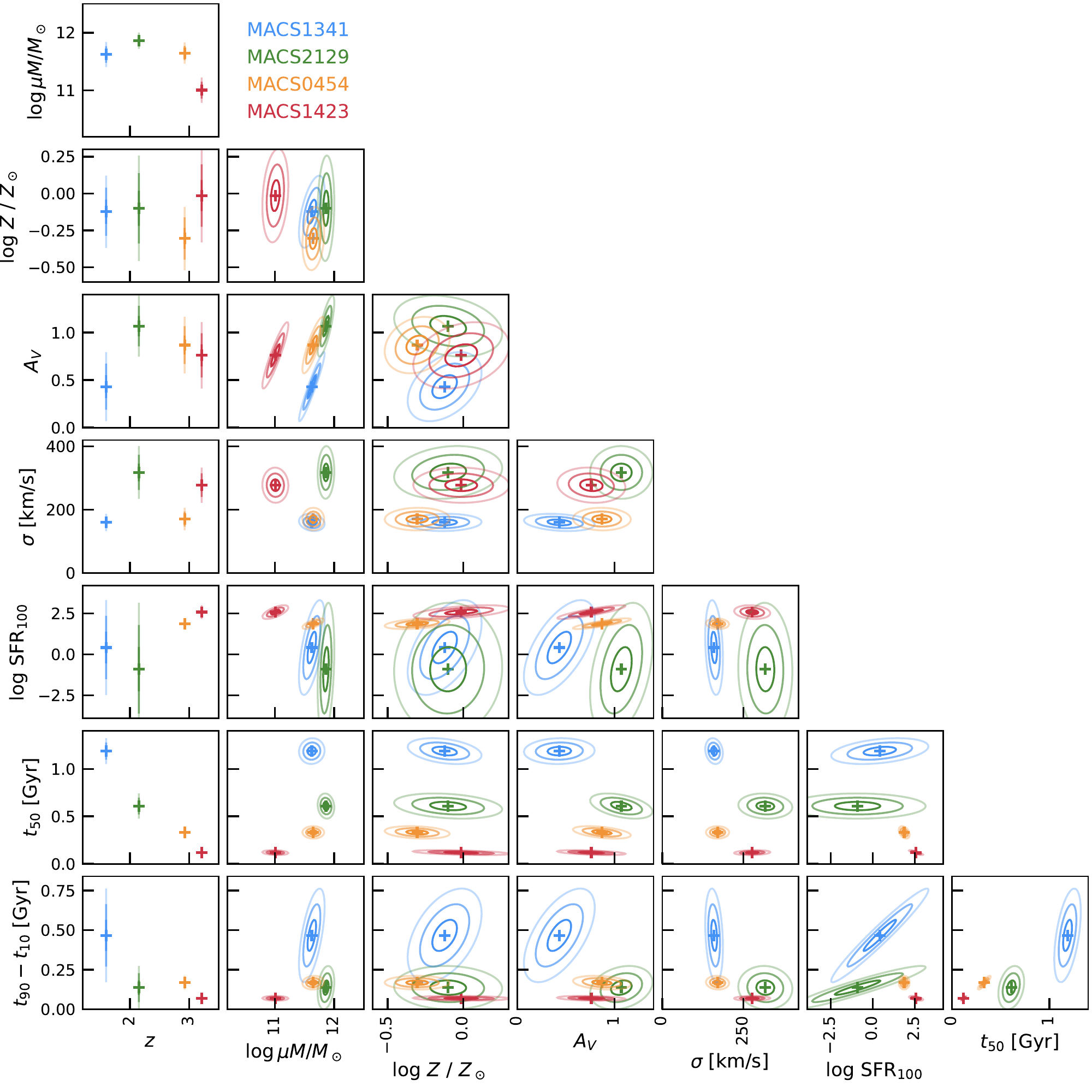}
      \caption{
      Covariance chain for the stellar population fitting.
      More details can be found in \S\ref{sec:specfit}.
      }
    \label{fig:corner_spop}
   \end{figure*}

\begin{figure*}[htbp!]
   \centering
   \includegraphics[width=\linewidth]{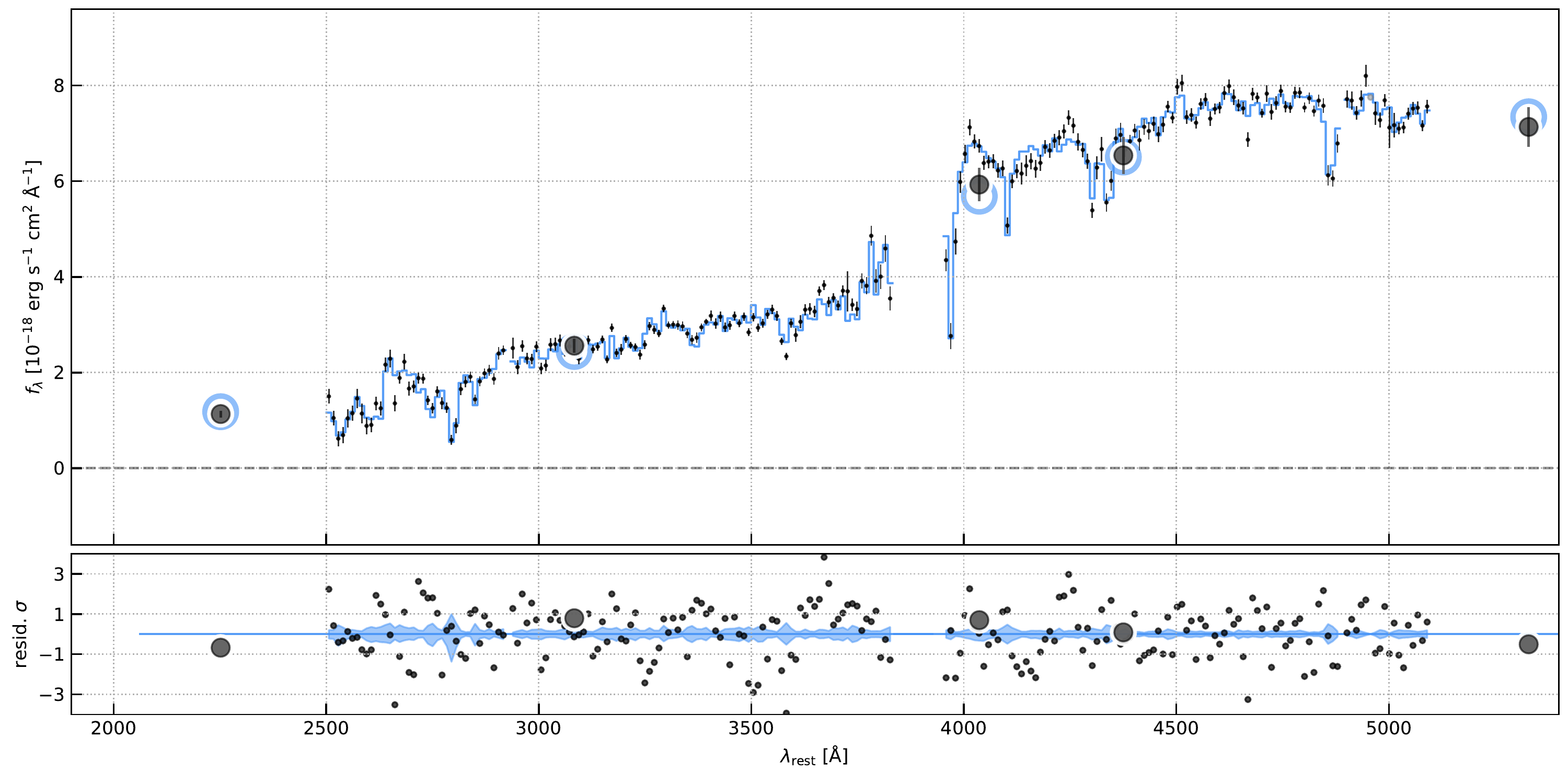}
    \caption{
    Photometry and spectrum of \m1341.
      Photometric data are shown as large black circles.
      The binned VLT/X-SHOOTER spectrum is shown as small black squares.
      The best-fit stellar population model spectra and photometry are shown as colored lines and circles. The bottom panel shows the residual spectrum. 
      }
    \label{fig:sed_m1341}
   \end{figure*}

\begin{figure*}[htbp!]
  \centering
   \includegraphics[width=\linewidth]{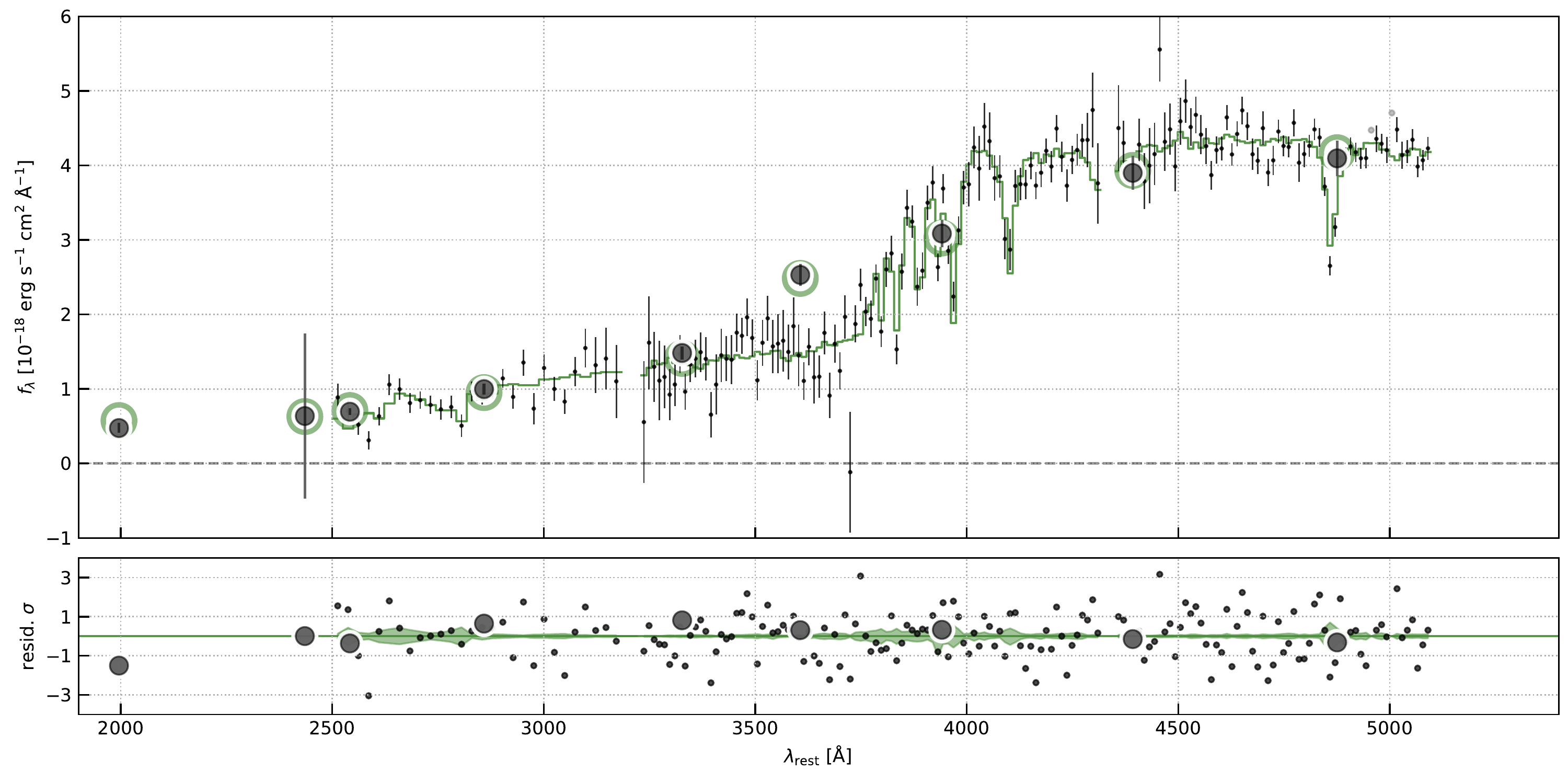}
    \caption{
    Photometry and spectrum of \m2129.
      Refer to Figure~\ref{fig:sed_m1341} for explanations of the symbols.
        }
     \label{fig:sed_m2129}
  \end{figure*}

\begin{figure*}[htbp!]
   \centering
   \includegraphics[width=\linewidth]{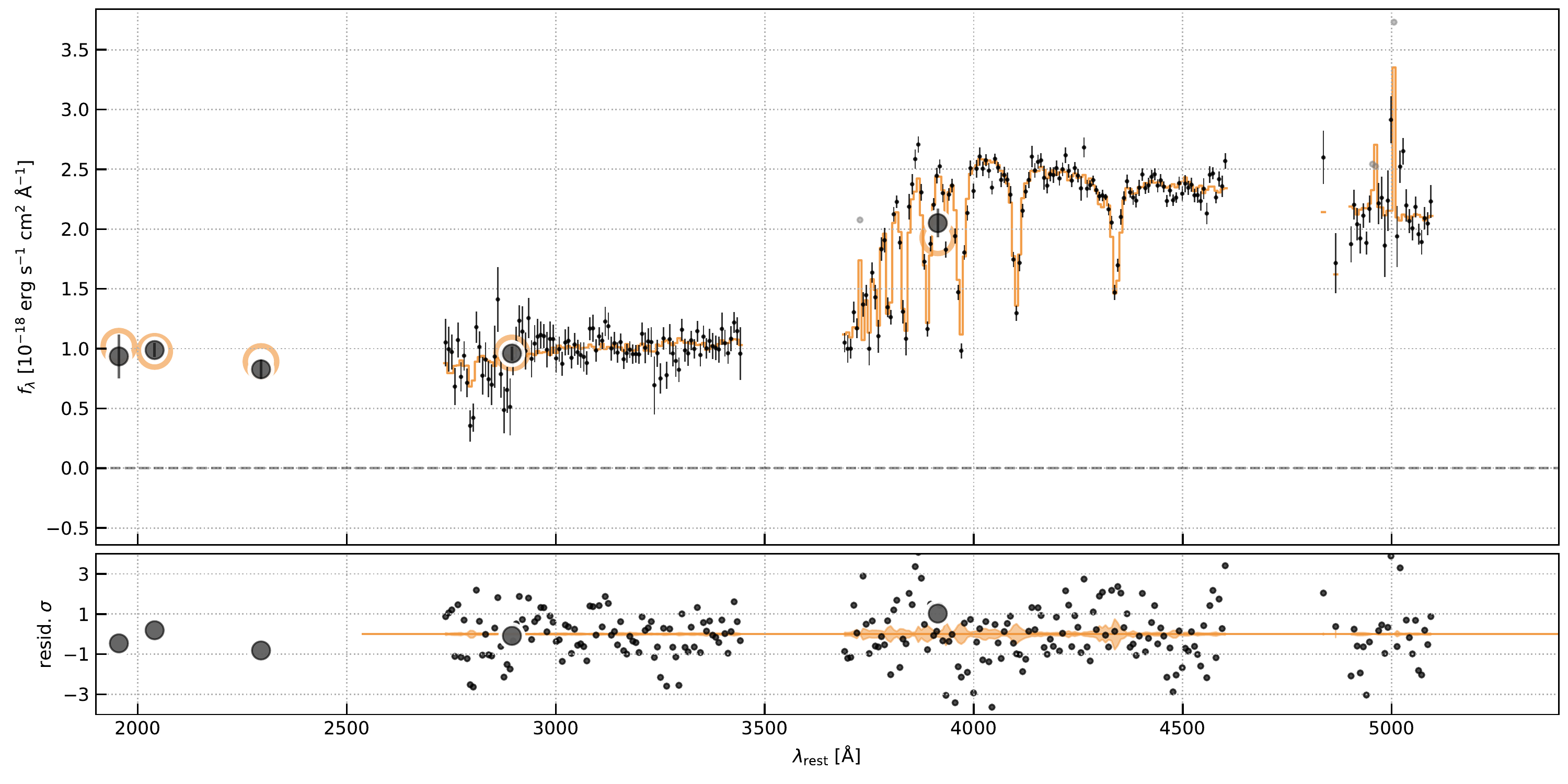}
    \caption{
    Photometry and spectrum of \m0451.
      Refer to Figure~\ref{fig:sed_m1341} for explanations of the symbols.}
     \label{fig:sed_m0454}
   \end{figure*}

\begin{figure*}[htbp!]
   \centering
   \includegraphics[width=\linewidth]{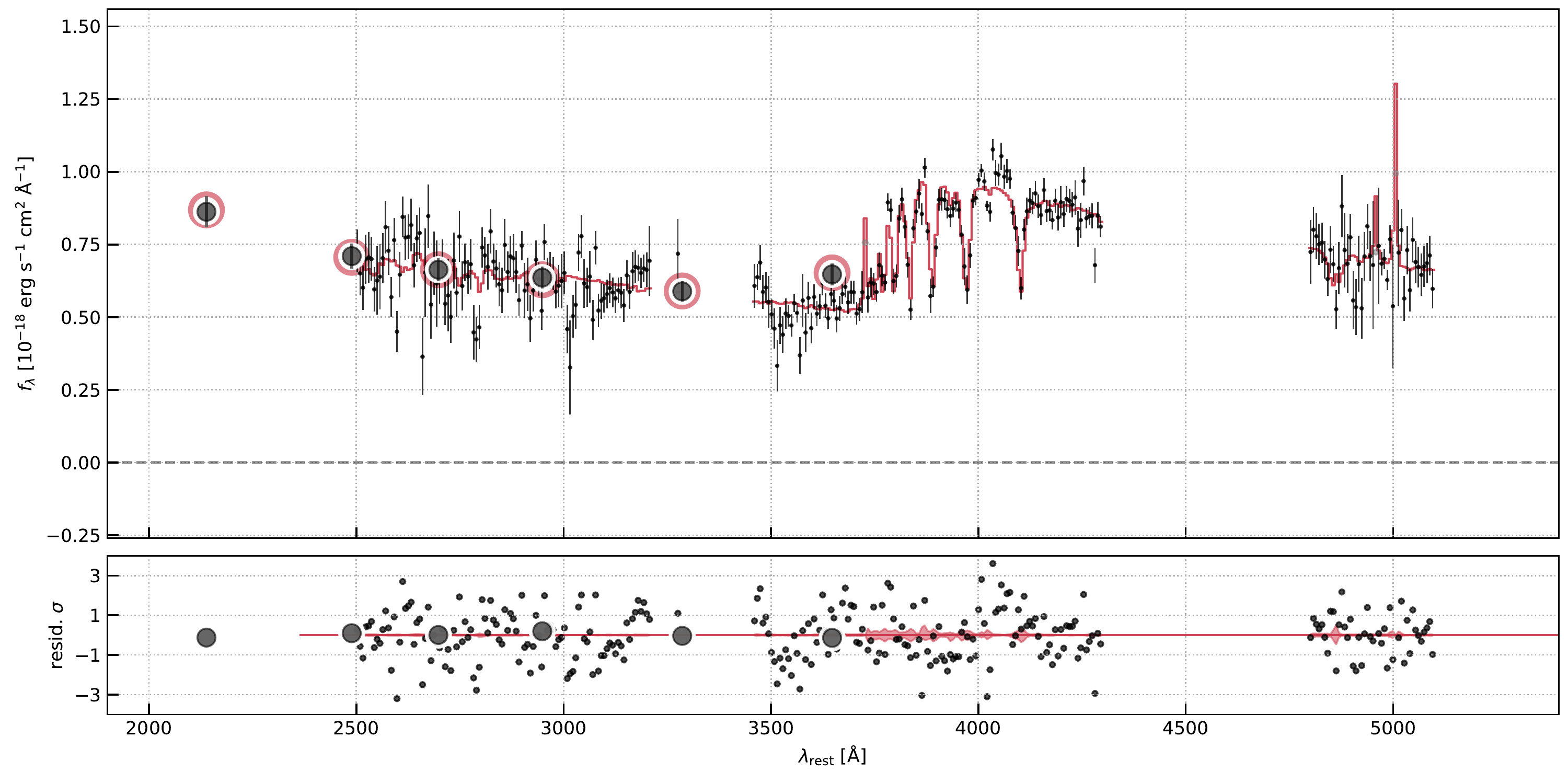}
   \caption{
    Photometry and spectrum of \m1423.
      Refer to Figure~\ref{fig:sed_m1341} for explanations of the symbols.
      }
    \label{fig:sed_m1423}
   \end{figure*}

\clearpage

\section{Comparison of parameters derived for m2129}\label{sec:m2129}

Table~\ref{table:m2129} shows a comparison of the stellar population and morphological parameters of \m2129\ derived in this work with published values from \citet{Toft2017} and \citet{Newman2018a}.
The stellar population parameters are in general agreement, except for the age and stellar metallicity.
The values presented in this work lie between those in \citet{Toft2017} and \citet{Newman2018a}. The substantial uncertainty in stellar metallicity across all three works implies that it is poorly constrained.
The measurement presented in \citet{Toft2017} was derived using a different stellar population fitting procedure:
the light-weighted stellar metallicity in the library varies with the star formation history as a function of stellar mass.
The error bars in \citet{Toft2017} are larger because a large range of model parameters were allowed:
for each parameter the quoted uncertainties are given by the width of the probability distribution function marginalized over all the other 
parameters. 

All three works uniformly point to a compact (\re$\approx 0\farcs3$) disk well-fitted by an exponential disk ($n\approx1$).
The \texttt{GALFIT} parameters reported in \citet{Toft2017} have been revised due to a mistake in the \texttt{GALFIT} input file:
a PSF fine sampling factor of 2 was erroneously used rather than 0.5.
This work reports the rectified \texttt{GALFIT} results.
The revised \re\ and $n$ are similar to the values previously reported in \citet{Toft2017}, and $q\approx0.4$ is lower and thus closer to $q\approx0.3$ reported in \citet{Newman2018a}.
The axis ratio $q$ is not as well-constrained and shows a larger variation across the multiple images, available for two other targets, as discussed in \S\ref{sec:morphology}.
Despite the differences in modeling methods and priors, the resulting stellar population and morphological parameters are in good qualitative agreement. These differences do not affect the conclusions drawn in this work.

\begin{table*}[h]
	\centering
	\caption{
    Comparison of derived parameters of \m2129}
	\label{table:m2129}
	\begin{tabular}{lccc}
		\hline
		& This work & \citet{Toft2017} & \citet{Newman2018a} \\
	    \hline
       \zspec & $2.1487\pm0.0002$ & $2.1478\pm0.0006$ & 2.1487 \\
       $\mu$ & $4.6\pm0.2$ & $4.6\pm0.2$ & 4.5 \\
       log(\mstar/\msun) & $11.20\pm0.05$ & $11.10^{+0.26}_{-0.20}$ & $10.96\pm0.10$ \\
       log(\Zstar/\Zsun) & $-0.10\pm0.11$ & $-0.6\pm0.5$ & $0.16\pm0.13$ \\
       \tage\ (Gyr) &$0.61\pm0.05$ (\thalf) & $1.05^{+0.81}_{-0.46}$ (light-weighted) & $0.80\pm0.10$ \\
       SFR (\msun\peryr) & $0.03^{+0.61}_{-0.03}$ & $0.0^{+0.2}_{-0.0}$ & $0.6\pm0.4$ \\
       \re ($\arcsec$) & $0.26\pm0.01$ & $\dagger0.27^{+0.05}_{-0.04}$ & $0.29\pm0.02$ \\
       $n$ & $1.17\pm0.01$ & $\dagger1.01^{+0.12}_{-0.06}$ & $1$ \\
       $q$ & $0.42\pm0.01$ & $\dagger0.59^{+0.03}_{-0.09}$ & $0.29\pm0.03$ \\
		\hline
	\end{tabular}
	\tablecomments{
	This work assumes the \citet{Kroupa2001} IMF, while \citet{Chabrier2003} was assumed in both \citet{Toft2017} and \citet{Newman2018a}.
	Library B, Fit 1 (delayed-tau model and fitted over entire spectrum and photometry) is taken for \citet{Toft2017} values. Single \sersic\ model value of \citet{Newman2018a} is shown here. The dagger symbols mark published values affected by an error as described in the text.
    }
\end{table*}

\section{Details on morphological analysis}

\begin{figure*}[htbp!]
    \begin{minipage}[b]{0.48\linewidth}
	\centering
	\includegraphics[width=\linewidth]{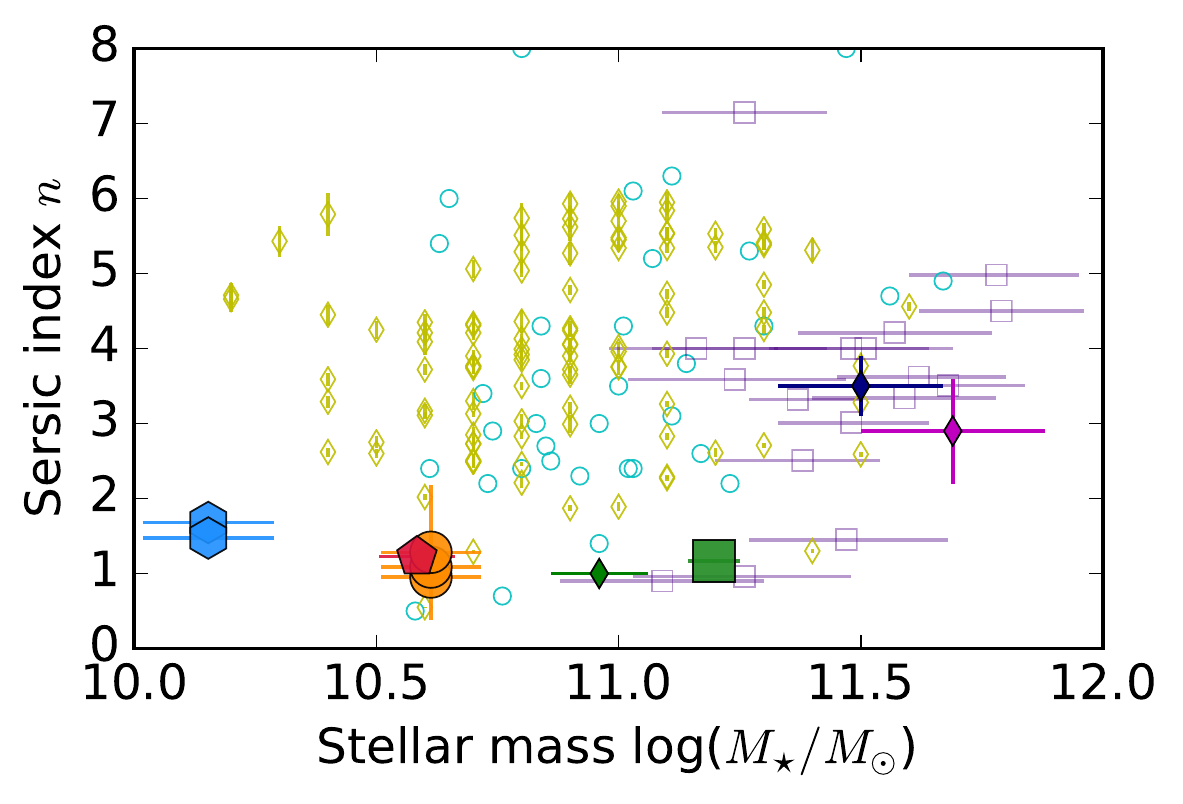}
	\end{minipage}
    \begin{minipage}[b]{0.5\linewidth}
	\centering
	\includegraphics[width=\linewidth]{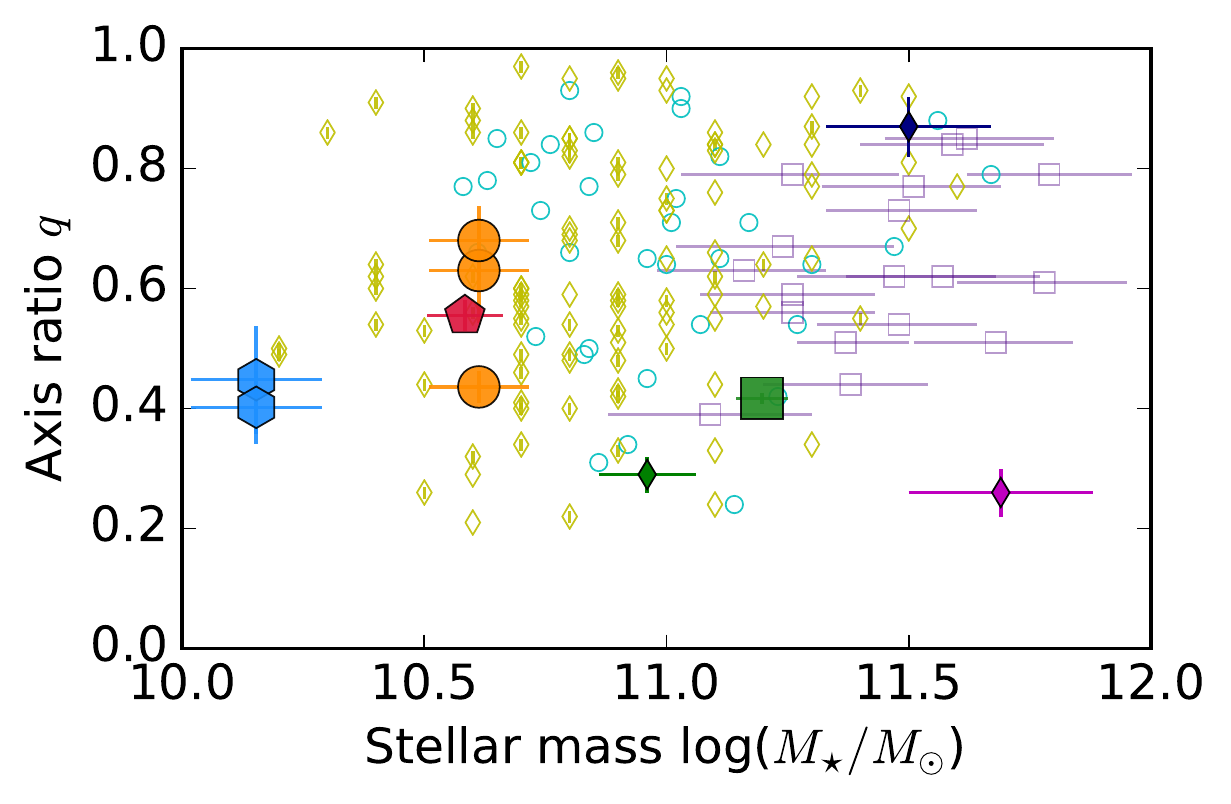}
	\end{minipage}
    \caption{
    The stellar mass dependence of \sersic\ indices and axis ratios.
    The symbols follow those of Figure~\ref{fig:n_q}.
    }
    \label{fig:n_q_Mstar}
   \end{figure*}

\bibliographystyle{aasjournal}

\end{document}